\documentclass[twocolumn]{aastex631}
\usepackage{hyperref}
\usepackage{graphicx}
\usepackage{natbib}
\usepackage{upgreek}
\usepackage{xurl}
\shorttitle{WASP-17\,{\rm b}'s JWST NIRISS SOSS Emission Spectrum}
\shortauthors{A. Gressier, et al.}
\accepted{for publication in the Astronomical Journal, October 4, 2024.}

\begin{document}

% \title{JWST-TST DREAMS: Detection of H$_2$O and Evidence of FeH in WASP-17\,b's Day-side Atmosphere from NIRISS SOSS Eclipse Spectroscopy}

\title{JWST-TST DREAMS: A Super-Solar Metallicity %and Evidence of FeH 
in WASP-17\,b's Dayside Atmosphere \\ from NIRISS SOSS Eclipse Spectroscopy}

%IMPORTANT NOTE ON AUTHORSHIP: Please refer to the authorship guidelines presented in the TelSci GTO Team Policies: 
%https://drive.google.com/file/d/1X_6jihO8Y3u98glUefXXcs7VU_5HH3t9/view?usp=share_link
%And authorship clarifications/expansions for the Transiting Exoplanet Spectroscopy subteam outlined here:
%https://docs.google.com/document/d/16eXlmx4QMDNEkNbtTLZiuAytTikqzhLWb1EkRZCP5IM/edit?usp=sharing
%To be included as a co-author on a paper, an individual shall at a minimum: (a) review the paper; (b) agree explicitly to be listed as a co-author; and (c) provide any relevant feedback for consideration by the other authors.

%Project Leads and project/science team members who led/contributed significantly to work presented
%Group 0
\author[0000-0003-0854-3002]{Am\'{e}lie Gressier}
\affiliation{Space Telescope Science Institute, 3700 San Martin Drive, Baltimore, MD 21218, USA}

\author[0000-0003-4816-3469]{Ryan J. MacDonald}
\affiliation{Department of Astronomy, University of Michigan, 1085 S. University Ave., Ann Arbor, MI 48109, USA}

\author[0000-0001-9513-1449]{N\'{e}stor Espinoza}
\affiliation{Space Telescope Science Institute, 3700 San Martin Drive, Baltimore, MD 21218, USA}
\affiliation{William H. Miller III Department of Physics and Astronomy, Johns Hopkins University, Baltimore, MD 21218, USA}

\author[0000-0003-4328-3867]{Hannah R. Wakeford}
\affiliation{University of Bristol, HH Wills Physics Laboratory, Tyndall Avenue, Bristol, UK}

\author[0000-0002-8507-1304]{Nikole K. Lewis}
\affiliation{Department of Astronomy and Carl Sagan Institute, Cornell University, 122 Sciences Drive, Ithaca, NY 14853, USA}

%Group 1
\author[0000-0002-8515-7204]{Jayesh Goyal}
\affiliation{School of Earth and Planetary Sciences (SEPS), National Institute of Science Education and Research (NISER), HBNI, Odisha, India}

\author[0000-0002-2457-272X]{Dana R. Louie}
\affiliation{Catholic University of America, Department of Physics, Washington, DC, 20064, USA}
\affiliation{Exoplanets and Stellar Astrophysics Laboratory (Code 667), NASA Goddard Space Flight Center, Greenbelt, MD 20771, USA}
\affiliation{Center for Research and Exploration in Space Science and Technology II, NASA/GSFC, Greenbelt, MD 20771, USA}

\author[0000-0002-3328-1203]{Michael Radica}
\affiliation{Institut Trottier de recherche sur les exoplanètes and Département de Physique, Université de Montréal, 1375 Avenue Thérèse-Lavoie-Roux, Montréal, QC, H2V 0B3, Canada}

%Group 2
\author[0000-0003-1240-6844]{Natasha E. Batalha}
\affiliation{NASA Ames Research Center, Moffett Field, CA, 94035, USA}

\author[0000-0002-2508-9211]{Douglas Long}
\affiliation{Space Telescope Science Institute, 3700 San Martin Drive, Baltimore, MD 21218, USA}

\author[0000-0002-2739-1465]{Erin M. May}
\affiliation{Johns Hopkins APL, 11100 Johns Hopkins Rd, Laurel, MD 20723, USA}

\author[0000-0003-0814-7923]{Elijah Mullens}
\affiliation{Department of Astronomy and Carl Sagan Institute, Cornell University, 122 Sciences Drive, Ithaca, NY 14853, USA}

\author[0000-0002-6892-6948]{Sara Seager}
\affiliation{Department of Earth, Atmospheric and Planetary Sciences, Massachusetts Institute of Technology, Cambridge, MA 02139, USA}
\affiliation{Kavli Institute for Astrophysics and Space Research, Massachusetts Institute of Technology, Cambridge, MA 02139, USA} 
\affiliation{Department of Aeronautics and Astronautics, MIT, 77 Massachusetts Avenue, Cambridge, MA 02139, USA}

\author[0000-0002-7352-7941]{Kevin B. Stevenson}
\affiliation{Johns Hopkins APL, 11100 Johns Hopkins Rd, Laurel, MD 20723, USA}

\author[0000-0003-3305-6281]{Jeff A. Valenti}
\affiliation{Space Telescope Science Institute, 3700 San Martin Drive, Baltimore, MD 21218, USA}

%Group 3
\author[0000-0001-8703-7751]{Lili Alderson}
\affiliation{University of Bristol, HH Wills Physics Laboratory, Tyndall Avenue, Bristol, UK}

\author[0000-0002-0832-710X]{Natalie H. Allen}
\affiliation{William H. Miller III Department of Physics and Astronomy, Johns Hopkins University, Baltimore, MD 21218, USA}

\author[0000-0003-4835-0619]{Caleb I. Ca\~{n}as}
\affiliation{NASA Goddard Space Flight Center, Greenbelt, MD 20771, USA}

\author[0000-0002-8211-6538]{Ryan C. Challener}
\affiliation{Department of Astronomy and Carl Sagan Institute, Cornell University, 122 Sciences Drive, Ithaca, NY 14853, USA}

\author[0000-0001-8020-7121]{Knicole Col\'{o}n}
\affiliation{NASA Goddard Space Flight Center, Greenbelt, MD 20771, USA}

\author[0000-0002-5322-2315]{Ana Glidden}
\affiliation{Department of Earth, Atmospheric and Planetary Sciences, Massachusetts Institute of Technology, Cambridge, MA 02139, USA}

\author[0000-0001-5878-618X]{David Grant}
\affiliation{University of Bristol, HH Wills Physics Laboratory, Tyndall Avenue, Bristol, UK}

\author[0000-0001-5732-8531]{Jingcheng Huang}
\affiliation{Department of Earth, Atmospheric and Planetary Sciences, Massachusetts Institute of Technology, Cambridge, MA 02139, USA}

\author[0000-0003-0525-9647]{Zifan Lin}
\affiliation{Department of Earth, Atmospheric and Planetary Sciences, Massachusetts Institute of Technology, Cambridge, MA 02139, USA}

\author[0000-0002-2643-6836]{Daniel Valentine}
\affiliation{University of Bristol, HH Wills Physics Laboratory, Tyndall Avenue, Bristol, UK}

%Group 4
\author{C. Matt Mountain}
\affiliation{Association of Universities for Research in Astronomy, 1331 Pennsylvania Avenue NW Suite 1475, Washington, DC 20004, USA}

\author{Laurent Pueyo}
\affiliation{Space Telescope Science Institute, 3700 San Martin Drive, Baltimore, MD 21218, USA}

\author[0000-0002-3191-8151]{Marshall D. Perrin}
\affiliation{Space Telescope Science Institute, 3700 San Martin Drive, Baltimore, MD 21218, USA}

\author[0000-0001-7827-7825]{Roeland P. van der Marel}
\affiliation{Space Telescope Science Institute, 3700 San Martin Drive, Baltimore, MD 21218, USA}
\affiliation{William H. Miller III Department of Physics and Astronomy, Johns Hopkins University, Baltimore, MD 21218, USA}

\begin{abstract}

We present the first emission spectrum of the hot Jupiter WASP-17\,b using one eclipse observation from the JWST Near Infrared Imager and Slitless Spectrograph (NIRISS) Single Object Slitless Spectroscopy (SOSS) mode. Covering a wavelength range of 0.6 to 2.8\,µm, our retrieval analysis reveals a strong detection of H$_2$O in WASP-17\,b's dayside atmosphere (6.4$\sigma$). Our retrievals consistently favor a super-solar dayside H$_2$O abundance and a non-inverted temperature-pressure profile over a large pressure range. Additionally, our examination of the brightness temperature reveals excess emission below 1\,µm, suggesting the possibility of a high internal temperature (600 to 700 K) and/or contributions from reflected light. We highlight that JWST emission spectroscopy retrieval results can be sensitive to whether negative eclipse depths are allowed at optical wavelengths during light curve fitting. Our findings deepen our understanding of WASP-17\,b's atmospheric composition while also highlighting the sensitivity of our results to pressure-temperature profile parameterizations. This work is part of a series of studies by our JWST Telescope Scientist Team (JWST-TST), in which we will use Guaranteed Time Observations to perform Deep Reconnaissance of Exoplanet Atmospheres through Multi-instrument Spectroscopy (DREAMS).
\end{abstract}

\section{Introduction} \label{sec:intro}

Observing hot Jupiter exoplanets during the secondary eclipse allows the measurement of the dayside thermal flux \citep{Charbonneau_2005, Deming_2005}. Most of the thermal emission observed from exoplanets originates from re-radiated light from the star. Thus, we can estimate the flux of the planet relative to the flux of the star using the blackbody equation: \( \frac{F_p}{F_s}=\frac{B(T_p)}{B(T_s)}\left(\frac{R_p}{R_s}\right)^2 \). This ratio is higher at longer wavelengths, as the planet typically has a lower temperature than the star. Deviations from blackbody fluxes allow the identification of emission features in the spectrum. The emission observed during the exoplanet secondary eclipse originates from the planet's lower atmosphere; however, at longer wavelengths with larger optical depths, the photosphere can be located at higher altitudes where the temperatures may vary from the equilibrium temperature. The thermal flux, but also the size and shape of the features, are driven by the temperature-pressure profile of the atmosphere. Observations of planetary occultations are thus highly meaningful and valuable, as they provide a significant amount of information about the thermal structure of the planet's atmosphere, which inherently depends on the geometric albedo, the atmospheric opacities, and the heat re-distribution. 

Space telescopes, such as the Spitzer and Hubble Space Telescopes, have played a central role in characterizing hot Jupiter daysides via emission spectra. Initially, single studies and surveys focused on investigating thermal inversions in hot Jupiters by comparing the eclipse depths found at the two Spitzer IRAC channels at 3.6 and 4.5\,$\upmu$m \citep[e.g.,][]{Anderson_2011, Stevenson_2014a, Deming_2015, Baxter_2020, Goyal2021}. Subsequently, the HST WFC3 G141 grism complemented Spitzer's studies by enabling spectroscopic emission observations. This allowed for the detection of H$_2$O at 1.4\,$\upmu$m for many hot giant exoplanets \citep{Kreidberg_2014, Stevenson_2014b, Pluriel_2020, Changeat_2022}, as well as strong optical absorbers such as TiO and VO, which are indicative of thermal inversions \citep{haynes_2015, Mansfield_2018, Arcangeli_2018, Mikal_Evans_2019, Edwards_2020, Changeat_2021}. We can now complement these observations using JWST instrument modes to explore a wider range of wavelengths. Specifically, the precision of the JWST NIRISS SOSS mode \citep{Albert2023,Doyon2023} in the optical could build on the pioneering insights from Spitzer and Hubble by revealing reflected light at shorter wavelengths where it dominates over thermal emission (with a potential signal amplitude of 100\,ppm; \citealt{Angerhausen_2015}).

% For a typical hot Jupiter, like HD\,209458\,b, the flux ratio during the secondary eclipse is around 50\,ppm in the optical and reaches 1000\,ppm at 4.5\,$\upmu$m \citep{Line_2016, Kreidberg_2018}. This subtle signal presents a challenge in terms of detectability. 

The NIRISS SOSS instrument's mode \citep{Albert2023,Doyon2023} has proven highly effective in advancing our understanding of hot Jupiters in transmission \citep[e.g.,][]{Feinstein_2023, Radica2023} and Ultra-Hot Jupiters in emission \citep{Coulombe2023}. Its precision has facilitated the robust detection of H$_2$O, along with tentative identifications of TiO and VO, while also enabling the mapping of the dayside brightness temperature in WASP-18\,b \citep{Coulombe2023}. Expanding on these studies, we now present observations of WASP-17\,b's atmosphere using NIRISS SOSS, introducing the first emission spectrum obtained through a single NIRISS SOSS eclipse observation for a classic Hot Jupiter, which shifts our focus to a planet with a lower temperature compared to previously studied Ultra-Hot Jupiters.

WASP-17\,b is a hot Jupiter with an equilibrium temperature of approximately 1700\,K. Tidally locked, the planet exhibits a significantly high scale height of around 2,000\,km and an inflated atmosphere, making it a perfect target for transmission and emission spectroscopy. The permanent dayside irradiation suggests a significant day-to-night gradient in temperature and atmospheric composition. The planet is also in a retrograde orbit to its rotation \citep{Triaud_2010}, suggesting a high internal temperature. While well-studied in transmission \citep{Mandell_2013, Sing_2015, Alderson2022, grant_2023}, the emission characteristics of WASP-17\,b are less explored. \citet{Mandell_2013} first identified water absorption using HST WFC3 G141 transit observations. This was then complemented by HST STIS G430L and G750L and Spitzer photometric measurements, integrated into the hot Jupiter survey led by \citet{Sing_2015}. Sodium and water were identified in a cloudless atmosphere. The amplitude of the feature in the water band reaches 90$\%$, presenting one of the largest amplitudes of the sample. A re-analysis of the Hubble, TESS, and Spitzer transit observations of WASP-17\,b led by \citet{Alderson2022} confirmed the detection of water at 7$\sigma$ and found evidence of carbon dioxide at 3$\sigma$ but did not confirm the presence of potassium and sodium. The study also highlights the bimodal distribution in the water's abundance, leading to a degeneracy in the inferred atmospheric metallicity, which we hope to solve using JWST's observations. The addition of the first observations of WASP-17\,b using JWST MIRI LRS by \citet{grant_2023} favored a depletion of water in a super-solar atmosphere. This low abundance of water coupled with a super-solar C/O ratio is indicative of the formation of high-temperature aerosols consistent with the detection of quartz condensation in MIRI observations. Oxygen may have been depleted from the gas phase and condensed into SiO$_2$(s) clouds.

The present paper introduces the spectroscopic emission spectrum of WASP-17\,b, obtained through a single NIRISS SOSS eclipse observation covering the wavelength range from 0.6 to 2.8\,µm. Until now, the dayside of WASP-17\,b's atmosphere has been characterized with only two photometric points using Spitzer, revealing brightness temperatures of 1881$\pm$50 K and 1580$\pm$150\,K at 4.5 and 8\,µm, respectively. These temperatures indicate a low-albedo planet, efficiently redistributing heat from its day side to its night side \citep{Anderson_2011}.

Our paper is organized as follows. We first present the observations in Section\,\ref{sec:obs}. To ensure the accuracy of the final emission spectrum, we use three different reduction pipelines, providing a comprehensive overview of the steps involved in the reduction and the light curve fitting. We detail the steps of data reduction for the \text{\tt transitspectroscopy} \citep{espinoza_nestor_2022}, the \text{\tt Supreme-SPOON} \citep{Feinstein_2023, Radica2023}, and the \text{\tt Ahsoka} pipeline (Louie et al. 2024, submitted), in respectively, Sections\,\ref{sec:2.1},\ref{sec:2.2}, and \ref{sec:2.3}.
The interpretation of the emission spectrum involves a combination of forward model grid-fitting using \text{\tt ATMO} \citep{Tremblin2015, Amundsen2014,Drummond2016, Goyal2018}, and, free retrieval analysis presented respectively in Section\,\ref{sec:atmo} and Section\,\ref{sec:retrieval}. \text{\tt TauREx} \citep{Al_Refaie_2021} retrieval analysis method is presented in Section\,\ref{sec:retrieval:taurex}, while \text{\tt POSEIDON}'s \citep{MacDonald_2017, MacDonald_2023} is in Section\,\ref{sec:retrieval:poseidon}. Subsequently, the results of the forward modeling, retrieved molecular abundances, and temperature/pressure profile are discussed in Section\,\ref{sec:discussion}. The planet's flux over the star's flux is converted into brightness temperature to illustrate the deviation of our results from a blackbody. We finally discuss the observed emission excess and its implications in terms of interior temperature and albedo.

This paper is part of a series by the JWST Telescope Scientist Team (JWST-TST)\footnote{\url{https://www.stsci.edu/$\sim$marel/jwsttelsciteam.html}} which uses Guaranteed Time Observer (GTO) time awarded by NASA in 2003 (PI Matt Mountain) for studies in three different subject areas: (a) Transiting Exoplanet Spectroscopy (lead: N. Lewis); (b) Exoplanet and Debris Disk Coronagraphic Imaging (lead: M. Perrin); and (c) Local Group Proper Motion Science (lead: R. van der Marel). Here we present new results in the first area; previously reported results across these areas include \citet{grant_2023}; \citet{ruffio_2023} and \citet{rebollido_2024}; and \citet{Libralato_2023}, respectively.
A common theme of these investigations is the desire to pursue and demonstrate science for the astronomical community at the limits of what is made possible by the exquisite optics and stability of JWST. %The team, led by M. Mountain, was convened in 2002, following a competitive NASA selection process. In addition to providing scientific support for observatory development through launch and commissioning \citep[e.g.][]{Perrin2014}, the team was awarded 210 hours of Guaranteed Time Observer (GTO) time. This time is being used over the first three JWST observing cycles for studies in three different subject areas: (a) Transiting Exoplanet Spectroscopy (lead: N. Lewis); (b) Exoplanet and Debris Disk Coronagraphic Imaging (lead: M. Perrin); and (c) Local Group Proper Motion Science (lead: R. van der Marel). A common theme of these investigations is the desire to pursue and demonstrate science for the astronomical community at the limits of what is made possible by the exquisite optics and stability of JWST. 
The present paper is part of our work on Transiting Exoplanet Spectroscopy, which focuses on detailed exploration of three transiting exoplanets representative of key exoplanet classes: Hot Jupiters (WASP-17\,b, GTO~1353), Warm Neptunes (HAT-P-26b, GTO~1312), and Temperate Terrestrials (TRAPPIST-1e, GTO~1331). 
Here we present WASP-17\,b's NIRISS SOSS emission spectrum.

\section{Observations} \label{sec:obs}

WASP-17\,b was observed during the eclipse using the NIRISS SOSS \citep{Doyon2023, Albert2023} mode as part of the JWST GTO cycle 1 Program 1353, led by PI N. Lewis.  The observations started on UTC March 22$^{\rm nd}$ 2023, covered the full eclipse (4.42 hours) and sufficient baseline for a total of 9.89 hours, to account for the eclipse time uncertainty and the observation start window. The time series observation comprised 720 integrations, each composed of 8 groups, with an exposure time of 49.46 seconds per integration. It used the SUBSTRIP256 subarray (2048$\times$256 pixels) with the GR700XD/CLEAR filter and the detector displaying three diffraction orders of the target (see Figure\,\ref{fig:frames}). An additional exposure was performed using the F277W filter, limiting the SOSS wavelength range above 2.6\,µm. This exposure lasted 0.137 seconds using 10 integrations and the same exposure time. 

The reduction of WASP-17\,b's NIRISS SOSS eclipse observations was done with three pipelines, among them, \text{\tt transitspectroscopy} and \text{\tt supreme-SPOON} have already been benchmarked for the JWST Early Release Science (ERS) program \citep{Feinstein_2023} and used in recent studies \citep{Radica2023, Coulombe2023}.

\subsection{Data reduction with \text{\tt transitspectroscopy}} \label{sec:2.1}
We carried out a data reduction using the \texttt{transitspectroscopy} pipeline \citep{espinoza_nestor_2022}\footnote{\url{https://github.com/nespinoza/transitspectroscopy}}. This pipeline uses the stage 1 \textit{.rateints.fits} files obtained from the \texttt{jwst} pipeline v.1.8.2\footnote{\url{https://jwst-pipeline.readthedocs.io/en/latest/jwst/pipeline/calwebb_detector1.html}}. The first steps and corrections are applied during this process with default settings, including the group scale correction, data quality initialization, saturation detection, superbias subtraction, reference pixel correction, linearity correction, dark current subtraction, jump detection, ramp-fitting step, and gain scale correction.

\texttt{transitspectroscopy} is then used to produce Order 1 and Order 2 trace positions, subtract the zodiacal background, correct for the 1/f noise and order 0 contaminations, extract the stellar spectrum and produce light curves. These steps are detailed below.

\begin{figure*}[htpb!]
\centering
  \includegraphics[width=\textwidth]{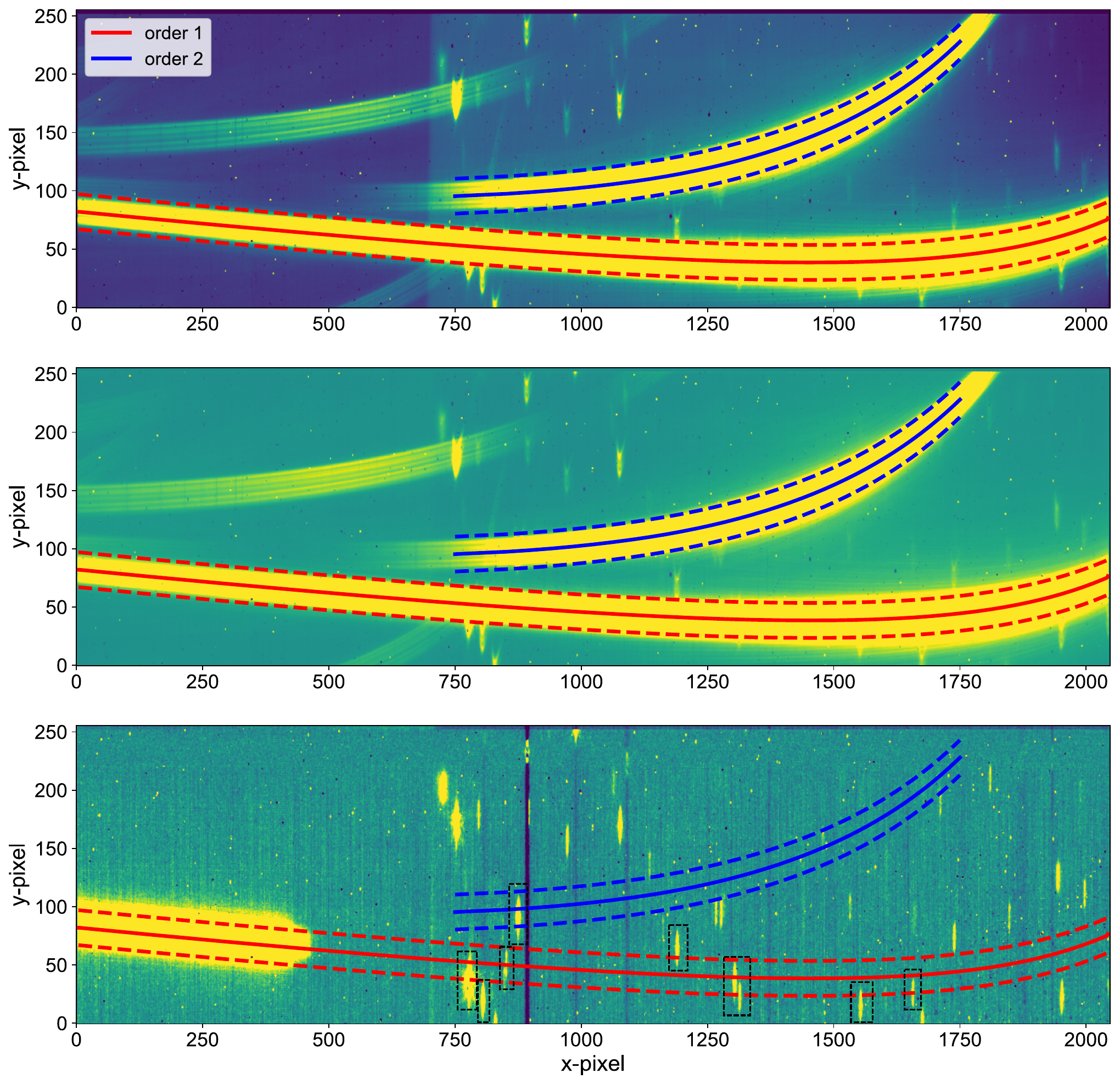}
 \caption{Top: Out-of-transit median frame of the NIRISS SOSS CLEAR exposure from \text{\tt transitspectroscopy}'s reduction. The positions of the traces for Order 1 in blue and Order 2 in red are shown along with the 15 pixels radius extraction aperture. \textit{Middle}: The same out-of-transit median frame of the NIRISS SOSS CLEAR exposure but now corrected for the zodiacal background. Bottom: Median frame of the NIRISS SOSS F277W exposure. Order 0 contaminants corrected in our analysis are framed. Around pixel 750  we can distinguish three field star contaminants impacting Order 1.}
 \label{fig:frames}
\end{figure*}

\textbf{Trace order positions.}
The positions of the traces for NIRISS SOSS Order 1 and order 2 are obtained using the \texttt{transitspectroscopy.trace\_spectrum} routine. This function returns the center position of the trace by getting the maximum of a cross-correlation input function with each column of the detector. For tracing the NIRISS SOSS order profiles, we use a double Gaussian input function with the following parameters: $\upmu_1 = -7.5$, $\upsigma_1 = 3.0$, $\upmu_2 = 7.5$, and $\upsigma_2 = 3.0$. These parameter values were derived from the NIRISS SOSS observations of HAT-P-14\,b (JWST Program ID 1541; PI Espinoza). To trace the position, we consider the x-pixels range of 4 to 2043 for Order 1 and 750 to 1750 for Order 2. Due to the low SNR of order 2 the trace does not extend as far as Order 1 on the detector, we also truncate the order 2 to avoid the inclusion of contaminants in the spectrum. %Order 2 is shorter due to the low throughput and thus the method not confidently tracing the order but also to avoid the inclusion of contaminants. 
Next, we apply smoothing to the trace positions using spline functions. For Order 1, the spline parameters are set as x${knots, 1}$ = [[6, 1200-5], [1200, 1500-5], [1500, 1700-5], [1700, 2041]] and n${knots, 1}$ = [4, 2, 3, 4]. Similarly, for Order 2, the spline parameters are x${knots, 2}$ = [[751, 850-5], [850, 1100-5], [1100, 1749]] and n${knots, 2}$ = [2, 2, 5]. The smoothed trace positions are shown in the top panel of Figure\,\ref{fig:frames} along with the 15 pixels radius extraction. \\

\textbf{Correct zodiacal background.}
The NIRISS SOSS detector background exhibits a brightness step near pixel 750 in the spectral direction, (i.e., in the x-direction) (see top panel Figure\,\ref{fig:frames}). This characteristic shape arises from the zodiacal background decreasing at the 'pick-off mirror' region. To eliminate the zodiacal background, we scale the model background provided in the STScI JDox User Documentation to match the observations. We use a small portion of the median integrations from the NIRISS SOSS subarray [x\,=\,220:250, y\,=\,500:850]. This chosen portion encompasses the `pick-off mirror' region and has no order 0 contaminants. We determine the scaling factor by computing the ratio between the pixels in this portion and those in the background model. We assess the median ratio of pixels within the second quartile. This analysis yields a scaling factor of 0.821, which we subtract from all integrations to achieve consistent background correction.  \\

\begin{figure}[htpb!]
  \includegraphics[width=\columnwidth]{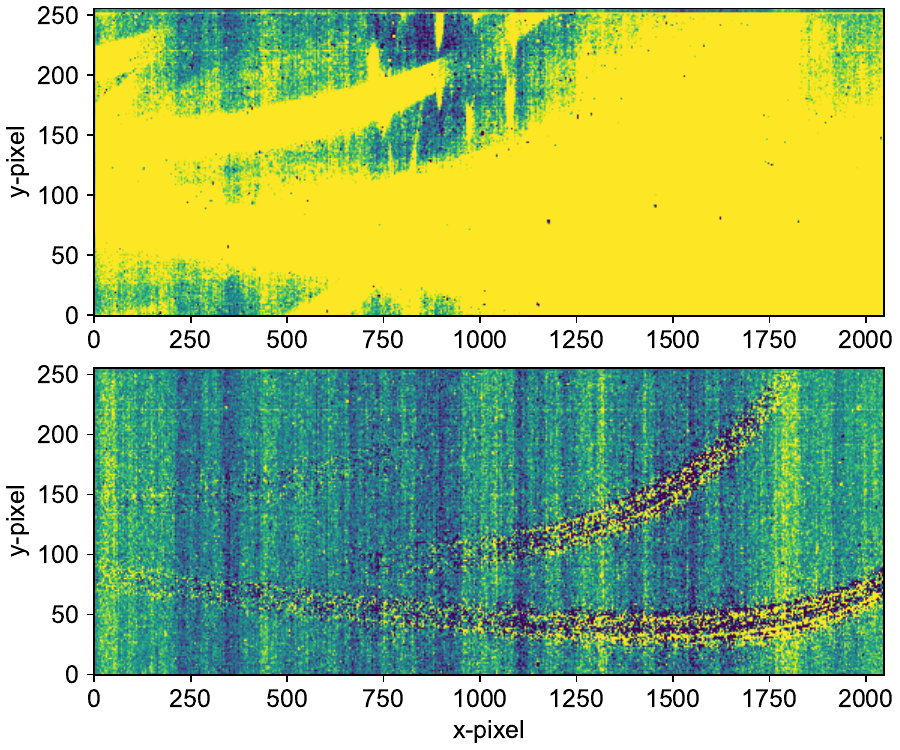}
 \caption{Top: Individual frame of the NIRISS SOSS CLEAR exposure corrected for the zodiacal background from \text{\tt transitspectroscopy}'s reduction. Bottom: Same individual frame with the out-of-transit median frame subtracted. The residuals enable us to identify the 1/f noise banding. We use the out-of-transit median frame to correct each individual frame.}
 \label{fig:1_fnoise}
\end{figure}

\textbf{Correct the 1/f noise.} 
The 1/f noise is corrected in each individual integration frame through the following procedure. First, we create an out-of-transit median frame by combining integrations from 0 to 200 and 600 to 720 with the zodiacal background already corrected (see Figure\,\ref{fig:frames}, middle panel). This median out-of-transit frame is then subtracted from each frame, resulting in frames that are left with only detector-level effects such as the 1/f noise. We represent an individual frame before and after subtraction of the out-of-transit median frame in Figure\,\ref{fig:1_fnoise}. Next, the 1/f noise is assessed for each column of the residual frames. %This evaluation is performed column-by-column. 
To approximate the contribution of the 1/f noise in a given column, the median is computed for all pixels within a distance of 20 to 35 pixels from the center of the traces. This calculated median represents the estimated 1/f noise present in each column, which is subtracted. This process aims to remove the 1/f noise in each individual column of all integrations.\\

\textbf{Extract the 2D stellar spectrum.}
The stellar extraction step is performed on the background, 1/f noise and order 0 contamination corrected spectrum, using the \texttt{transitspectroscopy.spectroscopy.getSimpleSpectrum} routine with a 15 pixels radius aperture extraction, 30 pixels total, centered around the position of the traces. This function uses a box-extraction method and returns the added flux over the defined aperture radius. To address potential outliers in the 2D stellar spectra for Order 1 and Order 2, we apply a correction. Outliers exceeding a threshold of 5$\sigma$ are replaced using a 1D median version of the spectra. The 1D median spectrum is constructed by normalizing with the median for each integration. We compute the median and its error at each wavelength to look for outliers. When found, these are replaced by the re-scaled median 1D spectrum value. \\

\begin{figure}[htpb]
  \includegraphics[width=\columnwidth]{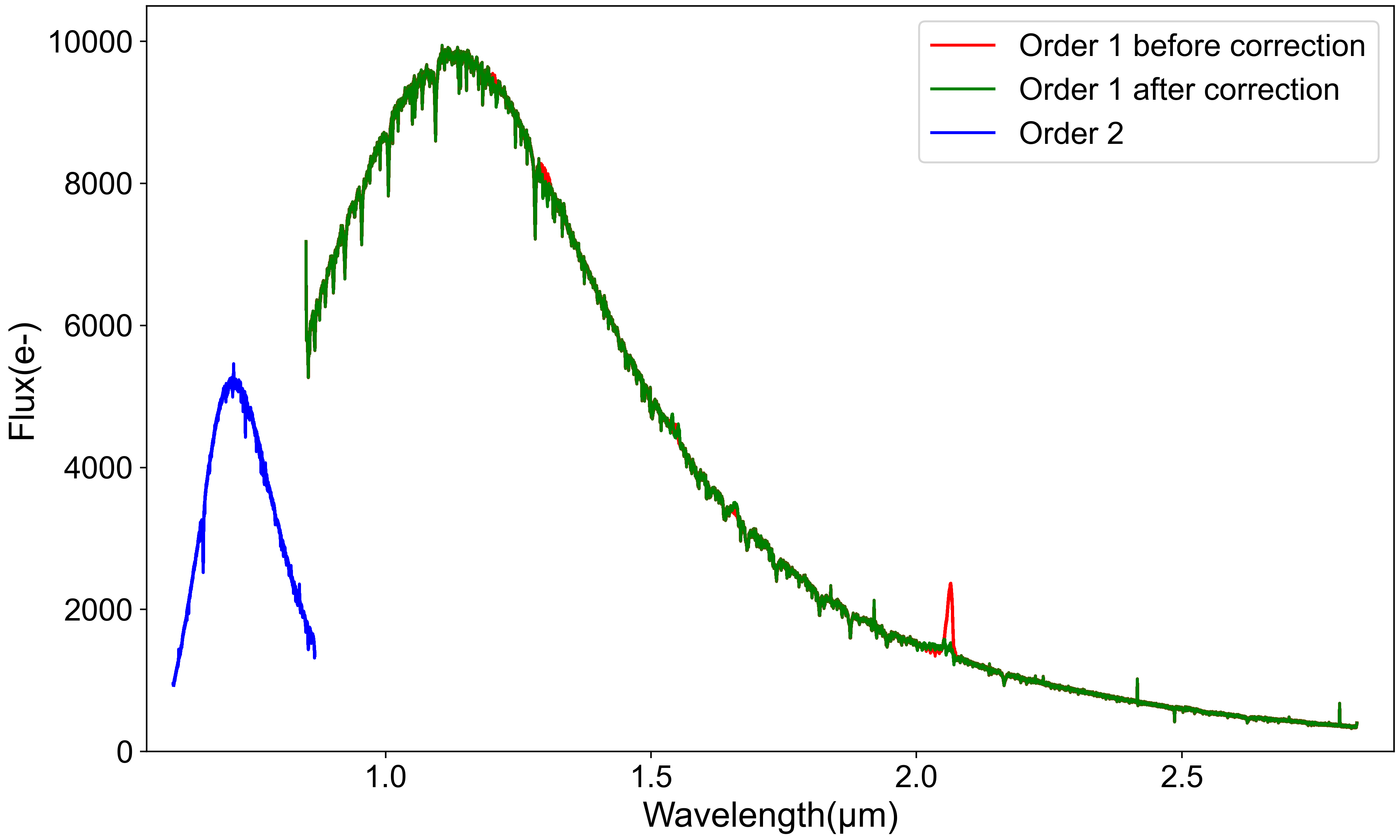}
 \caption{Extracted 2D stellar spectra for Order 1 (red) and Order 2 (blue) from \text{\tt transitspectroscopy}'s reduction. We overplot each spectrum for each integration. The Order 0 contamination-corrected Order 1 stellar spectra are in green. We can see that the feature around 2\,µm caused by the contaminant at $\sim$ 750 x-pixel (see Figure\,\ref{fig:frames} Bottom) has been corrected in the green spectra.  }
 \label{fig:spectra}
\end{figure}

\textbf{Correct Order 0 contamination.}
Order 0 contaminants manifest as concentrated bright spots on the detector subarray, and can be easily identified in the GR700XD/F277W exposure. Although field star contaminants are observable in the GR700XD/CLEAR science exposure, some may be masked behind the traces. These contaminants contribute to the dilution of the flux of the planet over the flux of the star at the corresponding wavelength, ultimately affecting the overall shape of the emission spectrum. However, this dilution is expected to be minimal, as it is a multiplying factor of the eclipse depth. We identify eight main field star contaminants in Order 1 and one in Order 2 (see Figure\,\ref{fig:frames} Bottom). To correct this dilution, we generate a "modelled super-profile" by stacking point spread function (PSF) columns immediately before and after the position of the contaminant. The "super-profiles" are constructed using 5 pixels before and after the contaminant's location. However, the three contaminants situated between 760 and 840 x-pixels are closely spaced, resulting in fewer uncontaminated columns available. The three "super-profiles" for this region are generated using 5 to 2 pixel columns before and after. The contaminant in Order 2 is corrected. However, it is located on the edge of the trace, where the spectrum throughput is low, and we decided not to incorporate these wavelengths in the Order 2 spectra. 

We apply linear interpolation to the "super-profile" across the contaminated columns and subtract this modelled profile.  The resulting residuals only represent the contamination profile. We create a smoothed median contaminant frame which models the contribution of the contaminants within a 60-pixels radius centred around the trace. We extract the spectrum of this contamination frame with a box-extraction method using the same aperture radius, 15-pixels, as for the total stellar spectrum. This 2D contamination spectrum is then subtracted from the 2D Order 1 stellar spectrum. Figure\,\ref{fig:spectra} shows the extracted 2D stellar spectra for Order 1 (green) and Order 2 (blue). For Order 1, we also represent the contaminated 2D stellar spectrum in red. We clearly see the correction applied around 2\,µm where the contaminants creates a prominent spike in the spectrum. 

\subsection{Data reduction with \text{\tt supreme-SPOON}} \label{sec:2.2}
In parallel, we perform a reduction using the \texttt{supreme-SPOON} pipeline \citep{Feinstein_2023, Radica2023, Coulombe2023, Lim2023, Radica_2024, Radica2024_exotedrf}, which performs the end-to-end reduction of NIRISS SOSS Time-Series Observations (TSOs). We closely follow the steps laid out in \citet{Radica2023} for Stages 1 -- 3. During the correction of column-correlated 1/$f$ noise, we mask all undispersed order 0 contaminants of background field stars, as well as the dispersed contaminant below the target Order 1 between columns 500 and 750. We perform the background subtraction in a ``piecewise" manner \citep[e.g.,][]{Lim2023, Fournier-Tondreau_2023}, whereby we separately scale the STScI background model on either side of the background ``step" near column 700. We find the optimal scaling values to be 0.86023 and 0.81722 respectively red-wards and blue-wards of the step.

We perform the spectral extraction using the \texttt{ATOCA} algorithm \citep{Darveau-Bernier2022, Radica2022} to explicitly model the self-contamination of the first two diffraction orders of the target spectra on the detector. If not properly taken into account, we estimate that the self-contamination would cause a maximum anti-dilution of $\sim$70\,ppm, peaking at $\sim$0.72\,µm. We use an extraction aperture of 32 pixels, as we find this minimises the scatter in the white light curves.

After the spectral extraction, we correct for the contamination from undispersed background sources following the methodology presented in \citet{Radica2023}. Briefly, we linearly interpolate over contaminated parts of the trace, using uncontaminated sections of the trace to either side. The interpolated profile is subtracted off, revealing the contaminant within the residuals. We then perform a spectral extraction on these residuals, using the same parameters as above, in order to estimate the flux contribution of the contaminant, which is then removed from the extracted spectra of the target. In this way, we account for the dilution caused by all eight contaminants highlighted in the bottom panel of Figure~\ref{fig:frames}. However, we are unable to correct for the effects of the dispersed contaminant intersecting the target Order 1, visible between columns 500 and 700. The reasons for this are two-fold: firstly, there is simply not enough of the contaminant trace visible on the detector to accurately estimate the spectrum of the contaminant star, and thus extrapolate the resulting contamination on the target trace \citep[e.g.,][]{Radica2023}. Furthermore, simulations using a custom SOSS contamination tool\footnote{\url{http://maestria.astro.umontreal.ca/niriss/SOSS_cont/SOSScontam.php}} show that this dispersed contaminant is actually the very bluest end of an Order 1 trace ($\lambda<0.75$\,µm). We do not see this segment of Order 1 in the target trace as it is truncated by the detector edge at $\sim$0.85\,µm. As such, the reference throughput is relatively poor in this region, all but precluding the necessary flux calibration of the contaminant.

\subsection{Data reduction with \text{\tt Ahsoka}} \label{sec:2.3}

We developed a new data analysis pipeline called \texttt{Ahsoka} (Louie et al. 2024 submitted) \footnote{\url{https://github.com/Witchblade101/ahsoka}}, which we use to perform a third reduction of the emission spectrum. Our \texttt{Ahsoka} pipeline is an amalgamation of the \texttt{jwst calibration} \citep{Bushouse2023}, \texttt{supreme-SPOON} \citep{Feinstein_2023, Radica2023}, \texttt{nirHiss} \citep{Feinstein_2023}, and \texttt{Eureka!} \citep{Bell2022} pipelines. We employed \texttt{jwst calibration} pipeline version 1.8.2 in our analysis. 

We begin our analysis using the \textit{uncal.fits} files downloaded from the Barbara A. Mikulski Archive for Space Telescopes (MAST),\footnote{\url{https://archive.stsci.edu/}} and perform the following \texttt{jwst} pipeline stage 1 detector-level\footnote{\texttt{calwebb\_detector1}, see \url{https://jwst-pipeline.readthedocs.io/en/stable/jwst/pipeline/calwebb_detector1.html}} steps: \texttt{dq\_init, saturation, superbias,} and \texttt{refpix.} Following these initial pipeline steps, we apply the \texttt{supreme-SPOON} background subtraction and 1/f noise removal techniques at the group level. We employ a slightly modified version of the \texttt{supreme-SPOON} background subtraction algorithm to scale the STScI background model\footnote{\url{https://jwst-docs.stsci.edu/jwst-near-infrared-imager-and-slitless-spectrograph/niriss-observing-strategies/niriss-soss-recommended-strategies##NIRISSSOSSRecommendedStrategies-SOSSBackgroundObservations}} to group-level median frames of our observations. The 8 median frames (one for each group) are created from the out-of-transit integrations for each of the 8 groups \citep{ Radica2023}. The background level intensity changes abruptly near column (or ``x-pixel’’) 700. As discussed by \cite{ Albert2023}, we apply separate scalings to each side of this break in the background intensity. On the left side of the break, we scale the STScI background model to the following small region in the upper left side of our median frames: x $\in$ [250,500], y $\in$ [210,250]. On the right side of the break, we use x $\in$ [750,850], y $\in$ [210,250]. We then apply the \texttt{supreme-SPOON} group-level 1/f noise subtraction algorithm, masking the field star contaminants and spectral traces. The 1/f subtraction algorithm adds the background noise back into the frames following 1/f noise removal \citep{Radica2023}. We complete our detector-level reduction with the following \texttt{jwst} pipeline steps: \texttt{linearity, jump, ramp\_fitting,} and \texttt{gain\_scale.} 

We next apply stage 2 spectroscopic processing\footnote{\texttt{calwebb\_spec2}, see \url{https://jwst-pipeline.readthedocs.io/en/latest/jwst/pipeline/calwebb_spec2.html}} to our stage 1 outputs, beginning with the following \texttt{jwst} pipeline steps: \texttt{assign\_wcs, srctype,} and \texttt{ flat\_field.} We then employ the \texttt{supreme-SPOON} background subtraction algorithm, using the same method (to include scaling regions) as with the detector-level application, except that only one median frame is constructed during stage 2. Finally, we perform the \texttt{supreme-SPOON BadPix} custom cleaning step to flag and correct outlying/hot pixels \citep{Radica2023}. 

In \texttt{Ahsoka} stage 3, we apply the simple box extraction algorithm developed as part of the \texttt{nirHiss} pipeline \citep{Feinstein_2023} to the \texttt{BadPix} step output files, thus generating a time series of 1D stellar spectra for NIRISS SOSS Orders 1 and 2. We use an extraction width of 24 pixels, and we define the Order 1 and Order 2 spectral traces using JWST Calibration Data Reference System (CRDS) file \textit{jwst\_niriss\_spectrace\_0023.fits}. We export our stage 3 outputs to \texttt{HDF5} files in an \texttt{Xarray} format that are compatible with \texttt{Eureka!} stage 4. We apply \texttt{Eureka!} to generate and fit white light and spectroscopic light curves.

\subsection{Light curve fitting with \text{\tt juliet} from \text{\tt transitspectroscopy} reduction} \label{sec:2.4}

\textbf{White light curves analysis.}
\begin{figure*}[htpb]
\centering
\includegraphics[width=\textwidth]{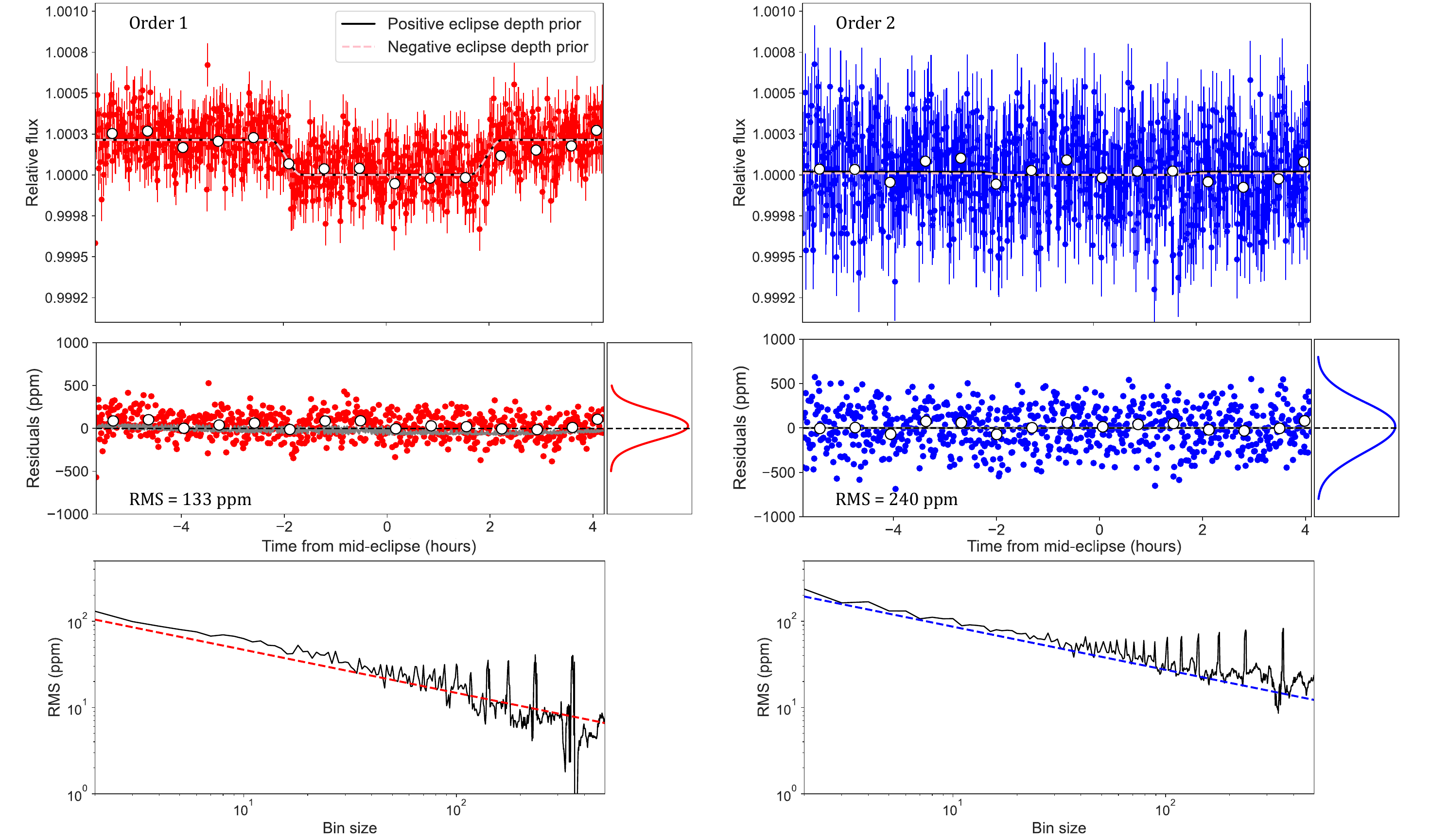}
   \caption{Top: White light curves for Order 1 (left) and Order 2 (right) with the corresponding best-fit eclipse model (black line for positive eclipse depth prior and pink dotted for negative eclipse depth prior) from \text{\tt transitspectroscopy}'s reduction using \text{\tt juliet}. White data points represent binned observations over time for visualization. Middle:  Eclipse model-subtracted light curve residuals, and GP model (light grey). The Root Mean Squared scatter of the residuals (RMS) indicated for each panel. Bottom :  Allan variance plots for the \text{\tt transitspectroscopy}'s reduction with variations of the white light-curves with longer bin timescales.}
% \caption{Top: White light curves for Order 1 (left) and Order 2 (right) with the corresponding best-fit eclipse model (black line) from \text{\tt transitspectroscopy}'s reduction using \text{\tt juliet}. The shaded regions represent the 1-sigma uncertainty. White data points represent binned observations over time for visualization. Bottom: Residuals of the eclipse fit and the Root Mean Squared scatter of the residuals (RMS) indicated for each panel.}
 \label{fig:wlc_analysis}
\end{figure*}
We create white light curves and pixel-level light curves, one for each column on the detector, from the 2D stellar spectra using \text{\tt transitspectroscopy}'s reduction.\\
White light curves are generated by integrating over all wavelengths for each order. Order 1 covers the range of 0.85 to 2.83 $\upmu$m, while Order 2 covers 0.60 to 0.87 $\upmu$m. To fit the light curves, we use the \text{\tt juliet} Python package \citep{Espinoza_2019}, which incorporates the \text{\tt batman} \citep{Kreidberg_2015} package to model the eclipse. We use nested sampling implemented in \text{\tt dynesty} \citep{Speagle_2020} to fit the light curve with \text{\tt juliet}. We fit separately the white light curve of each order. We fixed the mid-transit time to 60024.197373797 BJD-2400000, the period to 3.73548546 days \citep{Alderson2022} and the eccentricity is set to zero. The rest of the orbital parameters are fixed to the values using \citet{Alderson2022}'s results. The flux of the planet over the flux of the star is fitted using a uniform distribution between 0 and 0.01.
The eclipse is visible in the Order 1 but not in the Order 2 white light curve (see Figure\,\ref{fig:wlc_analysis}). We set a normal prior to the mid-eclipse time $\sim \mathcal{N}(2460026.1, 0.1^{2})$ for the Order 1 white light curve and use the result of this fit to fix it for Order 2 analysis. We also fit for a mean-out-of eclipse offset ($M_{SOSS}$) using a normal prior $\sim \mathcal{N}(0, 0.1)$ and include a jitter parameter (white noise, $\sigma_{w, SOSS}$) with a log-uniform distribution between 0.001 and 1000 ppm. \\

To compare the different de-trending models and select our best-fit model we use bayesian evidences. The Bayesian evidence is computed using Bayes'theorem for a set of $\theta$ parameters in a model H for the data D \citep{Feroz_2009}
\begin{equation}
    \rm P(\theta|D, H)=\rm \frac{P(D|\theta, H)P(\theta|H)}{P(D|H)}
,\end{equation}
where P($\theta|$D, H)$\equiv$P($\theta$) is the posterior probability distribution, P(D$|\theta$, H) $\equiv$L($\theta$) is the likelihood, P($\theta|$H)$\equiv$ $\pi$($\theta$) is the prior, and P(D$|$H)$\equiv$ $\mathcal{Z}$ is the Bayesian evidence. The nested sampling method estimates the Bayesian evidence of a given likelihood volume, and the evidence can be expressed as follows:
\begin{equation}
 \mathcal{Z}= \rm \int L(\theta) \pi(\theta) \, \mathrm{d}\theta
.\end{equation}
To compare the two H$_0$ and H$_1$ models, we can compute the respective posterior probabilities, given the observed data set D,
\begin{equation}
 \rm \frac{P(H_1|D)}{P(H_0|D)}=  \rm \frac{P(D|H_1)P(H_1)}{P(D|H_0)P(H_0)}=\frac{E_1 P(H_1)}{E_0 P(H_0)}
,\end{equation}
where P(H$_1$)/P(H$_0$) is the a priori probability ratio for the two models, which can often be set to unity \citep{Feroz_2009}. We used the logarithm version of the model selection to compute the Bayes factor. A model is considered statistically statistically favored over another if the positive difference in logarithmic evidence, i.e., the Bayes factor, exceeds three.

We implement linear de-trending models against time (ln$\mathcal{Z}$ = 5346 for Order 1, and  4952 for Order 2 ), the full-width-half-maximum (fwhm) (ln$\mathcal{Z}$ = 5361 for Order 1 and 4952 for Order 2), and their quadratic terms (ln$\mathcal{Z}$ = 5346 for Order 1 and 4938 for Order 2). We find ln$\mathcal{Z}$ = 5347 for Order 1 and 4939 for Order 2 while including the time and the fwhm with their quadratic term. \\

We also test Gaussian Processes (GP) to account for systematic trends using the package \text{\tt george} \citep{Ambikasaran_2015, Foreman_Mackey_2017} and a Matèrn 3/2 kernel to model the systematics. We define a log-uniform prior for the GP amplitude ($\sigma_{{GP}_{SOSS}}$) between 0.001 and 100 ppm and an exponential prior for the length-scale of the GP processes ($\rho_{0_{{GP}_{SOSS}}}$, $\rho_{1_{{GP}_{SOSS}}}$) using standardized time and fwhm as GP regressors. Our analysis yielded the following results: For Order 1, we obtained ln$\mathcal{Z}$ = 5368, and for Order 2, ln$\mathcal{Z}$ = 4962. On the other hand, when utilizing only time as a GP regressor, we achieved ln$\mathcal{Z}$ of 5358 and 4961 for Order 1 and Order 2 respectively. Fitting an eclipse model alone results in ln$\mathcal{Z}$ = 5350 for Order 1 and 4962 for Order 2. The statistical analysis strongly favours the use of the GP light curve fitting method, particularly when including time and the fwhm as regressors. This observation holds especially true for Order 1,the change in ln$\mathcal{Z}$ is 18 which is statistically significant at over 6$\sigma$, and, we apply the same fitting approach to Order 2. Figure\,\ref{fig:wlc_analysis} presents the corrected light curves along with the best-fit eclipse models. We plot also the eclipse model-subtracted light curve residuals, and GP detrending model. For Order 2, the GP model is close to zero as expected. There is no evidence of remaining trends from the residuals distribution and Allan variance plots. Details on the light curve fitting process with \text{\tt juliet} and results for both orders are in Table\,\ref{table:wlc_fitting}.\\

\textbf{Spectral light curves analysis.}
The spectral light curves are generated by integrating over 4 pixel-level light curves. To model the binned spectral light curves, we use \text{\tt juliet} with a setup similar to that used for the Order 2 white light curve. This involves fixing the planetary parameters and mid-eclipse time, while allowing fitting for the eclipse depth, instrumental parameters, as well as the amplitude and length scale of the GP process. Consequently, this results in the fitting of 509 and 150 light curves for Order 1 and Order 2 respectively.

\begin{figure*}[htpb]
    \includegraphics[width=\textwidth]{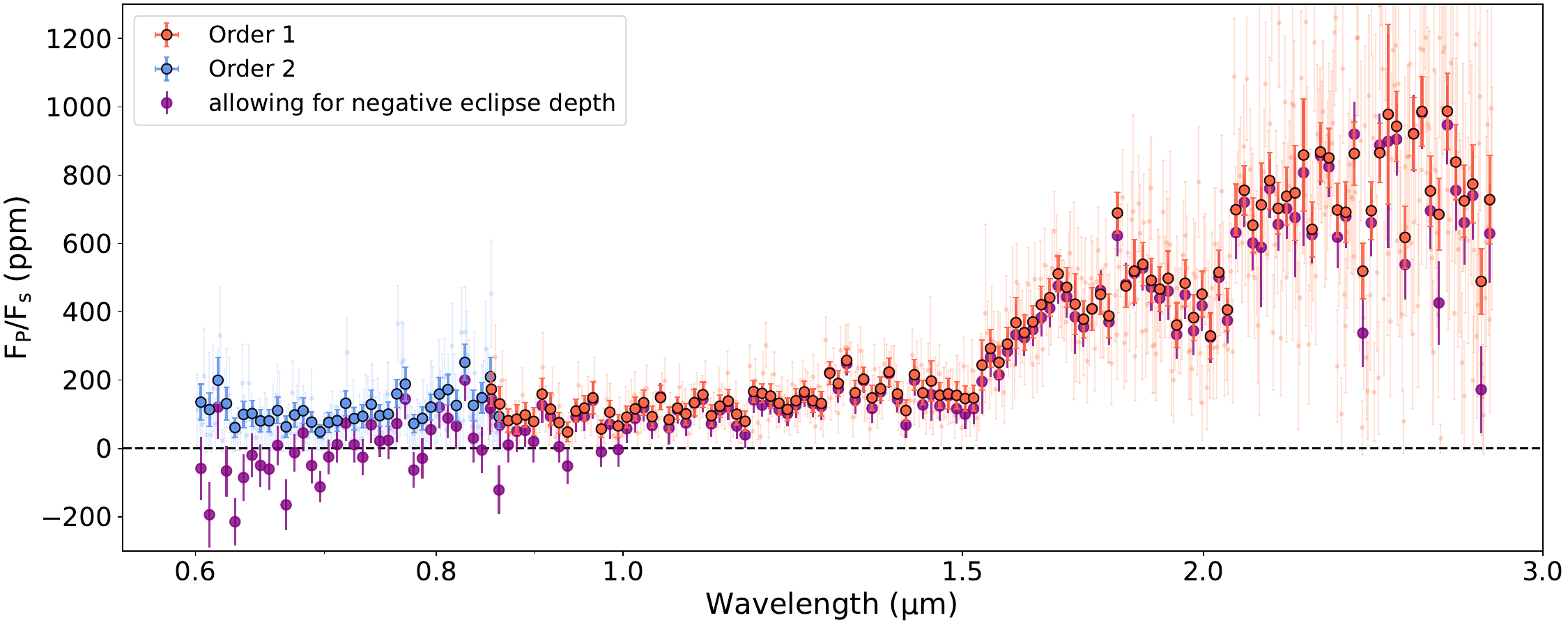}
 \caption{WASP-17\,b NIRISS SOSS emission spectra obtained using \text{\tt transitspectroscopy} data reduction and fitted using \text{\tt juliet}. NIRISS SOSS Order 1 (red) and Order 2 (blue) emission spectra from the 4-pixels bins (light) and at R$\sim$100 (solid). Spectrum's results allowing for negative eclipse depth fitting are in purple for both orders.  }
 \label{fig:emission_spec}
\end{figure*}

To create a spectrum with a resolution of approximately R$\sim$100, we extract the posterior results from each of the wavelength bins within the larger bin. We create a mean posterior by sampling points from these bins. From this mean posterior distribution, we then calculate the quantiles, median, and standard deviation. This method ensures that we correctly combine the spectral light curves and do not over-estimate the flux of the planet over the flux of the star. We check that the mean posterior is not consistent with zero especially for wavelength bins below 1.5$\upmu$m. We over-plot in Figure\,\ref{fig:emission_spec} Order 1 (red) and Order 2 (blue) emission spectra from the 4-pixels bins (light) and at R$\sim$100 (solid).

We observed small eclipse depth values, particularly at shorter wavelengths, especially on Order 2 (see Figure\,\ref{fig:emission_spec}). To ensure the detection of eclipses, we conduct an additional fitting procedure that allowed for negative eclipse depth values. We now fit the eclipse depth using a uniform prior between -0.01 and 0.01. The rest of the orbital parameters and the GP de-trending method are identical. The most significant disparities were observed in the shorter wavelengths of Order 2.

\begin{deluxetable*}{lCCC}
\label{table:wlc_fitting}
\tablecaption{White light curve fitting priors and values from the \texttt{transitspectroscopy} reduction. }
\tablehead{ 
    \colhead{Parameters} & \colhead{Priors} & \colhead{Order 1} & \colhead{Order 2} 
}
\startdata
\hline
\multicolumn{4}{l}{\textbf{Parameters fixed in the fit}} \\
P[days]& \textnormal{fixed \citep{Alderson2022} }& \multicolumn{2}{c}{3.73548546} \\
t$_0$ [BJD$_{\rm TDB}$ - 2400000] &\textnormal{ fixed }& \multicolumn{2}{c}{60024.197373797}  \\
Impact parameter, $b$ & \textnormal{fixed} & \multicolumn{2}{c}{0.481}  \\
a/R$_{\star}$ & \textnormal{fixed \citep{Alderson2022}} & \multicolumn{2}{c}{7.025}  \\
e & \textnormal{fixed} & \multicolumn{2}{c}{0}  \\
$\omega$ & \textnormal{fixed} &\multicolumn{2}{c}{90} \\
R$_{\rm P}$/R$_{\rm s}$ & \textnormal{fixed \citep{Alderson2022}} &\multicolumn{2}{c}{0.1228} \\
\hline
\multicolumn{4}{l}{\textbf{Results of the fit using positive eclipse depth priors}} \\
t$_{secondary}$ [BJD$_{\rm TDB}$ - 2400000] & $\mathcal{N}(60026.1, 0.1)$ & $60026.06695^{+0.00152}_{-0.00153}$ & 60026.06695 \textnormal{(fixed)}\\
f$_{\rm P}$/f$_{\rm s}$ (ppm)& $\mathcal{U}(0, 10^4)$ & $205.56^{+21.37}_{-19.55}$ &  $17.00 ^{+16.30}_{-11.14}$ \\
$\sigma_{w, SOSS}$ (ppm)&$\mathcal{LU}(0.001, 1000)$ & $122.22^{+4.01}_{-3.78}$ & $220.90 ^{+6.94}_{-6.63}$ \\
$M_{SOSS}$  & $\mathcal{N}(0, 0.1)$ & $1.923 ^{+0.343}_{-0.378}\times 10^{-4}$ & $1.079 ^{+0.135}_{-0.122}\times 10^{-4}$ \\
$\sigma_{{GP}_{SOSS}}$ &$\mathcal{LU}(0.001, 100)$ & $36.91 ^{+15.14}_{-9.40}$ & $0.24 ^{+8.36}_{-0.24}$ \\
$\rho_{0_{{GP}_{SOSS}}}$& Exp($\lambda$ = 1) &  $0.559 ^{+0.756}_{-0.358}$ &  $0.725 ^{+1.201}_{-0.533}$ \\
$\rho_{1_{{GP}_{SOSS}}}$ & Exp($\lambda$ = 1) &  $0.386 ^{+0.524}_{-0.245}$ &  $0.669 ^{+1.146}_{-0.511}$\\
\hline
\multicolumn{4}{l}{\textbf{Results of the fit allowing for negative eclipse depth}} \\ 
t$_{secondary}$ [BJD$_{\rm TDB}$ - 2400000] & $\mathcal{N}(60026.1, 0.1)$ & $60026.06693^{+0.00154}_{-0.00147}$ & 60026.06695 \textnormal{(fixed)}\\
f$_{\rm P}$/f$_{\rm s}$ (ppm)& $\mathcal{U}(-10^4, 10^4)$ & $206.28^{+21.49}_{-20.29}$ &  $9.80 ^{+19.21}_{-19.55}$ \\
$\sigma_{w, SOSS}$ (ppm)&$\mathcal{LU}(0.001, 1000)$ & $122.23^{+3.92}_{-3.81}$ & $220.89 ^{+6.96}_{-6.70}$ \\
$M_{SOSS}$ & $\mathcal{N}(0, 0.1)$ & $1.93 ^{+0.353}_{-0.379}\times 10^{-4}$ & $0.057 ^{+0.155}_{-0.163}\times 10^{-4}$ \\
$\sigma_{{GP}_{SOSS}}$ &$\mathcal{LU}(0.001, 100)$ & $37.34 ^{+14.97}_{-9.58}$ & $0.240^{+9.617}_{-0.233}$ \\
$\rho_{0_{{GP}_{SOSS}}}$& Exp($\lambda$ = 1) &  $0.603 ^{+0.812}_{-0.387}$ &  $0.699 ^{+1.166}_{-0.517}$ \\
$\rho_{1_{{GP}_{SOSS}}}$ & Exp($\lambda$ = 1) &  $0.373 ^{+0.508}_{-0.230}$ &  $0.657 ^{+1.100}_{-0.499}$\\
\hline
\enddata
\tablecomments{The fit is performed using the Python package \texttt{juliet} \citep{Espinoza_2019} \footnote{https://github.com/nespinoza/juliet}
}
\end{deluxetable*}

%\begin{figure*}[htpb]
%  \includegraphics[width=\textwidth]{w17b_emission_spectrum_comparison_plot.pdf}
% \caption{WASP-17\,b NIRISS SOSS at R$\sim$100 for a positive (pink) and negative (blue) eclipse depth fitting.  }
% \label{fig:emission_spec_comp}
%\end{figure*}

\subsection{Data reductions intercomparison} \label{sec:data_reduction_comparison}

\begin{figure*}[htpb]
  \includegraphics[width=\textwidth]{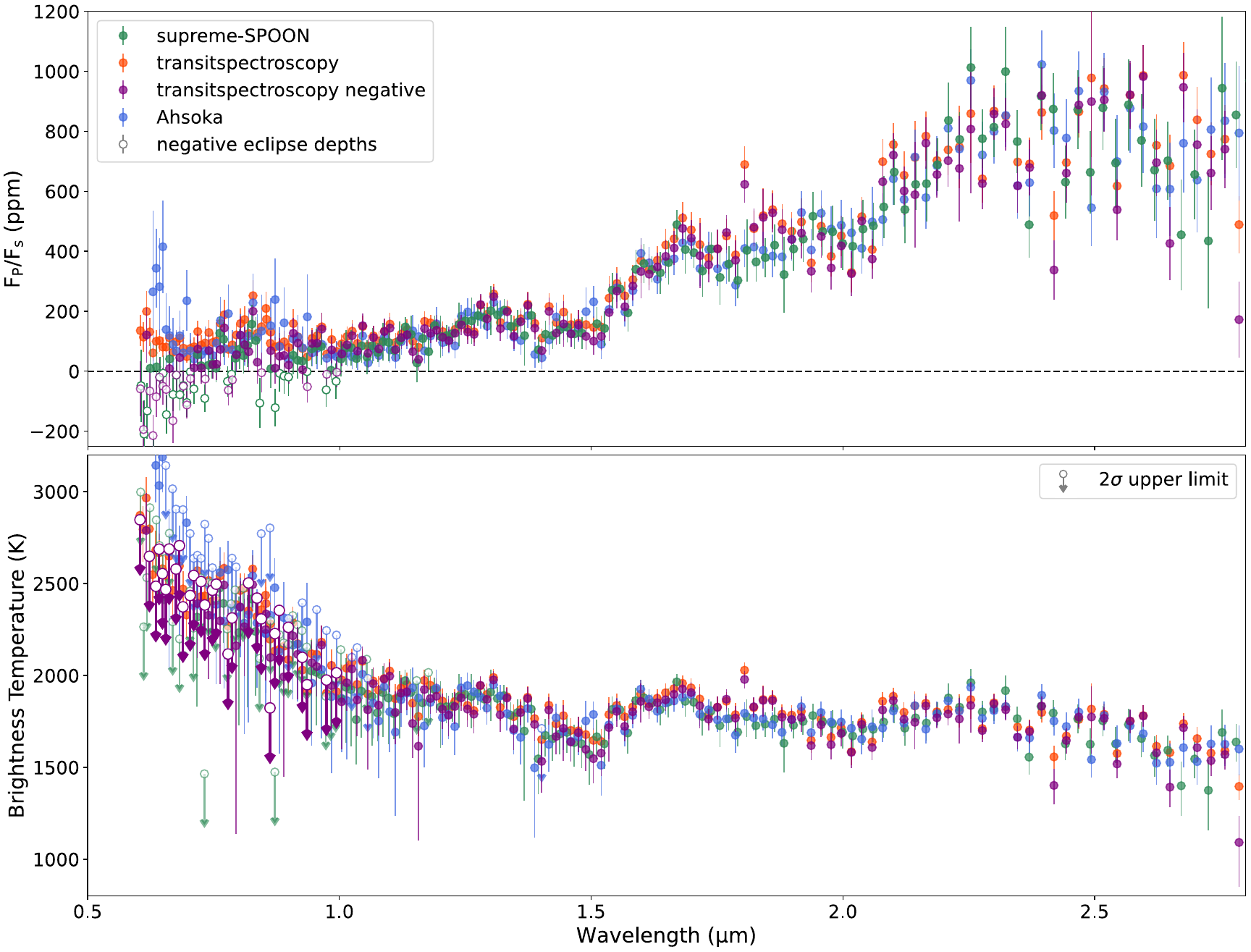}
 \caption{Top : WASP-17\,b NIRISS SOSS emission spectra obtained using \text{\tt transitspectroscopy} (orange), \text{\tt supreme-SPOON} (green) and \text{\tt Ahsoka} (blue). We represent both positive (orange) and negative (purple) eclipse depth fitting results from \text{\tt transitspectroscopy}'s reduction. White data points represent actual eclipse depths constrained to negative values. Bottom : Brightness Temperature of WASP-17\,b as a function of wavelength for the three reductions. For values consistent with zero, we represent the 2$\sigma$ upper limit (downward arrow) computed using the confidence interval. } % Second to fourth: Residuals between reductions. White data points correspond to negative eclipse depth values. }
 \label{fig:emission_spec_intercomparison}
\end{figure*}

Light curves obtained using the \text{\tt supreme-SPOON} and \text{\tt Ahsoka} reductions were independently fitted. The top panel in Figure\,\ref{fig:emission_spec_intercomparison}, displays all three reduction emission spectra overlaid. For \text{\tt transitspectroscopy} reductions we plot both reductions allowing for positive and negative eclipse depths fitting.\\

Order 1 and Order 2 white light curves from the \text{\tt supreme-SPOON} analysis are jointly fitted using \text{\tt juliet}. Orbital and planetary parameters are fixed to the values in Table\,\ref{table:wlc_fitting}. The fit includes the following free parameters: mid-eclipse time, eclipse depth, and scatter. The mid-eclipse time is shared between both orders, while the eclipse depth and scatter are fitted separately for each order. The systematics model consists of a linear trend with time, distinct for each order, as well as a Simple Harmonic Oscillator (SHO) Gaussian Process (GP). The periodicity of the GP is shared between the orders, and the amplitude is fitted separately. Spectral light curves were binned by 5 pixels to slightly enhance the signal-to-noise ratio and then refitted with \text{\tt juliet} with the mid-eclipse time fixed to the best-fitting value from the white light curve. The systematics model includes the linear trend and a scaling factor for the best-fitting white light curve GP model. The eclipse depths are fitted, allowing them to be negative, and the scatter is also fitted. \\

For the \texttt{Ahsoka} pipeline, we fit the white light curves for Order 1 and Order 2 separately using the publicly available \texttt{Eureka!} \citep{Bell2022} package \footnote{\url{https://eurekadocs.readthedocs.io/en/latest/}}, employing \texttt{emcee} \citep{foreman-mackey2013, foreman-mackey2019} in our fits. We modeled the eclipse using the \texttt{batman} software package \citep{Kreidberg_2015}, and modeled systematics with a polynomial comprised of constant and linear terms. We fixed the period, eccentricity, and omega to the values listed in Table\,\ref{table:wlc_fitting}, fixed the mid-transit time to our values obtained for each order in the WASP-17 b transit white light curve fits, and then fit for both $a/R_{\rm s}$ and inclination, with priors taken from the WASP-17 b transit white light curve fit. We also fit for the mid-eclipse time. We used wide uniform priors to fit $F_{\rm p}/F_{\rm s}$, with values ranging between 0 and 0.05. 

The \texttt{Ahsoka} spectral light curve fits employed the same techniques, but we fixed $a/R_{\rm s}$ and inclination to the values obtained during our white light curve fits for each order. We again made use of wide uniform priors with values ranging between 0 and 0.05 to fit for $F_{\rm p}/F_{\rm s}$. Rather than fit the eclipses at the pixel-level, we used \texttt{Eureka!} to bin across pixels and directly fit the light curves for each $R=100$ channel.\\

The second to third panels in Figure\,\ref{fig:emission_spec_residuals} in the Appendix represent the residuals between reductions. The comparison between \text{\tt transitspectroscopy} and \text{\tt supreme-SPOON} is made using negative eclipse depths fitting while the comparison between \text{\tt transitspectroscopy} and \text{\tt Ahsoka} uses positive eclipse depths fitting results. %All three reductions agree. 
\text{\tt Ahsoka}'s reductions are showing a deep increase at shorter wavelengths, which might be due to a poor fit of order 2  wavelength bins where the eclipse is close to zero. \text{\tt transitspectroscopy} shows more deviation from \text{\tt Ahsoka} and \text{\tt supreme-SPOON} reductions above 2\,µm which could be explained by several factors. \text{\tt Ahsoka} shares data reduction steps with \text{\tt supreme-SPOON}'s pipeline, while \text{\tt transitspectroscopy} follows the \text{\tt jwst pipeline} for the first steps treating the \textit{uncal.fits} files. The treatment of the 1/f noise is done at the group level for \text{\tt Ahsoka} and \text{\tt SupremeSPOON} pipelines  \citep{Radica_2024} whereas at the integration level for \text{\tt transitspectroscopy} which could explained the differences above 2\,µm. \\ %First, the extraction aperture is 30 pixels, compared to 32 and 24 for respectively \text{\tt supreme-SPOON} and \text{\tt Ahsoka}. 

%We decided to use \text{\tt transitspectroscopy} positive fitting using Order 1 and Order 2 to continue the interpretation analysis. The latter displays the lowest MAD-based standard deviation: 120 ppm compared to \text{\tt supreme-SPOON} 240 ppm and \text{\tt Ahsoka} 190 ppm. 

While the existence of negative eclipse depths is unphysical, choosing a data reduction approach that restricts eclipse depths from formally going below zero might introduce a Lucy-Sweeney-type bias \citep{Eastman_2013, Deming_2023}. Allowing negative eclipse depths can mitigate this bias by defining the likelihood differently. Several factors could contribute to the appearance of negative eclipse depths, such as changes in baseline flux due to stellar activity, where a peak in stellar flux might coincide with the eclipse. Alternatively, instrument systematics could cause what should be negative eclipse depths to appear positive \citep{Bell2017_W12}. We note some discrepancies between the white light curve eclipse depth values and the eclipse spectra, with spectra constrained to positive eclipse depths displaying inflated values. Additionally, a systematic vertical offset between spectra with and without negative depths could suggest potential bias. Since the eclipse depth values at short wavelengths depend on the fitting techniques, choice of priors, and binning method, we present both the positive and negative eclipse depth fittings from the \text{\tt transitspectroscopy} reduction and apply the interpretation methodology to both spectra.\\

In the bottom panel of Figure\,\ref{fig:emission_spec_intercomparison} we convert our $F_{\rm p}/F_{\rm s}$ to brightness temperatures \citep[e.g.][]{Seager2010, Goyal2021} using a PHOENIX stellar model \citep{Allard2012, Husser_2013} for WASP-17\,b (T$_{\rm eff} = 6550$\,K, [M/H] = -0.25, $\log g = 4.2$; \citealt{southworth2012homogeneous}). We consider brightness temperatures derived from $F_{\rm p}/F_{\rm s}$ measurements that both allow and disallow negative eclipse depths. We use a ``bootstrap" methodology to determine the errors on our brightness temperature estimates. We draw 100,000 random samples at each wavelength point assuming a normal distribution in the $F_{\rm p}/F_{\rm s}$ errors. We use confidence intervals to determine the 1, 2, and 3 sigma errors on our brightness temperature estimates. In cases where the brightness temperature is consistent with zero to within 1$\sigma$, we note that values only represent upper limits. For all of our reductions, assuming both positive-only and negative eclipse depths, robust estimates of brightness temperature can be made across both order 1 and order 2.
Each reduction, including those allowing for negative eclipse depths, shows a consistent shape with a temperature increase toward shorter wavelengths, primarily driven by order 2 with an uptick at shorter wavelengths in order 1. While the upper limits from negative eclipse depths do not conclusively confirm a temperature rise, 30\% of the negative eclipse depth data still support this trend. However, the pronounced rise in brightness temperature is observed most strongly in the positive eclipse depth spectrum. We discuss the implications of the brightness temperature in Section\,\ref{sec:discussion}.

%\section{Methods} \label{sec:methods}

%\section{Results} \label{sec:results}

%\section{Discussion} \label{sec:discuss}

\section{Atmospheric Forward Modelling with \text{\tt ATMO} }\label{sec:atmo}

We focus our effort on the interpretation of WASP-17\,b's Order 1 and Order 2 emission spectra using the \text{\tt transitspectroscopy} data reduction. Fist we present the forward-grid modelling results using \text{\tt ATMO} \citep{Tremblin2015, Amundsen2014,Drummond2016, Goyal2018}. Then we proceed to an atmospheric retrieval analysis using \text{\tt TauREx} \citep{Al_Refaie_2021} and \text{\tt POSEIDON} in Section\,\ref{sec:retrieval}.  \\

We compute a grid of emission spectra for WASP-17\,b using the 1D-2D planetary atmosphere model \text{\tt ATMO} \citep{Tremblin2015, Amundsen2014,Drummond2016, Goyal2018}. This grid is computed using radiative-convective equilibrium $P$-$T$ profiles consistent with equilibrium chemistry \citep{Goyal2020, Goyal2021}. As compared to the grid presented in \citet{Goyal2021} we now use a denser grid for this work. Our parameters for the grid are the same with addition of the internal temperature (T$_{\rm int}$) of the planet as one more grid parameter. We reduce the spacing between subsequent grid parameters, with the aim to obtain more precise constraints if permitted by the observations. This new grid has been computed for 8 re-circulation factors (0.2 - 1.33), representing the extent to which heat is redistributed from a planet's starward hemisphere to its night side, 22 metallicities (0.01x - 100x), 14 C/O ratios (0.01 - 2) and 4 T$_{\rm int}$ values (100\,K - 400\,K). The new model grid for the target planet have been computed at the spectral resolution of R$\sim$1000 with correlated-k opacities. We include H$_{2}$O, CO$_2$, CO, CH$_4$, NH$_3$, Na, K, Li, Rb, Cs, TiO, VO, FeH, CrH, PH$_3$, HCN, C$_{2}$H$_{2}$, H$_{2}$S, SO$_{2}$, H$^-$, and Fe opacities in our model, along with H$_2$-H$_2$ and H$_2$-He collision induced absorption. The details of the opacities, their treatment and model choices for computing the grid are all detailed in \citet{Goyal2020}.

We fitted this grid of forward models to observations reduced using \text{\tt transitspectroscopy}. We obtain the best-fit model having a re-circulation factor of 0.5, metallicity of 100x solar (super-solar) and C/O ratio of 0.2 (sub-solar), with reduced $\chi^2$ value of 3.28 (DOF =155).
% , as shown in Figure \ref{fig:Tbright} converted to brightness temperature.
The re-circulation factor is constrained between 0.4 and 0.6, the metallicity to greater than 30x solar and C/O ratio to less than 0.3, with all the models within the 3$\sigma$ $\chi^2$ of the best-fit model. This fit is discussed further in Section\,\ref{sec:discussion} along with results from our detailed retrieval analysis below. 
% The best fit model $P$-$T$ profile is without any temperature inversion as shown in Figure \ref{fig:Tbright}.

\section{Atmospheric Retrieval analysis}\label{sec:retrieval}

We now turn to Bayesian atmospheric retrievals to infer the atmospheric composition and temperature structure of WASP-17\,b's dayside. As with the previous section, we focus primarily on the \text{\tt transitspectroscopy} data reduction.

\subsection {\text{\tt TauREx 3.1} Retrieval Configuration}\label{sec:retrieval:taurex}

We first performed a retrieval of WASP-17\,b's atmospheric properties using \text{\tt TauREx 3.1} \citep{Al_Refaie_2021}\footnote{\url{https://github.com/ucl-exoplanets/TauREx3_public}} with free chemistry. We use the nested sampling algorithm \text{\tt MultiNest} \citep{Feroz_2009, Buchner_2014} to explore the parameter space. The atmosphere is simulated using 100 layers, uniformly distributed in log-space, ranging from 10$^{-2}$ and 10$^{6}$ Pa (10$^{-7}$--10\,bar). We adopted an evidence tolerance of 0.5 and 1500 live points. The stellar parameters are set to 1.49 R$_{\odot}$, T$_{\rm star}$ to 6550\,K and the metallicity to -0.25. The stellar emission spectrum is read from the PHOENIX library \citep{Husser_2013} and interpolated to the correct temperature. The planetary radius (R$_{\rm P, ref}$), set at the base of the atmosphere (10\,bar), is fitted between 1.12 and 2.15 R$_{\rm Jup}$. We include molecular line lists and continuum from the ExoMol project \citep{Tennyson_2016,Chubb_2021}, HITEMP \citep{Tennyson_2018}, and HITRAN \citep{Rothman_1987, Rothman_2013}. 

In our retrieval analysis, we consider the following active molecules, including H$_2$O \citep{Polyansky_2018}, CO \citep{Li_2015}, CO$_2$ \citep{Rothman_2010}, CH$_4$ \citep{Yurchenko_2017}, NH$_3$ \citep{Yurchenko_2011}, FeH \citep{Wende_2010}, HCN \citep{Barber_2014}, TiO \citep{McKemmish_2019}, VO \citep{McKemmish_2016}, and H$^-$. We fit the abundance of each molecule (log(X$_{\rm VMR}$)) between 10$^{-12}$ and 10$^{-1}$. The ratio of helium to hydrogen is set to the solar value of 0.17. The temperature-pressure profile is fitted using, first, the formalism described in \citet{Madhusudhan_2009} as outlined in Bieger et al. (submitted) and is accessible through the \text{\tt MadhuSeager2009 plugin}\footnote{\url{https://github.com/michellebieger/MadhuSeager2009}}. The temperature-pressure parameters included in the fit are the temperature at the top of the atmosphere (T$_{\rm top}$), the pressure at Layers 1, 2, and 3 of the atmosphere in Pascals (log(P$_{\rm i}$)), and the multiplicative factors $\alpha_1$ and $\alpha_2$. The temperature is fitted between 1000 and 3500 K, while the pressure priors are 10$^{-2}$ and 10$^{6}$ Pa (10$^{-7}$--10\,bar). $\alpha_1$ and $\alpha_2$ are fitted between 0.02 and 1.2.

Besides, we have designed a new temperature parameterization in \text{\tt TauREx}, based on a brown dwarf-type temperature-pressure profile, utilizing the custom option in the code. We retrieve a temperature at the top of the atmosphere, defined by the chosen pressure, and fit 5 $\Delta T$ parameters corresponding to the temperature difference between fixed pressure nodes ($\log (P / \mathrm{Bar}) = [-7, -4, -2, -1, 0, 1]$). The priors on the $\Delta T$ parameters are set between 0 and 1000\,K except for the first $\Delta T$, which is between -500 and 1000 K. We interpolate the temperature-pressure (T-P) profile to the pressure nodes and then smooth the interpolated profile using a moving average filter. The corresponding file for use in the code is openly accessible and can be easily modified and incorporated into further \text{\tt TauREx} retrieval analyses for hot Jupiters\footnote{\url{https://github.com/agressier/taurex-BD-TP.git}}. We also test higher numbers of pressure nodes (not presented here). The priors for all parameters are listed in the Appendix, in Table~\ref{table:retrieval_priors}. % The results of this specific retrieval are presented in Appendix\,\ref{fig:BD_Taurex} to avoid overloading the results section \ref{sec:retrieval_results}.

\subsection {\text{\tt POSEIDON} Retrieval Configuration}\label{sec:retrieval:poseidon}

Additionally, we run retrievals with the \text{\tt POSEIDON} atmospheric retrieval code \citep{MacDonald_2017,MacDonald_2023}. \text{\tt POSEIDON} is an open-source\footnote{\url{https://github.com/MartianColonist/POSEIDON}} Python package for exoplanet transmission and emission spectra. The secondary eclipse emission spectrum radiative transfer model is described in \citet{Coulombe2023}. Our \text{\tt POSEIDON} retrievals also sample the parameter space using \text{\tt MultiNest}, here with 2,000 live points.

Our \text{\tt POSEIDON} retrievals adopt a similar model configuration to {\text{\tt TauREx}. First, we consider the \citet{Madhusudhan_2009} T-P profile, as above, but modified with the reference temperature parameter at 10\,mbar. Second, we investigate a flexible free temperature profile based on the P-T profile prescription used for a brown dwarf in \citet{Piette2020} (referred to as the `BD' T-P profile henceforth). Our BD T-P profile retrieves for a photosphere proxy temperature (placed at $\log (P / \mathrm{bar}) = -1.5$) and 9 $\Delta T$ parameters corresponding to the temperature difference between fixed pressure nodes ($\log (P / \mathrm{bar}) = [-6, -5, -4, -3, -2, -1.5, -1, 0, 1, 2]$). The retrieved temperatures at the nodes are interpolated with a spline to create a smooth T-P profile monotonically increasing with pressure (i.e. inversions are explicitly ruled out). The priors for both T-P profile parameterizations are listed in the Appendix in Table~\ref{table:retrieval_priors}. For both T-P profiles, the model atmospheres cover $10^{-7}$--$100$\,bar with 100 layers uniformly distributed in log-pressure.

We calculate model emission spectra at a spectral resolution of $R =$ 20,000 from 0.55--3.0\,µm. On this wavelength grid, we sample opacities from high-resolution pre-computed cross sections \citep{MacDonald_2022} for the following gases (line lists in parentheses): H$_2$O \citep{Polyansky_2018}, CO \citep{Li_2015}, CO$_2$ \citep{Tashkun_2011}, CH$_4$ \citep{Yurchenko_2017}, NH$_3$ \citep{Coles_2019}, FeH \citep{Wende_2010}, HCN \citep{Barber_2014}, TiO \citep{McKemmish_2019}, and VO \citep{McKemmish_2016}. We also include continuum absorption from H$_2$ and He collision-induced absorption \citep{Karman_2019}, H$^{-}$ bound-free absorption \citep{John_1988}, and H$_2$ Rayleigh scattering \citep{Hohm_1994}. We calculate the observed planetary flux by weighting the emergent planetary flux by $R_{\mathrm{p, \, eff}}^2 / d^2$, where $R_{\mathrm{p, \, eff}}$ is the wavelength-dependent effective planetary radius at the $\tau = 2/3$ surface \citep[e.g.][]{Fortney_2019}. We model the stellar flux by interpolating PHOENIX models \citep{Husser_2013}, using the \texttt{PyMSG} package \citep{Townsend_2023}, to T$_{\rm eff} = 6550$\,K, [M/H] = -0.25, and $\log g = 4.2$. The planetary and stellar fluxes are respectively convolved with the NIRISS PSF and binned down to the data resolution, before we finally take their ratio to calculate the $F_p/F_s$. We compare this planet-star flux ratio to the data during the retrieval likelihood evaluation.

\subsection {Retrieval Results: Positive Eclipse Depth Priors} \label{sec:retrieval_results_pos}

\begin{figure*}[ht]
    \includegraphics[width=\textwidth]{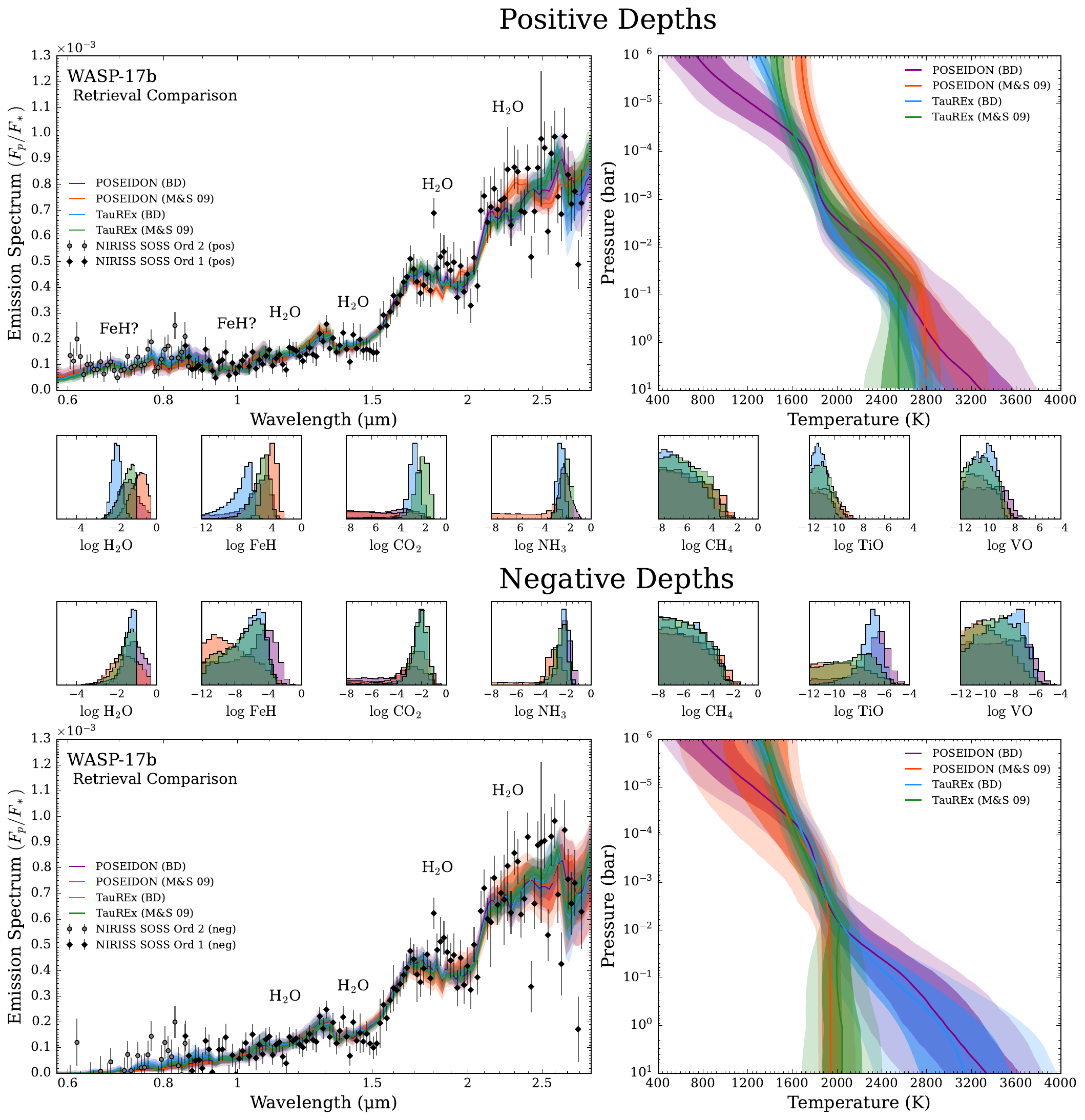}
    \caption{Atmospheric retrieval results for WASP-17\,b's dayside emission spectrum. Top left: Retrieved model emission spectra from our JWST/NIRISS data using the positive eclipse depth prior (error bars). Four retrieval configurations are shown: \texttt{POSEIDON} with the \citet{Piette2020} T-P profile (purple); \texttt{POSEIDON} with the \citet{Madhusudhan_2009} T-P profile (orange); \texttt{TauREx} with the BD T-P profile (blue); and \texttt{TauREx} with the \citet{Madhusudhan_2009} T-P profile (green). For each retrieval, the median retrieved spectrum (solid lines), 1\,$\sigma$, and 2\,$\sigma$ confidence intervals (shaded contours) are overplotted. Absorption features of prominent molecular bands are annotated. Top right: Corresponding retrieved T-P profiles. Middle histograms: Posterior probability distributions for several molecules with strong opacity in the NIRISS wavelength range. Lower half: Same retrievals for the JWST/NIRISS data with the negative eclipse depth prior. Full quantitative retrieval results are provided in the Appendix, Table~\ref{table:retrieval_results}. A super-solar H$_2$O abundance is favored across all 8 retrievals, irrespective of the eclipse depth prior, T-P profile parameterization, and for both retrieval codes.
    }
\label{fig:retrieval_comparison}
\end{figure*}

%MadhuSeager T/P profile : consistent with Tau-REx
%Brown dwarf T/P profile : better fit detection of H2O and FeH lower abundances consistent with solar/slightly super-solar values

We summarize our retrieval results from both codes in Figure~\ref{fig:retrieval_comparison} and Table\,\ref{table:retrieval_results}. We focus first on the results for the eclipse depths constrained to be positive during the light curve fitting, with the results allowing for negative depths presented in Section~\ref{sec:retrieval_results_neg}. We find that WASP-17\,b's NIRISS SOSS emission spectrum can be primarily explained with multiple prominent H$_2$O absorption features in the infrared, with the potential presence of several FeH bands sculpting the spectrum at optical wavelengths. Through Bayesian model comparisons with \texttt{TauREx}, we report a detection of H$_2$O at 6.4$\sigma$ and evidence of FeH at 3$\sigma$. While some retrieval model configurations suggest the presence of CO$_2$ and/or NH$_3$ (with implausibly high abundances), these are spurious signals driven by fits to data reduction-specific features. We established this by running additional \texttt{POSEIDON} retrievals on the \texttt{supreme-SPOON} and \texttt{Ahsoka} reductions, which yielded only upper limits on the CO$_2$ and NH$_3$ abundances but with otherwise consistent atmospheric properties.

% Based on this comprehensive, cloudless atmospheric retrieval analysis,  with a ln$\mathcal{Z}$ = 1256.58.
%We remove each molecule individually from the total atmospheric fit and use the Bayes factor to assess the significance of each detection. Thus, we report the detection of H$_2$O, NH$_3$, and, FeH with 6.4$\sigma$, 3.6$\sigma$ and 3$\sigma$ significance respectively, and the tentative detection of CO$_2$ at 2.7$\sigma$. 

%FeH is predicted in hot Jupiter atmospheres due to their similar equilibrium temperatures to M and L dwarfs. Metal hydride molecules are expected to exhibit dominant opacity features in these hot atmospheres \citep{Kirkpatrick_1999}; however, their detection has proven challenging. Atomic iron (Fe) has been identified in several hot Jupiter atmospheres, including WASP-76\,b \citep{Ehrenreich_2020, Kesseli_2021} and WASP-121\,b \citep{Sing_2019}. FeH, on the other hand, has only been suggested in hot Jupiter atmospheres, as observed in WASP-33\,b, MASCARA-2\,b, WASP-127\,b, and WASP-121\,b through high-resolution spectroscopy \citep{Kesseli_2020} and space-based low-resolution spectroscopy using Hubble \citep{Evans_2016, Sotzen_2019, Skaf_2020}. This tentative detection in the day-side atmosphere of WASP-17\,b provides evidence of the presence of this anticipated molecule in the day-side of a hot Jupiter.

% \textcolor{red}{We need to split this paragraph into several smaller ones and expand on the retrieval results/comparison -- Ryan working here}.

% H2O abundances from three codes
% Upper priors with TauREx

\begin{figure*}[htpb]
    \includegraphics[width=\textwidth]{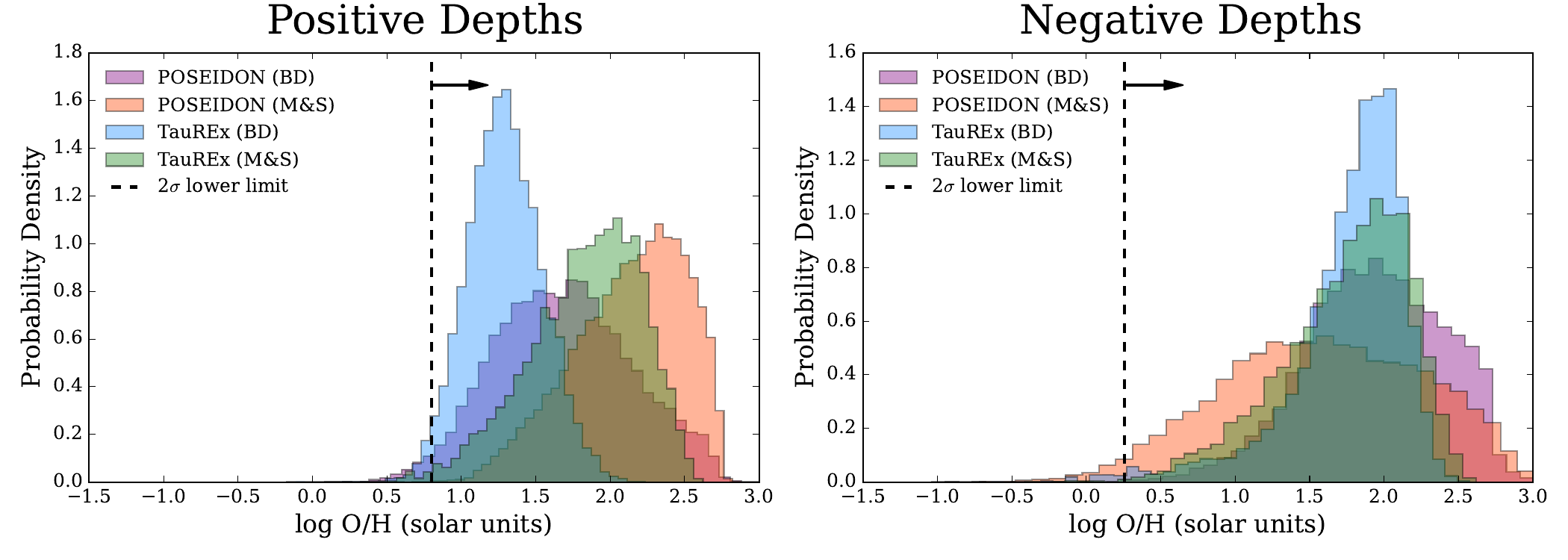}
    \caption{Retrieved oxygen-to-hydrogen ratio of WASP-17\,b's dayside atmosphere. Left: O/H constraints using the positive eclipse depth fitting method. Right: O/H constraints using the eclipse depth fitting method allowing negative depths. The O/H ratio posteriors are derived from the full set of posterior samples from each of the 8 retrievals shown in Figure~\ref{fig:retrieval_comparison}. The 2$\sigma$ lower limits for WASP-17b's dayside O/H ratio ($> 6.6\times$\,solar for the positive eclipse depth prior and $> 1.8\times$\,solar for the prior allowing negative eclipse depths) are overplotted for reference (dashed lines).}
\label{fig:metallicity}
\end{figure*}

Our retrievals suggest a highly metal-enriched dayside atmosphere for WASP-17\,b (Figure~\ref{fig:retrieval_comparison}). Our strongest result is a lower limit on the dayside H$_2$O abundance, seen across all four retrieval configurations,
which is robustly super solar. Compared to a solar composition atmosphere, with log H$_2$O = -3.3, we find log H$_2$O = $-1.46^{+0.31}_{-0.45}$ (\texttt{TauREx}, M\&S 2009 T-P profile),$-2.01^{+0.24}_{-0.23}$ (\texttt{TauREx}, BD T-P profile), $-0.89^{+0.35}_{-0.46}$ (\texttt{POSEIDON}, M\&S 2009 T-P profile), and $-1.40^{+0.94}_{-0.85}$ (\texttt{POSEIDON}, BD T-P profile). Across our four retrievals, the lowest allowed H$_2$O abundance to within 3\,$\sigma$ is log H$_2$O = -2.7 (from the \texttt{TauREx}  BD T-P profile retrieval). We similarly see retrieved FeH abundances enhanced by $\sim 10^3$ compared to thermochemical equilibrium solar metallicity expectations at 2000\,K (log FeH $\sim -8$; \citealt{Woitke2018}). Figure~\ref{fig:metallicity} shows the inferred probability distribution for the O/H ratio of WASP-17\,b's dayside atmosphere (calculated from the posterior samples of all the oxygen- and hydrogen-bearing molecules in our retrievals). We see that all four retrievals have a maximum likelihood O/H ratio $\sim 100 \times$ enhanced over a solar composition atmosphere. We report a 3\,$\sigma$ lower limit of O/H $>$ 3$\times$ solar, implying an enrichment of metals, in particular oxygen, in WASP-17\,b's atmosphere.

The temperature-pressure profiles from our retrievals are shown in Figure~\ref{fig:retrieval_comparison} (top right panel). Our retrievals find that WASP-17\,b's atmosphere has an $\sim$ 800--1000\,K temperature gradient between 10$^{-4}$--10$^{-1}$\,bar, with temperature monotonically increasing with pressure (even for the \citet{Madhusudhan_2009} T-P profile prescription, which allows thermal inversions). Given that our NIRISS data contains only absorption features (Figure~\ref{fig:retrieval_comparison}, top left panel), the retrieved T-P profiles rule out a temperature inversion over the pressure range probed by our NIRISS emission spectrum. While our retrieved T-P profiles are in broad agreement in the pressure range where spectral features form, we find that the two retrievals assuming the \citet{Madhusudhan_2009} T-P profile parameterization favor nearly isothermal atmospheres at pressures near the top ($P < 10^{-4}$\,bar) and bottom of the atmosphere ($P > 10^{-1}$\,bar), while our modified `BD T-P' prescription --- from \citet{Piette2020} for \texttt{POSEIDON}, and from this work for \texttt{TauREx} (see Section\,\ref{sec:retrieval:taurex}) --- allows for the possibility of an adiabatic T-P profile in the deep atmosphere. This difference arises from the additional flexibility of the `free temperature' BD T-P retrieval method, while the \citet{Madhusudhan_2009} profile is constrained to a specific isothermal functional form in the deep atmosphere.

% Additionally, the temperature/profile retrieved using the \citet{Madhusudhan_2009} formalism appears to be favoring a hot isothermal profile in the adiabatic region (see Figure\,\ref{fig:Tbright}).

\begin{figure}[htpb]
    \includegraphics[width=\columnwidth]{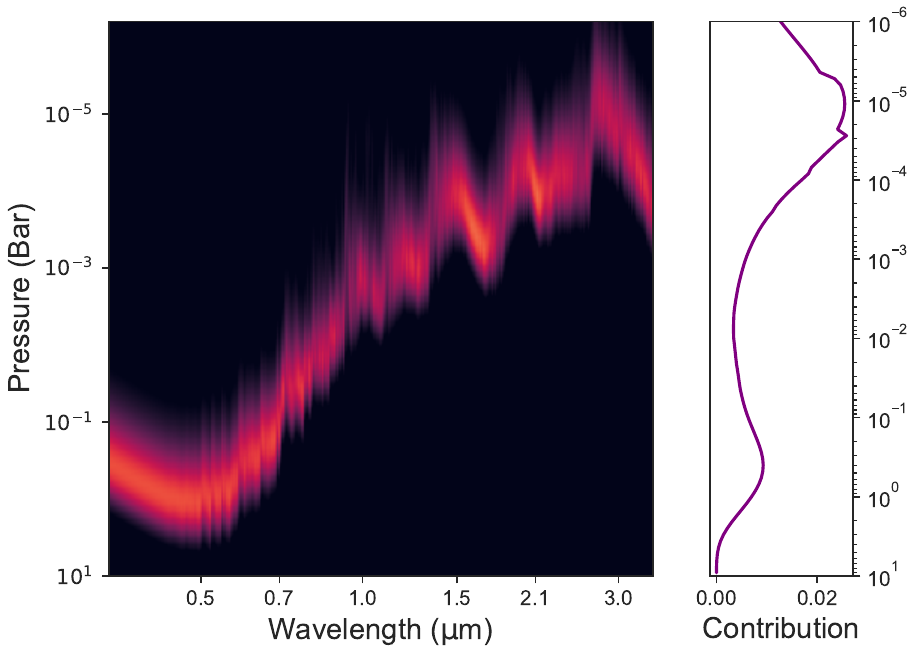}
    \caption{Contribution functions corresponding to the best-fitting \text{\tt TauREx} model emission spectrum using the \citet{Madhusudhan_2009} T-P profile and the observed emission spectrum obtained with the positive eclipse depth fitting method. Left: spectral contributions (color scale) showing the pressure range probed by our NIRISS observations at each wavelength. Right: integrated average pressure contribution function.}
\label{fig:contribution}
\end{figure}

We further demonstrate the wide pressure sensitivity of our NIRISS emission spectrum in Figure~\ref{fig:contribution}. The emission contribution function highlights the relative importance of a given layer to the emergent flux from the atmosphere (i.e. the contribution color is a visual proxy for the pressures where the spectrum forms at a given wavelength). Here, we show the contribution function calculated from the best-fitting model from the \texttt{TauREx} retrieval using the \citet{Madhusudhan_2009} profile. We see that the shortest visible wavelengths covered by the NIRISS 2nd order provide constraints on the deep atmosphere (approaching 1\,bar at 0.6\,$\micron$), while the longest wavelengths in the NIRISS 1st order probe pressures in the upper atmosphere (nearly reaching 10$^{-5}$\,bar at 2.8\,$\micron$). The different pressure sensitivities are primarily driven by the wavelength-variation in the H$_2$O opacity --- since the longer wavelengths in the near-infrared have stronger H$_2$O bands --- causing the $\tau \sim 1$ optical depth surface to form higher in the atmosphere.

% Besides, in Figure\,\ref{fig:contribution}, we represent the variation of the optical depth across both wavelength and pressure (left), and the average optical depth of the atmosphere as a function of pressure (right). The optical depth refers to the atmospheric extinction along the line of sight, indicating the origin of emission. The map emphasises the need for the emission to encompass a broad pressure range to explain the spectrum. These three aspects suggest that the solution found by the retrieval is influenced by the temperature/pressure profile, imposing a hot isothermal adiabatic region, which is counterbalanced by the chemistry pushing toward a high atmospheric metallicity. % This aspect will be explored more deeply in the Discussion section\,\ref{sec:discussion}.

\subsection{Retrieval Results: Negative Eclipse Depth Priors} \label{sec:retrieval_results_neg}

We repeated our retrieval analysis with the \text{\tt transitspectroscopy} data reduction allowing for negative eclipse depths using both \text{\tt TauREx} and \text{\tt POSEIDON}. As motivated in Section~\ref{sec:data_reduction_comparison}, considering the negative depths, while unphysical, allows one to mitigate statistical biases from imposing a prior restricting eclipse depths to be positive. However, we note that the retrieval models always have positive eclipse depths since the planetary flux cannot be negative. We proceed to quantify how negative eclipse depths shortwards of 1\,$\micron$ propagate into our retrieved atmospheric properties of WASP-17b's dayside.

\begin{figure*}[htpb]
  \includegraphics[width=\textwidth]{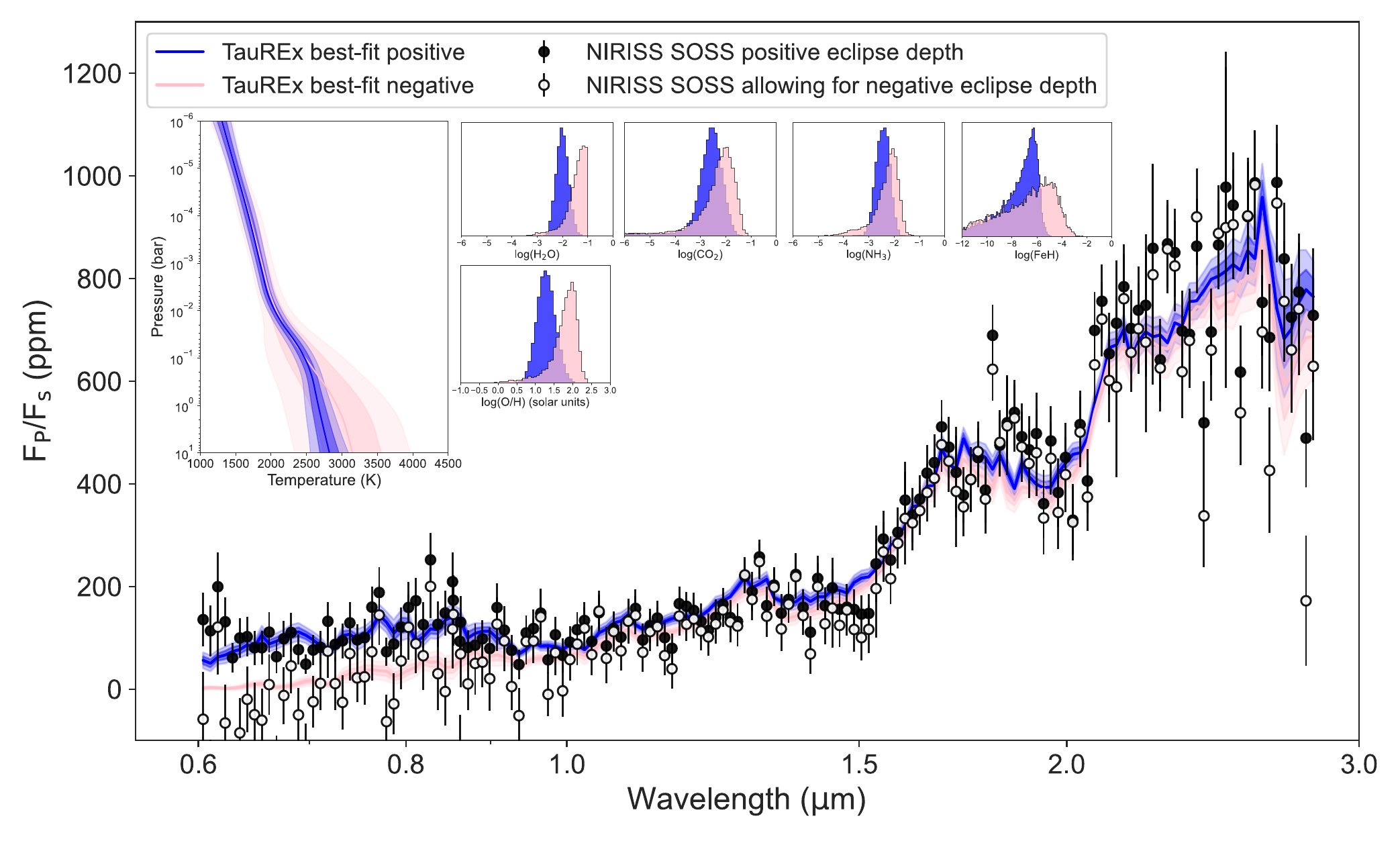}
  \includegraphics[width=\textwidth]{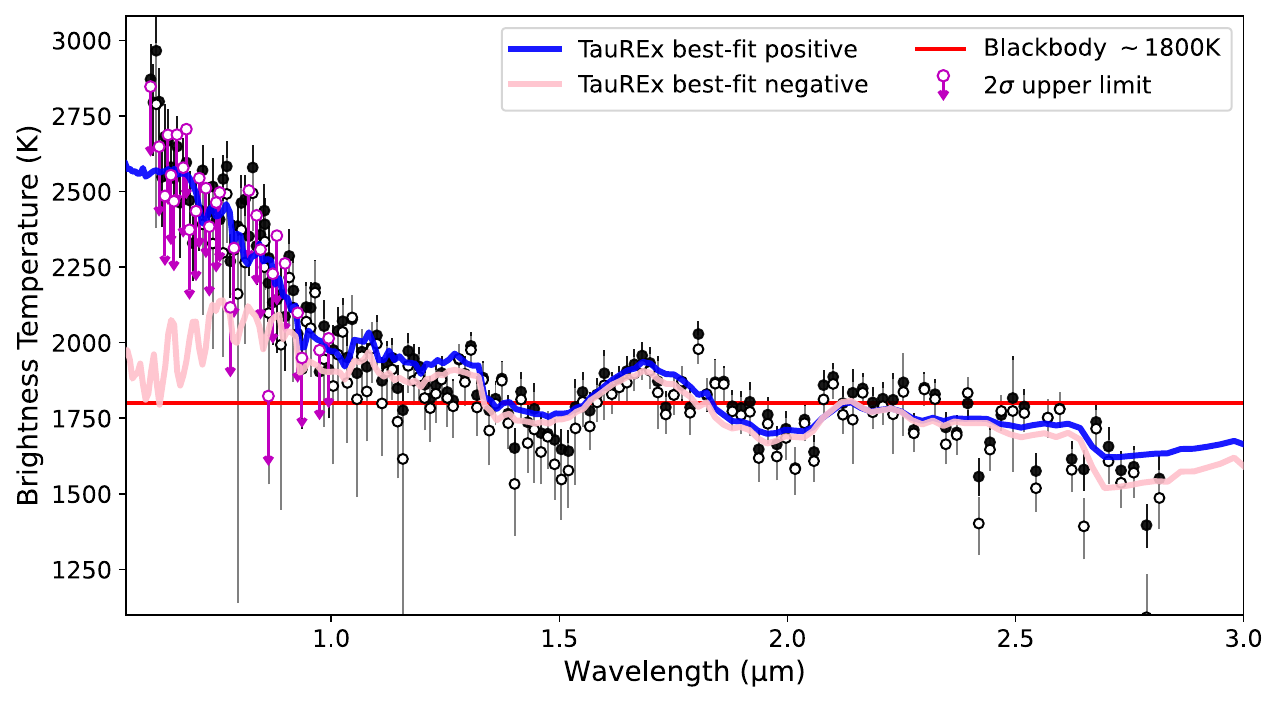}
 \caption{Top: Atmospheric retrieval results for WASP-17\,b's emission spectra using free chemistry and the BD-TP profile in \text{\tt TauREx}. The emission spectra are from \text{\tt transitspectroscopy}'s data reduction obtained using positive (black) and negative (white) priors for both Orders. Bottom: Brightness temperatures of WASP-17\,b as a function of wavelength for the \texttt{transitpectroscopy} reductions plotted against the best fit models from the retrieval analysis. Brightness temperatures consistent with zero within the confidence interval are represented as 2$\sigma$ upper limits.} %The T-P profile is designed to have increasing temperature with pressure.  H$_2$O and FeH abundances are lower using the BD-TP profile than using the Madhu-Seager formalism. Only an upper limit can be put in FeH abundance, while using the spectrum created allowing for negative eclipse depths. H$_2$O is still strongly detected with abundance consistent with a super-solar atmospheric metallicity in both cases.}  }
 \label{fig:BD_Taurex}
\end{figure*}

Figure\,\ref{fig:retrieval_comparison} compares the \text{\tt TauREx} and \texttt{POSEIDON} retrieval results with negative eclipse depths allowed to those discussed above for positive depths only. We also provide a specific comparison between the two \text{\tt TauREx} BD T-P retrievals with positive and negative depths in Figure\,\ref{fig:BD_Taurex}. We see that H$_2$O remains strongly detected at more than 6$\sigma$, with a 2$\sigma$ lower limit of $\log \mathrm{H_2 O} = -2.94$ from the \texttt{POSEIDON} retrieval with the \citet{Madhusudhan_2009} T-P profile. The retrieved H$_2$O abundance of WASP-17b's dayside atmosphere is consistent across all 8 retrievals presented in Figure\,\ref{fig:retrieval_comparison}. While hints of FeH are still present in the posterior distributions for the negative eclipse depths, only an upper limit be formally assigned on the FeH abundance (i.e., $\log \mathrm{FeH} < -2.4$, see Table~\ref{table:retrieval_results}). Therefore, the inference of FeH is contingent on whether the eclipse depths below 1\,$\micron$ are allowed to be negative during light curve fitting. However, we confirm the super-solar metallicity and the enrichment in oxygen in WASP-17,b's atmosphere, even with negative eclipse depths allowed, as shown in Figure~\ref{fig:metallicity}. Similarly, our negative depths retrieval analysis confirms a non-inverted T-P profile over the pressure range probed by our NIRISS observations. We provide the full quantitative comparison between our retrievals, for both positive and negative eclipse depths, in the Appendix in Table~\ref{table:retrieval_results}.

\section{Discussion} \label{sec:discussion}

The three reduction pipelines yielded consistent solutions; however, some differences persist in the spectroscopic emission spectrum, particularly above 2\,µm between \text{\tt transitspectroscopy} and the other two pipelines (see Figure\,\ref{fig:emission_spec_residuals}). Below, we will elaborate on the factors contributing to these variations. The \text{\tt Ahsoka} and \text{\tt SupremeSPOON} pipelines share similarities in the initial steps of reduction, progressing from \textit{uncal.fits} to the creation of light curves, while \text{\tt transitspectroscopy} follows the \text{\tt jwst pipeline} with additional custom steps, i.e. bakcground subtraction, 1/f noise correction, Order 0 contaminants correction. Especially the treatment of the 1/f noise is done at the group level for \text{\tt Ahsoka} and \text{\tt SupremeSPOON} pipelines whereas at the integration level for \text{\tt transitspectroscopy} \citep{Radica_2024}. Additionally, the light curve fitting processes differ among procedures, with \text{\tt Ahsoka} using linear detrending models, while \text{\tt transitspectroscopy} and \text{\tt SupremeSPOON} incorporate GP to model the noise. An important consideration is the choice of priors for the eclipse depth value. Allowing negative eclipse depth may result in negative emissions for certain wavelength bins below 1\,µm,  but we consider both positive and negative emission spectra in our interpretation. The results from this wavelength range must be approached cautiously. It is noteworthy that the robust detection of H$_2$O from the retrieval analysis is driven by the emission from the Order 1 wavelength range and thus remains unaffected by challenges in this specific range (see Appendix\,\ref{fig:BD_Taurex}).

The retrieval analyses conducted by \text{\tt TauREx} and \text{\tt POSEIDON} consistently indicate a substantial presence of water in WASP-17b's dayside atmosphere for both explored temperature profile prescriptions.
% while employing the \citet{Madhusudhan_2009} temperature/pressure profile. However, the high metallicity in a hot adiabatic region is unexpected. The adoption of an alternative temperature/pressure profile, such as the Brown-dwarf model (BD T-P profile), favored by the Bayesian evidence, helps break the degeneracies between metallicity and molecular abundances. 
%
The water abundance aligns with what is expected for a hot Jupiter with a slightly super-solar atmospheric metallicity. 
%With the brown dwarf-inspired T-P profile, H$_2$O is still strongly detected at over 6$\sigma$, but FeH is now categorised as tentative detection at 2$\sigma$. 
While we infer evidence of FeH at 2$\sigma$ significance using the positive eclipse depth fitting method, we find only an upper limit on the FeH abundance using the eclipse depths allowed to become negative. However, H$_2$O remains strongly detected ($> 6\,\sigma$) regardless of whether the positive or negative eclipse depth treatment. While the existence of FeH in WASP-17b's dayside atmosphere is contingent on the eclipse depth fitting method, we note that FeH is predicted in hot Jupiter atmospheres due to their similar equilibrium temperatures to M- and L-dwarfs. Metal hydride molecules are expected to exhibit dominant opacity features in these hot atmospheres \citep{Kirkpatrick_1999}, even though their detection has proven challenging. Atomic iron (Fe) has been identified in several hot Jupiter atmospheres, including WASP-76\,b \citep{Ehrenreich_2020, Kesseli_2021} and WASP-121\,b \citep{Sing_2019}. FeH, on the other hand, has only been suggested in hot Jupiter atmospheres, such as WASP-33b, MASCARA-2b, WASP-127b, and WASP-121\,b through high-resolution ground-based spectroscopy \citep{Kesseli_2020} and space-based low-resolution spectroscopy using Hubble \citep{Evans_2016, Sotzen_2019, Skaf_2020}. Our results demonstrate the challenge in robustly detecting molecules with visible wavelength opacity in hot Jupiter dayside emission spectra using JWST observations, motivating additional work to establish best practices for eclipse depth fitting at optical wavelengths.

% in the \textbf{dayside} atmosphere of WASP-17\,b provides preliminary evidence of this predicted molecule in the \textbf{dayside} of a hot Jupiter using JWST observations. %Our tentative detection in the day-side atmosphere of WASP-17\,b provides the first evidence of this predicted molecule in the day-side of a hot Jupiter using JWST observations.

% We note that while the retrieval analysis performed by \text{\tt TauREx} also suggests the presence of NH$_3$ and CO$_2$ (see Figure~\ref{fig:retrieval_comparison}), these constraints are not confirmed across \text{\tt POSEIDON}'s retrievals using the \text{\tt transitspectroscopy}'s emission spectrum nor the other reductions. These tentative detections lack confirmation compared to water and iron hydride, which consistently emerge in all retrieval and forward modeling analyses. \\
The JWST NIRISS SOSS emission spectrum of WASP-17\,b longward of 1\,µm indicates that photospheric temperatures hover roughly around 1800~K with $\pm\sim$100\,K variations around absorption features (Figures~\ref{fig:BD_Taurex} and \ref{fig:Tbright}). Given the WASP-17 system parameters (Table~\ref{table:wlc_fitting}) we would predict a global average equilibrium temperature of $\sim$1750~K \citep{Seager2010}. The daysides of hot Jupiters are typically much hotter than this predicted global average. Using the formulations presented in \citet{Cowan2011}, we estimate that WASP-17b's average dayside temperature would be $\sim$2200~K assuming no recirculation of heat to the nightside of the planet and a Bond albedo of zero. In estimating hot Jupiter dayside temperatures there is a well-known degeneracy between assumptions of the planetary Bond albedo and day-night heat redistribution efficiency \citep[e.g.,][]{Cowan2011, Schwartz2015}. Our observed photospheric temperatures of 1800~K for WASP-17b's dayside could be indicative of either highly efficient longitudinal (but no latitudinal) transport of heat from day to night or a non-zero Bond Albedo.
\begin{figure*}
    \centering
    \includegraphics[width=\textwidth]{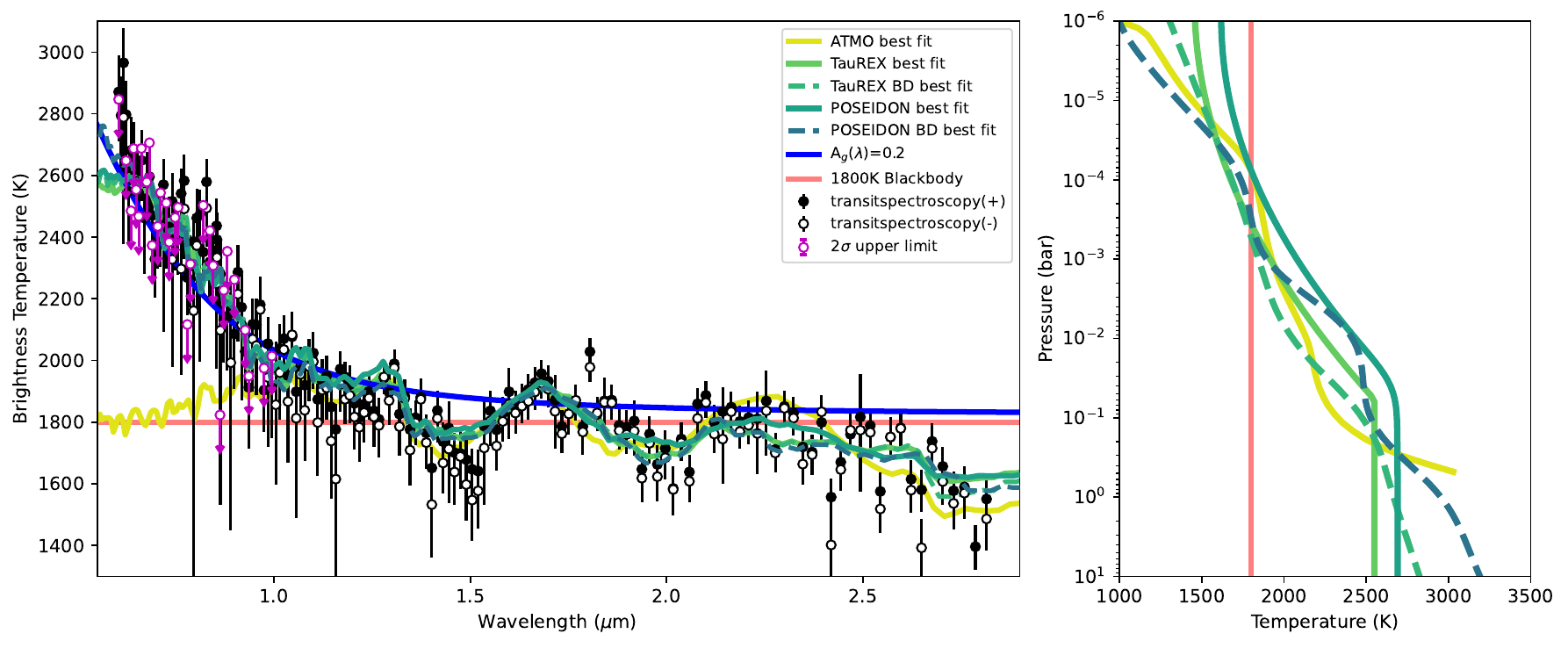}
    \caption{Left: brightness temperature of WASP-17\,b as a function of wavelength for the \texttt{transitpectroscopy} reductions plotted against the best fit models from our forward model and retrieval analysis. We additionally show a 1800~K blackbody (red line) and a simple parametric function assuming geometric albedo of 0.2 (blue line). Right: associated pressure-temperature (P-T) profiles for each of the best fit models. The red line indicates 1800~K blackbody temperature.}
    \label{fig:Tbright}
\end{figure*}
Shortward of 1\,µm, the estimated photospheric brightness temperatures for WASP-17\,b rise, indicating a change in contributions to the planet's spectrum at those wavelengths. We indicate 2$\sigma$ upper limits for brightness temperatures that were consistent with zero, noting that while these upper limits are concentrated in order 2, they are interspersed with statistically significant measures of the planetary brightness temperature across the wavelength range. However, the pronounced rise in brightness temperature is observed most strongly in the positive eclipse depth spectrum  with only 30$\%$ of negative eclipse depth data supporting a similar trend. Future work will consider optimal binning strategies to increase the SNR of order 2 $F_{\rm p}/F_{\rm s}$ and brightness temperature estimates. Figure~\ref{fig:BD_Taurex} (bottom panel) shows the TauREx best-fit model in pink, based on the negative eclipse depth fitting. While the model above 1\,µm remains similar to the positive eclipse depth fit, it diverges below 1\,µm. Although the upper limits in the brightness temperature from the negative eclipse depths naturally trend upwards at shorter wavelengths they cannot definitively confirm the temperature increase. Our retrievals considering negative eclipse depths still suggest a thermal structure with significantly elevated temperatures at the 100\,mbar level and below to match the observed flux levels (see Figure~\ref{fig:retrieval_comparison}). \\

Wavelengths shortward of 1\,µm will be sensitive to any light that is reflected from the dayside of the planet. It is also possible that shortward of 1\,µm an opacity window occurs that allows us to peer into deeper hotter regions of WASP-17\,b's atmosphere. We here discuss these two possibilities in the context of both positive and negative eclipse depths further. Measuring exoplanet dayside reflected light spectra directly links to the planet's geometric albedo (A$_\mathrm{g}$) and the efficiency of the atmosphere at reflecting light. 
\citet{Cahoy2010} show that temperature is the primary driver of changes in albedo through the formation of clouds such that higher clouds cause a higher reflectivity and albedo. 
For temperatures below 750\,K these clouds are more likely photochemical in origin \citep[e.g.,][ showing the effect of sulfur hazes]{gao2017sulfur}, above this the condensate clouds formed predominantly of silicates are theorised to produce highly reflective atmospheres \citep[e.g.,][]{parmentier2016transitions,roman2019modeled}. 
UV measurements of the hot Jupiter HD\,189733b (T$\mathrm{eq}\,\sim$\,1,000\,K) show a rise in albedo from $<$0.12 at 450--530\,nm to 0.4 at 230--450\,nm suggesting the presence of highly scattering particles on the dayside \citep{Evans2013}. 
Unsurprisingly, broadband eclipse measurements of ultra-hot Jupiters (T$_\mathrm{eq}$\,$>$\,2,000\,K) in the optical with ground- and space-based observatories point to low geometric albedo values, often A$_\mathrm{g}\,<$\,0.2 suggesting little scattering in their upper atmospheres likely due to the absence of dayside cloud opacity \citep{Bell2017_W12,Mallonn2019}. 
From our JWST MIRI LRS tranmission observations of WASP-17\,b \citep{grant_2023}, we know that small particles of SiO$_2$ are present along the planet's terminator region. Since SiO$_2$ particles are highly reflective, this would result in a high albedo if they are also present on the dayside of WASP-17\,b. Using a parametric geometric albedo model similar to the form presented in \citet{Evans2013}, we find that the slope in our brightness temperature shortward of 1\,µm can be matched assuming A$_\mathrm{g}\,\sim$\,0.2 and underlying blackbody emission at 1800~K (Figure~\ref{fig:Tbright}). While this is inconsistent with A$_\mathrm{g}$ estimates for other hot Jupiters at similar wavelengths, it is not impossible to produce with some dayside cloud coverage on WASP-17\,b which could also contribute to a non-zero Bond albedo for the planet. %Going beyond this ``toy" model for A$_\mathrm{g}$ to investigate this possibility further will require broader wavelength coverage and retrievals that simultaneously consider reflected light and thermal emission, which we leave for a future study. 
Incorporating negative eclipse depths into our analysis, we recognize that while these depths may suggest a different reflection or emission profile, they do not rule out the possibility of a high albedo. The observed temperature trend, which aligns with a high albedo in the positive depth analysis, remains consistent with the trend observed in the upper limits and constrained values from negative eclipse depth findings. However, while the upper limits from negative eclipse depths naturally trend upwards at shorter wavelengths, they do not confirm a definitive temperature rise. To investigate the implications of negative eclipse depths and their effect on albedo interpretation beyond this ``toy'' model for A$_\mathrm{g}$, broader wavelength coverage and more comprehensive retrievals that account for both reflected light and thermal emission will be required.  We plan to address these aspects in future studies.

%We fit the slope towards the blue in the brightness temperature of WASP-17b with a parametric albedo model of the form presented in \citet{Evans2013}. The best fit geometric albedo is found to be 0.47$^{+0.25}_{-0.12}$, with a 3-sigma range stretching from 0.28--0.99. Despite larger uncertainties in the brightness temperature at short wavelengths in the other reductions the fit to the albedo maps to the upper edge of the slop resulting in a similar range of potential values and all models suggestive of an albedo greater than 0.2. However, this simple test is contradicted by the modeling which does not indicate a strong signal from cloud opacity on the dayside atmosphere. While partial limb coverage of silicate clouds would increase the albedo of the planet it would likely not explain the intensity of the temperature increase on its own. 

WASP-17\,b's atmosphere is exceptionally inflated 
($R_p\sim$1.9$R_J$, $M_p\sim$0.4$M_J$) compared to 
the bulk of the hot Jupiter population. The inflation of hot Jupiter radii is attributed to additional heating in the planet's interior that could occur from a variety of processes \citep[e.g.][]{Arras2006}. With an equilibrium temperature of $\sim$1750~K, WASP-17\,b sits near the peak of the inferred connection between $T_{eq}$ and internal luminosity \citep{Thorngren2018, Sarkis2021}. 
WASP-17\,b's retrograde orbit \citep{Anderson2010} also suggests that it had a unique dynamic history that may have provided additional opportunity for interior heating via tidal dissipation \citep[e.g.][]{Bodenheimer2001}. The low gravity ($g\sim3.16$~m~s$^{-2}$) of WASP-17\,b is additionally predicted to move the radiative convective boundary (RCB) to shallower pressures that are more likely to be probed with our eclipse observations, thus giving us a window into the deep convection region of the atmosphere \citep{Thorngren2019}. 

Our preferred forward model (with T$_{int}$=400\,K) and retrieved pressure-temperature profiles for WASP-17\,b all show a transition in their structure near 100~mbar (right panel, Figure~\ref{fig:Tbright}) consistent with RCB predictions from \citet{Thorngren2019} given WASP-17\,b's equilibrium temperature and gravity. Our free-chemistry atmospheric retrieval models are able to match the flux from WASP-17\,b shortward of 1\,µm by reducing optical opacities and thus probing deeper and hotter regions of the atmosphere at short wavelengths. We estimate that T$_{int}$ should be on the order of 600--700\,K for WASP-17\,b using the formulation to estimate presented in \citet{Thorngren2019}, which is consistent with the significant increase in temperature with pressure we predict in the photospheric region of our atmospheric models (Figure~\ref{fig:Tbright}). %Further modeling efforts will be needed in order to determine the processes giving rise to WASP-17\,b's emission spectrum shortward of 1\,µm, but it is clear that WASP-17\,b's dayside is offering us a unique and important look into hot Jupiter atmospheric physics and chemistry.  

\section{Conclusions} \label{sec:conclusions}

Our study of WASP-17\,b's dayside atmosphere using the NIRISS SOSS instrument has provided a highly precise emission spectrum, with robust spectral features of water vapor (6.4$\sigma$). This high-quality emission spectrum reveals the complexities of WASP-17\,b's atmospheric composition and thermal structure.

We explored different methods for reducing the NIRISS SOSS observations, employing three pipelines: \text{\tt transitspectroscopy}, \text{\tt supreme-SPOON}, and \text{\tt Ahsoka}. While different data reduction choices can lead to slight variations, the overall consistency across these methods confirms the reliability of hot Jupiter emission spectral extraction for NIRISS. This study also highlights the challenges of optical-range emission spectroscopy with JWST, particularly the impact of the Lucy-Sweeney bias. It emphasizes the importance of carefully considering eclipse depth priors, including negative values, to ensure accurate interpretation of transiting exoplanet spectra. This dataset sets a precedent for addressing this issue in emission spectroscopy, and we expect similar challenges in other JWST spectroscopic optical measurements.

Our atmospheric retrieval and forward models consistently favor high-metallicity solutions. Our \texttt{TauREx} and \texttt{POSEIDON} retrievals, using two different temperature-pressure profile parameterizations designed for hot Jupiters and brown dwarfs, reveal unexpectedly high abundances of H$_2$O. While both retrieval prescriptions derive consistent atmospheric compositions, the shape of the retrieved temperature-pressure profile is sensitive to the assumed parameterization in the deep atmosphere. Comparing different retrieval models provides valuable insights into the sensitivity of our results to the assumed shape of the planet's thermal structure, including the influence on inferred chemical abundances and the atmospheric structure.

Transforming the emission spectrum into brightness temperatures further emphasizes the sensitivity of our results to the assumed thermal structure of WASP-17\,b, given the broad pressure range probed by NIRISS SOSS. The atmospheric and internal constraints inferred here are derived from best-fit models to the brightness temperatures as a function of wavelength, initially based on the \text{\tt transitspectroscopy} reduction, which focused on positive eclipse depths. However, by incorporating a ``bootstrap'' methodology, we computed brightness temperature estimates with confidence intervals that account for both positive-only and negative eclipse depths. Despite the large uncertainties in the WASP-17\,b emission spectrum at wavelengths shorter than 1\,µm, the spectrum derived from positive eclipse depths remains inconsistent with an isothermal $\sim$1800 K blackbody. While the upper limits from the negative eclipse depths trend upwards at these shorter wavelengths, they do not confirm a definitive temperature rise. Nevertheless, the brightness temperatures derived from the positive eclipse depths suggest a potential flux excess below 1\,µm, with 30\% of the negative eclipse depth data supporting this trend.% This observation should be interpreted cautiously, as it cannot be definitively confirmed by the upper limits from negative eclipse depths.}%Robust brightness temperatures derived from both positive and negative eclipse depth fittings confirm the flux excess below 1\,µm. 

The excess emission below 1\,µm could be attributed to a high internal temperature, consistent with an inflated atmosphere for a planet in a retrograde orbit. Additionally, reflected light could also provide a reasonable explanation for the flux excess. Our current retrievals do not account for the possibility of reflected light, highlighting the need for further modeling efforts to fully understand the processes behind WASP-17\,b's emission spectrum shortward of 1\,µm. Nonetheless, it is clear that WASP-17\,b's dayside provides a unique and valuable perspective on hot Jupiter atmospheric physics and chemistry.

Looking ahead, a complete emission spectrum for WASP-17\,b spanning 0.5-12\,µm, as our JWST-TST DREAMS observations is constructing, will enhance our understanding of internal temperature contributions and reflected light. Our study paves the way for further investigations, highlighting the need to refine models and interpretations as we continue to enhance our understanding of hot Jupiter atmospheres through more precise observations with JWST.

%\begin{acknowledgments}
\section*{Acknowledgments}

%The team acknowledges useful conversations with A. Burgasser, others?
This paper reports work carried out in the context of the GTO Science Program (Co-PIs:
R.\,van der Marel, M.\,Perrin, N.\,Lewis) of the Team (see http://www.stsci.edu/$\sim$marel/jwsttelsciteam.html) of the JWST Telescope Scientist (M.\,Mountain).

The JWST data presented in this paper were obtained from the Mikulski Archive for Space Telescopes (MAST) at the Space Telescope Science Institute. The specific observations analyzed can be accessed via  \dataset[DOI: 10.17909/7d90-w905]{http://dx.doi.org/10.17909/7d90-w905}.

%This paper reports work carried out in the context of the JWST Telescope Scientist Team\footnote{\url{https://www.stsci.edu/~marel/jwsttelsciteam.htm}} (Co-PIs: R. van der Marel, M. Perrin, N. Lewis). 
Funding is provided to the team by NASA through grant 80NSSC20K0586. Based on observations with the NASA/ESA/CSA JWST, associated with program(s) JWST-GTO-1353 (PI: N.\,Lewis), obtained at the Space Telescope Science Institute, which is operated by AURA, Inc., under NASA contract NAS 5-03127.

R.J.M. is supported by NASA through the NASA Hubble Fellowship grant HST-HF2-51513.001, awarded by the Space Telescope Science Institute, which is operated by the Association of Universities for Research in Astronomy, Inc., for NASA, under contract NAS 5-26555.
H.R.W. was funded by UK Research and Innovation (UKRI) under the UK government’s Horizon Europe funding guarantee for an ERC Starter Grant [grant number EP/Y006313/1].
D.R.L. acknowledges research support by an appointment to the NASA Postdoctoral Program at the NASA Goddard Space Flight Center (GSFC), administered by Oak Ridge Associated Universities (ORAU) under contract with NASA. Additionally, DRL  acknowledges support from the GSFC Sellers Exoplanet Environments Collaboration (SEEC), which is supported by NASA's Planetary Science Division's Research Program. DRL also acknowledges support by NASA under award number 80GSFC21M0002.

We extend gratitude to the anonymous referee for their detailed feedback and fruitful dialogue that materially improved this study.

%\end{acknowledgments} 

\vspace{5mm}
\facilities{JWST(NIRISS SOSS)}

%\software{ }

\bibliography{WASP17_JWST}{}

\begin{thebibliography}{}
\expandafter\ifx\csname natexlab\endcsname\relax\def\natexlab#1{#1}\fi
\providecommand{\url}[1]{\href{#1}{#1}}
\providecommand{\dodoi}[1]{doi:~\href{http://doi.org/#1}{\nolinkurl{#1}}}
\providecommand{\doeprint}[1]{\href{http://ascl.net/#1}{\nolinkurl{http://ascl.net/#1}}}
\providecommand{\doarXiv}[1]{\href{https://arxiv.org/abs/#1}{\nolinkurl{https://arxiv.org/abs/#1}}}

\bibitem[{Al-Refaie {et~al.}(2021)Al-Refaie, Changeat, Waldmann, \&
  Tinetti}]{Al_Refaie_2021}
Al-Refaie, A.~F., Changeat, Q., Waldmann, I.~P., \& Tinetti, G. 2021, The
  Astrophysical Journal, 917, 37, \dodoi{10.3847/1538-4357/ac0252}

\bibitem[{{Albert} {et~al.}(2023){Albert}, {Lafreni{\`e}re}, {Ren{\'e}},
  {Artigau}, {Volk}, {Goudfrooij}, {Martel}, {Radica}, {Rowe}, {Espinoza},
  {Roy}, {Filippazzo}, {Darveau-Bernier}, {Talens}, {Sivaramakrishnan},
  {Willott}, {Fullerton}, {LaMassa}, {Hutchings}, {Rowlands}, {Vila}, {Zhou},
  {Aldridge}, {Maszkiewicz}, {Beaulieu}, {Cook}, {Piaulet}, {Roy},
  {Lamontagne}, {Morel}, {Frost}, {Salhi}, {Coulombe}, {Benneke}, {MacDonald},
  {Johnstone}, {Turner}, {Fournier-Tondreau}, {Allart}, \&
  {Kaltenegger}}]{Albert2023}
{Albert}, L., {Lafreni{\`e}re}, D., {Ren{\'e}}, D., {et~al.} 2023, \pasp, 135,
  075001, \dodoi{10.1088/1538-3873/acd7a3}

\bibitem[{{Alderson} {et~al.}(2022){Alderson}, {Wakeford}, {MacDonald},
  {Lewis}, {May}, {Grant}, {Sing}, {Stevenson}, {Fowler}, {Goyal}, {Batalha},
  \& {Kataria}}]{Alderson2022}
{Alderson}, L., {Wakeford}, H.~R., {MacDonald}, R.~J., {et~al.} 2022, \mnras,
  512, 4185, \dodoi{10.1093/mnras/stac661}

\bibitem[{{Allard} {et~al.}(2012){Allard}, {Homeier}, \&
  {Freytag}}]{Allard2012}
{Allard}, F., {Homeier}, D., \& {Freytag}, B. 2012, Philosophical Transactions
  of the Royal Society of London Series A, 370, 2765,
  \dodoi{10.1098/rsta.2011.0269}

\bibitem[{{Ambikasaran} {et~al.}(2015){Ambikasaran}, {Foreman-Mackey},
  {Greengard}, {Hogg}, \& {O'Neil}}]{Ambikasaran_2015}
{Ambikasaran}, S., {Foreman-Mackey}, D., {Greengard}, L., {Hogg}, D.~W., \&
  {O'Neil}, M. 2015, IEEE Transactions on Pattern Analysis and Machine
  Intelligence, 38, 252, \dodoi{10.1109/TPAMI.2015.2448083}

\bibitem[{{Amundsen} {et~al.}(2014){Amundsen}, {Baraffe}, {Tremblin},
  {Manners}, {Hayek}, {Mayne}, \& {Acreman}}]{Amundsen2014}
{Amundsen}, D.~S., {Baraffe}, I., {Tremblin}, P., {et~al.} 2014, \aap, 564,
  A59, \dodoi{10.1051/0004-6361/201323169}

\bibitem[{{Anderson} {et~al.}(2010){Anderson}, {Hellier}, {Gillon}, {Triaud},
  {Smalley}, {Hebb}, {Collier Cameron}, {Maxted}, {Queloz}, {West}, {Bentley},
  {Enoch}, {Horne}, {Lister}, {Mayor}, {Parley}, {Pepe}, {Pollacco},
  {S{\'e}gransan}, {Udry}, \& {Wilson}}]{Anderson2010}
{Anderson}, D.~R., {Hellier}, C., {Gillon}, M., {et~al.} 2010, \apj, 709, 159,
  \dodoi{10.1088/0004-637X/709/1/159}

\bibitem[{Anderson {et~al.}(2011)Anderson, Smith, Lanotte, Barman, Cameron,
  Campo, Gillon, Harrington, Hellier, Maxted, Queloz, Triaud, \&
  Wheatley}]{Anderson_2011}
Anderson, D.~R., Smith, A. M.~S., Lanotte, A.~A., {et~al.} 2011, Monthly
  Notices of the Royal Astronomical Society, 416, 2108–2122,
  \dodoi{10.1111/j.1365-2966.2011.19182.x}

\bibitem[{Angerhausen {et~al.}(2015)Angerhausen, DeLarme, \&
  Morse}]{Angerhausen_2015}
Angerhausen, D., DeLarme, E., \& Morse, J.~A. 2015, Publications of the
  Astronomical Society of the Pacific, 127, 1113–1130, \dodoi{10.1086/683797}

\bibitem[{Arcangeli {et~al.}(2018)Arcangeli, Désert, Line, Bean, Parmentier,
  Stevenson, Kreidberg, Fortney, Mansfield, \& Showman}]{Arcangeli_2018}
Arcangeli, J., Désert, J.-M., Line, M.~R., {et~al.} 2018, The Astrophysical
  Journal Letters, 855, L30, \dodoi{10.3847/2041-8213/aab272}

\bibitem[{{Arras} \& {Bildsten}(2006)}]{Arras2006}
{Arras}, P., \& {Bildsten}, L. 2006, \apj, 650, 394, \dodoi{10.1086/506011}

\bibitem[{{Barber} {et~al.}(2014){Barber}, {Strange}, {Hill}, {Polyansky},
  {Mellau}, {Yurchenko}, \& {Tennyson}}]{Barber_2014}
{Barber}, R.~J., {Strange}, J.~K., {Hill}, C., {et~al.} 2014, \mnras, 437,
  1828, \dodoi{10.1093/mnras/stt2011}

\bibitem[{Baxter {et~al.}(2020)Baxter, Désert, Parmentier, Line, Fortney,
  Arcangeli, Bean, Todorov, \& Mansfield}]{Baxter_2020}
Baxter, C., Désert, J.-M., Parmentier, V., {et~al.} 2020, Astronomy \&
  Astrophysics, 639, A36, \dodoi{10.1051/0004-6361/201937394}

\bibitem[{{Bell} {et~al.}(2017){Bell}, {Nikolov}, {Cowan}, {Barstow}, {Barman},
  {Crossfield}, {Gibson}, {Evans}, {Sing}, {Knutson}, {Kataria}, {Lothringer},
  {Benneke}, \& {Schwartz}}]{Bell2017_W12}
{Bell}, T.~J., {Nikolov}, N., {Cowan}, N.~B., {et~al.} 2017, \apjl, 847, L2,
  \dodoi{10.3847/2041-8213/aa876c}

\bibitem[{Bell {et~al.}(2022)Bell, Ahrer, Brande, Carter, Feinstein, {Guzman
  Caloca}, Mansfield, Zieba, Piaulet, Benneke, Filippazzo, May, Roy, Kreidberg,
  \& Stevenson}]{Bell2022}
Bell, T.~J., Ahrer, E.-M., Brande, J., {et~al.} 2022, Journal of Open Source
  Software, 7, 4503, \dodoi{10.21105/joss.04503}

\bibitem[{{Bodenheimer} {et~al.}(2001){Bodenheimer}, {Lin}, \&
  {Mardling}}]{Bodenheimer2001}
{Bodenheimer}, P., {Lin}, D.~N.~C., \& {Mardling}, R.~A. 2001, \apj, 548, 466,
  \dodoi{10.1086/318667}

\bibitem[{{Buchner} {et~al.}(2014){Buchner}, {Georgakakis}, {Nandra}, {Hsu},
  {Rangel}, {Brightman}, {Merloni}, {Salvato}, {Donley}, \&
  {Kocevski}}]{Buchner_2014}
{Buchner}, J., {Georgakakis}, A., {Nandra}, K., {et~al.} 2014, \aap, 564, A125,
  \dodoi{10.1051/0004-6361/201322971}

\bibitem[{{Bushouse} {et~al.}(2023){Bushouse}, {Eisenhamer}, {Dencheva},
  {Davies}, {Greenfield}, {Morrison}, {Hodge}, {Simon}, {Grumm}, {Droettboom},
  {Slavich}, {Sosey}, {Pauly}, {Miller}, {Jedrzejewski}, {Hack}, {Davis},
  {Crawford}, {Law}, {Gordon}, {Regan}, {Cara}, {MacDonald}, {Bradley},
  {Shanahan}, {Jamieson}, {Teodoro}, {Williams}, \&
  {Pena-Guerrero}}]{Bushouse2023}
{Bushouse}, H., {Eisenhamer}, J., {Dencheva}, N., {et~al.} 2023, {JWST
  Calibration Pipeline}, 1.12.0, Zenodo,  Zenodo,
  \dodoi{10.5281/zenodo.6984365}

\bibitem[{{Cahoy} {et~al.}(2010){Cahoy}, {Marley}, \& {Fortney}}]{Cahoy2010}
{Cahoy}, K.~L., {Marley}, M.~S., \& {Fortney}, J.~J. 2010, \apj, 724, 189,
  \dodoi{10.1088/0004-637X/724/1/189}

\bibitem[{Changeat \& Edwards(2021)}]{Changeat_2021}
Changeat, Q., \& Edwards, B. 2021, The Astrophysical Journal Letters, 907, L22,
  \dodoi{10.3847/2041-8213/abd84f}

\bibitem[{Changeat {et~al.}(2022)Changeat, Edwards, Al-Refaie, Tsiaras,
  Skinner, Cho, Yip, Anisman, Ikoma, Bieger, Venot, Shibata, Waldmann, \&
  Tinetti}]{Changeat_2022}
Changeat, Q., Edwards, B., Al-Refaie, A.~F., {et~al.} 2022, The Astrophysical
  Journal Supplement Series, 260, 3, \dodoi{10.3847/1538-4365/ac5cc2}

\bibitem[{Charbonneau {et~al.}(2005)Charbonneau, Allen, Megeath, Torres,
  Alonso, Brown, Gilliland, Latham, Mandushev, O’Donovan, \&
  Sozzetti}]{Charbonneau_2005}
Charbonneau, D., Allen, L.~E., Megeath, S.~T., {et~al.} 2005, The Astrophysical
  Journal, 626, 523–529, \dodoi{10.1086/429991}

\bibitem[{Chubb {et~al.}(2021)Chubb, Rocchetto, Yurchenko, Min, Waldmann,
  Barstow, Mollière, Al-Refaie, Phillips, \& Tennyson}]{Chubb_2021}
Chubb, K.~L., Rocchetto, M., Yurchenko, S.~N., {et~al.} 2021, Astronomy \&
  Astrophysics, 646, A21, \dodoi{10.1051/0004-6361/202038350}

\bibitem[{Coles {et~al.}(2019)Coles, Yurchenko, \& Tennyson}]{Coles_2019}
Coles, P.~A., Yurchenko, S.~N., \& Tennyson, J. 2019, Monthly Notices of the
  Royal Astronomical Society, 490, 4638–4647, \dodoi{10.1093/mnras/stz2778}

\bibitem[{{Coulombe} {et~al.}(2023){Coulombe}, {Benneke}, {Challener},
  {Piette}, {Wiser}, {Mansfield}, {MacDonald}, {Beltz}, {Feinstein}, {Radica},
  {Savel}, {Dos Santos}, {Bean}, {Parmentier}, {Wong}, {Rauscher}, {Komacek},
  {Kempton}, {Tan}, {Hammond}, {Lewis}, {Line}, {Lee}, {Shivkumar},
  {Crossfield}, {Nixon}, {Rackham}, {Wakeford}, {Welbanks}, {Zhang}, {Batalha},
  {Berta-Thompson}, {Changeat}, {D{\'e}sert}, {Espinoza}, {Goyal},
  {Harrington}, {Knutson}, {Kreidberg}, {L{\'o}pez-Morales}, {Shporer}, {Sing},
  {Stevenson}, {Aggarwal}, {Ahrer}, {Alam}, {Bell}, {Blecic}, {Caceres},
  {Carter}, {Casewell}, {Crouzet}, {Cubillos}, {Decin}, {Fortney}, {Gibson},
  {Heng}, {Henning}, {Iro}, {Kendrew}, {Lagage}, {Leconte}, {Lendl},
  {Lothringer}, {Mancini}, {Mikal-Evans}, {Molaverdikhani}, {Nikolov}, {Ohno},
  {Palle}, {Piaulet}, {Redfield}, {Roy}, {Tsai}, {Venot}, \&
  {Wheatley}}]{Coulombe2023}
{Coulombe}, L.-P., {Benneke}, B., {Challener}, R., {et~al.} 2023, \nat, 620,
  292, \dodoi{10.1038/s41586-023-06230-1}

\bibitem[{{Cowan} \& {Agol}(2011)}]{Cowan2011}
{Cowan}, N.~B., \& {Agol}, E. 2011, \apj, 729, 54,
  \dodoi{10.1088/0004-637X/729/1/54}

\bibitem[{{Darveau-Bernier} {et~al.}(2022){Darveau-Bernier}, {Albert},
  {Talens}, {Lafreni{\`e}re}, {Radica}, {Doyon}, {Cook}, {Rowe}, {Allart},
  {Artigau}, {Benneke}, {Cowan}, {Dang}, {Espinoza}, {Johnstone},
  {Kaltenegger}, {Lim}, {Pauly}, {Pelletier}, {Piaulet}, {Roy}, {Roy},
  {Splinter}, {Taylor}, \& {Turner}}]{Darveau-Bernier2022}
{Darveau-Bernier}, A., {Albert}, L., {Talens}, G.~J., {et~al.} 2022, \pasp,
  134, 094502, \dodoi{10.1088/1538-3873/ac8a77}

\bibitem[{Deming {et~al.}(2023)Deming, Line, Knutson, Crossfield, Kempton,
  Komacek, Wallack, \& Fu}]{Deming_2023}
Deming, D., Line, M.~R., Knutson, H.~A., {et~al.} 2023, The Astronomical
  Journal, 165, 104, \dodoi{10.3847/1538-3881/acb210}

\bibitem[{Deming {et~al.}(2005)Deming, Seager, Richardson, \&
  Harrington}]{Deming_2005}
Deming, D., Seager, S., Richardson, L.~J., \& Harrington, J. 2005, Nature, 434,
  740–743, \dodoi{10.1038/nature03507}

\bibitem[{Deming {et~al.}(2015)Deming, Knutson, Kammer, Fulton, Ingalls, Carey,
  Burrows, Fortney, Todorov, Agol, Cowan, Desert, Fraine, Langton, Morley, \&
  Showman}]{Deming_2015}
Deming, D., Knutson, H., Kammer, J., {et~al.} 2015, The Astrophysical Journal,
  805, 132, \dodoi{10.1088/0004-637x/805/2/132}

\bibitem[{{Doyon} {et~al.}(2023){Doyon}, {Willott}, {Hutchings},
  {Sivaramakrishnan}, {Albert}, {Lafreni{\`e}re}, {Rowlands}, {Bego{\~n}a
  Vila}, {Martel}, {LaMassa}, {Aldridge}, {Artigau}, {Cameron}, {Chayer},
  {Cook}, {Cooper}, {Darveau-Bernier}, {Dupuis}, {Earnshaw}, {Espinoza},
  {Filippazzo}, {Fullerton}, {Gaudreau}, {Gawlik}, {Goudfrooij}, {Haley},
  {Kammerer}, {Kendall}, {Lambros}, {Ignat}, {Maszkiewicz}, {McColgan},
  {Morishita}, {Ouellette}, {Pacifici}, {Philippi}, {Radica}, {Ravindranath},
  {Rowe}, {Roy}, {Roy}, {Saad}, {Sohn}, {Talens}, {Touahri}, {Thatte},
  {Taylor}, {Vandal}, {Volk}, {Wander}, {Warner}, {Zheng}, {Zhou}, {Abraham},
  {Beaulieu}, {Benneke}, {Ferrarese}, {Jayawardhana}, {Johnstone},
  {Kaltenegger}, {Meyer}, {Pipher}, {Rameau}, {Rieke}, {Salhi}, \&
  {Sawicki}}]{Doyon2023}
{Doyon}, R., {Willott}, C.~J., {Hutchings}, J.~B., {et~al.} 2023, \pasp, 135,
  098001, \dodoi{10.1088/1538-3873/acd41b}

\bibitem[{{Drummond} {et~al.}(2016){Drummond}, {Tremblin}, {Baraffe},
  {Amundsen}, {Mayne}, {Venot}, \& {Goyal}}]{Drummond2016}
{Drummond}, B., {Tremblin}, P., {Baraffe}, I., {et~al.} 2016, \aap, 594, A69,
  \dodoi{10.1051/0004-6361/201628799}

\bibitem[{Eastman {et~al.}(2013)Eastman, Gaudi, \& Agol}]{Eastman_2013}
Eastman, J., Gaudi, B.~S., \& Agol, E. 2013, Publications of the Astronomical
  Society of the Pacific, 125, 83–112, \dodoi{10.1086/669497}

\bibitem[{Edwards {et~al.}(2020)Edwards, Changeat, Baeyens, Tsiaras, Al-Refaie,
  Taylor, Yip, Bieger, Blain, Gressier, Guilluy, Jaziri, Kiefer,
  Modirrousta-Galian, Morvan, Mugnai, Pluriel, Poveda, Skaf, Whiteford, Wright,
  Zingales, Charnay, Drossart, Leconte, Venot, Waldmann, \&
  Beaulieu}]{Edwards_2020}
Edwards, B., Changeat, Q., Baeyens, R., {et~al.} 2020, The Astronomical
  Journal, 160, 8, \dodoi{10.3847/1538-3881/ab9225}

\bibitem[{Ehrenreich {et~al.}(2020)Ehrenreich, Lovis, Allart, Zapatero~Osorio,
  Pepe, Cristiani, Rebolo, Santos, Borsa, Demangeon, Dumusque,
  González~Hernández, Casasayas-Barris, Ségransan, Sousa, Abreu, Adibekyan,
  Affolter, Allende~Prieto, Alibert, Aliverti, Alves, Amate, Avila, Baldini,
  Bandy, Benz, Bianco, Bolmont, Bouchy, Bourrier, Broeg, Cabral, Calderone,
  Pallé, Cegla, Cirami, Coelho, Conconi, Coretti, Cumani, Cupani, Dekker,
  Delabre, Deiries, D’Odorico, Di~Marcantonio, Figueira, Fragoso, Genolet,
  Genoni, Génova~Santos, Hara, Hughes, Iwert, Kerber, Knudstrup, Landoni,
  Lavie, Lizon, Lendl, Lo~Curto, Maire, Manescau, Martins, Mégevand, Mehner,
  Micela, Modigliani, Molaro, Monteiro, Monteiro, Moschetti, Müller, Nunes,
  Oggioni, Oliveira, Pariani, Pasquini, Poretti, Rasilla, Redaelli, Riva,
  Santana~Tschudi, Santin, Santos, Segovia~Milla, Seidel, Sosnowska, Sozzetti,
  Spanò, Suárez~Mascareño, Tabernero, Tenegi, Udry, Zanutta, \&
  Zerbi}]{Ehrenreich_2020}
Ehrenreich, D., Lovis, C., Allart, R., {et~al.} 2020, Nature, 580, 597–601,
  \dodoi{10.1038/s41586-020-2107-1}

\bibitem[{Espinoza(2022)}]{espinoza_nestor_2022}
Espinoza, N. 2022, \dodoi{10.5281/zenodo.6960924}

\bibitem[{Espinoza {et~al.}(2019)Espinoza, Kossakowski, \&
  Brahm}]{Espinoza_2019}
Espinoza, N., Kossakowski, D., \& Brahm, R. 2019, Monthly Notices of the Royal
  Astronomical Society, 490, 2262, \dodoi{10.1093/mnras/stz2688}

\bibitem[{{Evans} {et~al.}(2013){Evans}, {Pont}, {Sing}, {Aigrain}, {Barstow},
  {D{\'e}sert}, {Gibson}, {Heng}, {Knutson}, \& {Lecavelier des
  Etangs}}]{Evans2013}
{Evans}, T.~M., {Pont}, F., {Sing}, D.~K., {et~al.} 2013, \apjl, 772, L16,
  \dodoi{10.1088/2041-8205/772/2/L16}

\bibitem[{Evans {et~al.}(2016)Evans, Sing, Wakeford, Nikolov, Ballester,
  Drummond, Kataria, Gibson, Amundsen, \& Spake}]{Evans_2016}
Evans, T.~M., Sing, D.~K., Wakeford, H.~R., {et~al.} 2016, The Astrophysical
  Journal, 822, L4, \dodoi{10.3847/2041-8205/822/1/l4}

\bibitem[{Feinstein {et~al.}(2023)Feinstein, Radica, Welbanks, Murray, Ohno,
  Coulombe, Espinoza, Bean, Teske, Benneke, Line, stamkulov, Saba, Tsiaras,
  Barstow, Fortney, Gao, Knutson, MacDonald, Mikal-Evans, Rackham, Taylor,
  Parmentier, Batalha, Berta-Thompson, Carter, Changeat, dos Santos, Gibson,
  Goyal, Kreidberg, L{\'{o}}pez-Morales, Lothringer, Miguel, Molaverdikhani,
  Moran, Morello, Mukherjee, Sing, Stevenson, Wakeford, Ahrer, Alam, Alderson,
  Allen, Batalha, Bell, Blecic, Brande, Caceres, Casewell, Chubb, Crossfield,
  Crouzet, Cubillos, Decin, D{\'{e}}sert, Harrington, Heng, Henning, Iro,
  Kempton, Kendrew, Kirk, Krick, Lagage, Lendl, Mancini, Mansfield, May, Mayne,
  Nikolov, Palle, dit de~la Roche, Piaulet, Powell, Redfield, Rogers, Roman,
  Roy, Nixon, Schlawin, Tan, Tremblin, Turner, Venot, Waalkes, Wheatley, \&
  Zhang}]{Feinstein_2023}
Feinstein, A.~D., Radica, M., Welbanks, L., {et~al.} 2023, Nature, 614, 670,
  \dodoi{10.1038/s41586-022-05674-1}

\bibitem[{Feroz {et~al.}(2009)Feroz, Hobson, \& Bridges}]{Feroz_2009}
Feroz, F., Hobson, M.~P., \& Bridges, M. 2009, Monthly Notices of the Royal
  Astronomical Society, 398, 1601–1614,
  \dodoi{10.1111/j.1365-2966.2009.14548.x}

\bibitem[{Foreman-Mackey {et~al.}(2017)Foreman-Mackey, Agol, Ambikasaran, \&
  th~Angus}]{Foreman_Mackey_2017}
Foreman-Mackey, D., Agol, E., Ambikasaran, S., \& th~Angus. 2017, The
  Astronomical Journal, 154, 220, \dodoi{10.3847/1538-3881/aa9332}

\bibitem[{{Foreman-Mackey} {et~al.}(2013){Foreman-Mackey}, {Hogg}, {Lang}, \&
  {Goodman}}]{foreman-mackey2013}
{Foreman-Mackey}, D., {Hogg}, D.~W., {Lang}, D., \& {Goodman}, J. 2013, \pasp,
  125, 306, \dodoi{10.1086/670067}

\bibitem[{{Foreman-Mackey} {et~al.}(2019){Foreman-Mackey}, {Farr}, {Sinha},
  {Archibald}, {Hogg}, {Sanders}, {Zuntz}, {Williams}, {Nelson}, {de
  Val-Borro}, {Erhardt}, {Pashchenko}, \& {Pla}}]{foreman-mackey2019}
{Foreman-Mackey}, D., {Farr}, W., {Sinha}, M., {et~al.} 2019, The Journal of
  Open Source Software, 4, 1864, \dodoi{10.21105/joss.01864}

\bibitem[{{Fortney} {et~al.}(2019){Fortney}, {Lupu}, {Morley}, {Freedman}, \&
  {Hood}}]{Fortney_2019}
{Fortney}, J.~J., {Lupu}, R.~E., {Morley}, C.~V., {Freedman}, R.~S., \& {Hood},
  C. 2019, \apjl, 880, L16, \dodoi{10.3847/2041-8213/ab2a10}

\bibitem[{{Fournier-Tondreau} {et~al.}(2023){Fournier-Tondreau}, {MacDonald},
  {Radica}, {Lafreni{\'e}re}, {Welbanks}, {Piaulet}, {Coulombe}, {Allart},
  {Morel}, {Artigau}, {Albert}, {Lim}, {Doyon}, {Benneke}, {Rowe},
  {Darveau-Bernier}, {Cowan}, {Lewis}, {Cook}, {Flagg}, {Genest}, {Pelletier},
  {Johnstone}, {Dang}, {Kaltenegger}, {Taylor}, \&
  {Turner}}]{Fournier-Tondreau_2023}
{Fournier-Tondreau}, M., {MacDonald}, R.~J., {Radica}, M., {et~al.} 2023,
  \mnras, \dodoi{10.1093/mnras/stad3813}

\bibitem[{Gao {et~al.}(2017)Gao, Marley, Zahnle, Robinson, \&
  Lewis}]{gao2017sulfur}
Gao, P., Marley, M.~S., Zahnle, K., Robinson, T.~D., \& Lewis, N.~K. 2017, The
  Astronomical Journal, 153, 139

\bibitem[{{Goyal} {et~al.}(2021){Goyal}, {Lewis}, {Wakeford}, {MacDonald}, \&
  {Mayne}}]{Goyal2021}
{Goyal}, J.~M., {Lewis}, N.~K., {Wakeford}, H.~R., {MacDonald}, R.~J., \&
  {Mayne}, N.~J. 2021, \apj, 923, 242, \dodoi{10.3847/1538-4357/ac27b2}

\bibitem[{{Goyal} {et~al.}(2018){Goyal}, {Mayne}, {Sing}, {Drummond},
  {Tremblin}, {Amundsen}, {Evans}, {Carter}, {Spake}, {Baraffe}, {Nikolov},
  {Manners}, {Chabrier}, \& {Hebrard}}]{Goyal2018}
{Goyal}, J.~M., {Mayne}, N., {Sing}, D.~K., {et~al.} 2018, \mnras, 474, 5158,
  \dodoi{10.1093/mnras/stx3015}

\bibitem[{{Goyal} {et~al.}(2020){Goyal}, {Mayne}, {Drummond}, {Sing},
  {H{\'e}brard}, {Lewis}, {Tremblin}, {Phillips}, {Mikal-Evans}, \&
  {Wakeford}}]{Goyal2020}
{Goyal}, J.~M., {Mayne}, N., {Drummond}, B., {et~al.} 2020, \mnras, 498, 4680,
  \dodoi{10.1093/mnras/staa2300}

\bibitem[{{Grant} {et~al.}(2023){Grant}, {Lewis}, {Wakeford}, {Batalha},
  {Glidden}, {Goyal}, {Mullens}, {MacDonald}, {May}, {Seager}, {Stevenson},
  {Valenti}, {Visscher}, {Alderson}, {Allen}, {Ca{\~n}as}, {Col{\'o}n},
  {Clampin}, {Espinoza}, {Gressier}, {Huang}, {Lin}, {Long}, {Louie},
  {Pe{\~n}a-Guerrero}, {Ranjan}, {Sotzen}, {Valentine}, {Anderson}, {Balmer},
  {Bellini}, {Hoch}, {Kammerer}, {Libralato}, {Mountain}, {Perrin}, {Pueyo},
  {Rickman}, {Rebollido}, {Sohn}, {van der Marel}, \& {Watkins}}]{grant_2023}
{Grant}, D., {Lewis}, N.~K., {Wakeford}, H.~R., {et~al.} 2023, \apjl, 956, L32,
  \dodoi{10.3847/2041-8213/acfc3b}

\bibitem[{Haynes {et~al.}(2015)Haynes, Mandell, Madhusudhan, Deming, \&
  Knutson}]{haynes_2015}
Haynes, K., Mandell, A.~M., Madhusudhan, N., Deming, D., \& Knutson, H. 2015,
  Spectroscopic Evidence for a Temperature Inversion in the Dayside Atmosphere
  of the Hot Jupiter WASP-33b.
\newblock \doarXiv{1505.01490}

\bibitem[{{Hohm}(1994)}]{Hohm_1994}
{Hohm}, U. 1994, Chemical Physics, 179, 533,
  \dodoi{10.1016/0301-0104(94)87028-4}

\bibitem[{Husser {et~al.}(2013)Husser, Wende-von Berg, Dreizler, Homeier,
  Reiners, Barman, \& Hauschildt}]{Husser_2013}
Husser, T.-O., Wende-von Berg, S., Dreizler, S., {et~al.} 2013, Astronomy \&
  Astrophysics, 553, A6, \dodoi{10.1051/0004-6361/201219058}

\bibitem[{{John}(1988)}]{John_1988}
{John}, T.~L. 1988, \aap, 193, 189

\bibitem[{{Karman} {et~al.}(2019){Karman}, {Gordon}, {van der Avoird},
  {Baranov}, {Boulet}, {Drouin}, {Groenenboom}, {Gustafsson}, {Hartmann},
  {Kurucz}, {Rothman}, {Sun}, {Sung}, {Thalman}, {Tran}, {Wishnow},
  {Wordsworth}, {Vigasin}, {Volkamer}, \& {van der Zande}}]{Karman_2019}
{Karman}, T., {Gordon}, I.~E., {van der Avoird}, A., {et~al.} 2019, \icarus,
  328, 160, \dodoi{10.1016/j.icarus.2019.02.034}

\bibitem[{Kesseli \& Snellen(2021)}]{Kesseli_2021}
Kesseli, A.~Y., \& Snellen, I. A.~G. 2021, The Astrophysical Journal Letters,
  908, L17, \dodoi{10.3847/2041-8213/abe047}

\bibitem[{Kesseli {et~al.}(2020)Kesseli, Snellen, Alonso-Floriano, Mollière,
  \& Serindag}]{Kesseli_2020}
Kesseli, A.~Y., Snellen, I. A.~G., Alonso-Floriano, F.~J., Mollière, P., \&
  Serindag, D.~B. 2020, The Astronomical Journal, 160, 228,
  \dodoi{10.3847/1538-3881/abb59c}

\bibitem[{Kirkpatrick {et~al.}(1999)Kirkpatrick, Reid, Liebert, Cutri,
  Beichman, Dahn, Monet, Gizis, \& Skrutskie}]{Kirkpatrick_1999}
Kirkpatrick, J.~D., Reid, I.~N., Liebert, J., {et~al.} 1999, The Astrophysical
  Journa, 519, 802, \dodoi{10.1086/307414}

\bibitem[{Kreidberg(2015)}]{Kreidberg_2015}
Kreidberg, L. 2015, Publications of the Astronomical Society of the Pacific,
  127, 1161, \dodoi{10.1086/683602}

\bibitem[{Kreidberg {et~al.}(2014)Kreidberg, Bean, Désert, Line, Fortney,
  Madhusudhan, Stevenson, Showman, Charbonneau, McCullough, Seager, Burrows,
  Henry, Williamson, Kataria, \& Homeier}]{Kreidberg_2014}
Kreidberg, L., Bean, J.~L., Désert, J.-M., {et~al.} 2014, The Astrophysical
  Journal, 793, L27, \dodoi{10.1088/2041-8205/793/2/l27}

\bibitem[{Li {et~al.}(2015)Li, Gordon, Rothman, Tan, Hu, Kassi, Campargue, \&
  Medvedev}]{Li_2015}
Li, G., Gordon, I.~E., Rothman, L.~S., {et~al.} 2015, The Astrophysical Journal
  Supplement Series, 216, 15

\bibitem[{Libralato {et~al.}(2023)Libralato, Bellini, van~der Marel, Anderson,
  Sohn, Watkins, Alderson, Allen, Clampin, Glidden, Goyal, Hoch, Huang,
  Kammerer, Lewis, Lin, Long, Louie, MacDonald, Mountain, Peña-Guerrero,
  Perrin, Pueyo, Rebollido, Rickman, Seager, Stevenson, Valenti, Valentine, \&
  Wakeford}]{Libralato_2023}
Libralato, M., Bellini, A., van~der Marel, R.~P., {et~al.} 2023, The
  Astrophysical Journal, 950, 101, \dodoi{10.3847/1538-4357/acd04f}

\bibitem[{{Lim} {et~al.}(2023){Lim}, {Benneke}, {Doyon}, {MacDonald},
  {Piaulet}, {Artigau}, {Coulombe}, {Radica}, {L'Heureux}, {Albert}, {Rackham},
  {de Wit}, {Salhi}, {Roy}, {Flagg}, {Fournier-Tondreau}, {Taylor}, {Cook},
  {Lafreni{\`e}re}, {Cowan}, {Kaltenegger}, {Rowe}, {Espinoza}, {Dang}, \&
  {Darveau-Bernier}}]{Lim2023}
{Lim}, O., {Benneke}, B., {Doyon}, R., {et~al.} 2023, \apjl, 955, L22,
  \dodoi{10.3847/2041-8213/acf7c4}

\bibitem[{MacDonald(2023)}]{MacDonald_2023}
MacDonald, R.~J. 2023, Journal of Open Source Software,,
  \dodoi{10.21105/joss.04873}

\bibitem[{{MacDonald} \& {Lewis}(2022)}]{MacDonald_2022}
{MacDonald}, R.~J., \& {Lewis}, N.~K. 2022, \apj, 929, 20,
  \dodoi{10.3847/1538-4357/ac47fe}

\bibitem[{MacDonald \& Madhusudhan(2017)}]{MacDonald_2017}
MacDonald, R.~J., \& Madhusudhan, N. 2017, Monthly Notices of the Royal
  Astronomical Society, 469, 1979–1996, \dodoi{10.1093/mnras/stx804}

\bibitem[{Madhusudhan \& Seager(2009)}]{Madhusudhan_2009}
Madhusudhan, N., \& Seager, S. 2009, The Astrophysical Journal, 707, 24–39,
  \dodoi{10.1088/0004-637x/707/1/24}

\bibitem[{{Mallonn} {et~al.}(2019){Mallonn}, {K{\"o}hler}, {Alexoudi}, {von
  Essen}, {Granzer}, {Poppenhaeger}, \& {Strassmeier}}]{Mallonn2019}
{Mallonn}, M., {K{\"o}hler}, J., {Alexoudi}, X., {et~al.} 2019, \aap, 624, A62,
  \dodoi{10.1051/0004-6361/201935079}

\bibitem[{Mandell {et~al.}(2013)Mandell, Haynes, Sinukoff, Madhusudhan,
  Burrows, \& Deming}]{Mandell_2013}
Mandell, A.~M., Haynes, K., Sinukoff, E., {et~al.} 2013, The Astrophysical
  Journal, 779, 128, \dodoi{10.1088/0004-637x/779/2/128}

\bibitem[{Mansfield {et~al.}(2018)Mansfield, Bean, Line, Parmentier, Kreidberg,
  Désert, Fortney, Stevenson, Arcangeli, \& Dragomir}]{Mansfield_2018}
Mansfield, M., Bean, J.~L., Line, M.~R., {et~al.} 2018, The Astronomical
  Journal, 156, 10, \dodoi{10.3847/1538-3881/aac497}

\bibitem[{McKemmish {et~al.}(2019)McKemmish, Masseron, Hoeijmakers,
  P{\'{e}}rez-Mesa, Grimm, Yurchenko, \& Tennyson}]{McKemmish_2019}
McKemmish, L.~K., Masseron, T., Hoeijmakers, H.~J., {et~al.} 2019, Monthly
  Notices of the Royal Astronomical Society, 488, 2836,
  \dodoi{10.1093/mnras/stz1818}

\bibitem[{McKemmish {et~al.}(2016)McKemmish, Yurchenko, \&
  Tennyson}]{McKemmish_2016}
McKemmish, L.~K., Yurchenko, S.~N., \& Tennyson, J. 2016, Monthly Notices of
  the Royal Astronomical Society, 463, 771, \dodoi{10.1093/mnras/stw1969}

\bibitem[{Mikal-Evans {et~al.}(2019)Mikal-Evans, Sing, Goyal, Drummond, Carter,
  Henry, Wakeford, Lewis, Marley, Tremblin, Nikolov, Kataria, Deming, \&
  Ballester}]{Mikal_Evans_2019}
Mikal-Evans, T., Sing, D.~K., Goyal, J.~M., {et~al.} 2019, Monthly Notices of
  the Royal Astronomical Society, 488, 2222–2234,
  \dodoi{10.1093/mnras/stz1753}

\bibitem[{Parmentier {et~al.}(2016)Parmentier, Fortney, Showman, Morley, \&
  Marley}]{parmentier2016transitions}
Parmentier, V., Fortney, J.~J., Showman, A.~P., Morley, C., \& Marley, M.~S.
  2016, The Astrophysical Journal, 828, 22

\bibitem[{{Piette} \& {Madhusudhan}(2020)}]{Piette2020}
{Piette}, A. A.~A., \& {Madhusudhan}, N. 2020, \mnras, 497, 5136,
  \dodoi{10.1093/mnras/staa2289}

\bibitem[{Pluriel {et~al.}(2020)Pluriel, Whiteford, Edwards, Changeat, Yip,
  Baeyens, Al-Refaie, Fabienne~Bieger, Blain, Gressier, Guilluy, Yassin~Jaziri,
  Kiefer, Modirrousta-Galian, Morvan, Mugnai, Poveda, Skaf, Zingales, Wright,
  Charnay, Drossart, Leconte, Tsiaras, Venot, Waldmann, \&
  Beaulieu}]{Pluriel_2020}
Pluriel, W., Whiteford, N., Edwards, B., {et~al.} 2020, The Astronomical
  Journal, 160, 112, \dodoi{10.3847/1538-3881/aba000}

\bibitem[{Polyansky {et~al.}(2018)Polyansky, Kyuberis, Zobov, Tennyson,
  Yurchenko, \& Lodi}]{Polyansky_2018}
Polyansky, O.~L., Kyuberis, A.~A., Zobov, N.~F., {et~al.} 2018, Monthly Notices
  of the Royal Astronomical Society, 480, 2597, \dodoi{10.1093/mnras/sty1877}

\bibitem[{{Radica}(2024)}]{Radica2024_exotedrf}
{Radica}, M. 2024, arXiv e-prints, arXiv:2407.17541,
  \dodoi{10.48550/arXiv.2407.17541}

\bibitem[{{Radica} {et~al.}(2022){Radica}, {Albert}, {Taylor},
  {Lafreni{\`e}re}, {Coulombe}, {Darveau-Bernier}, {Doyon}, {Cook}, {Cowan},
  {Espinoza}, {Johnstone}, {Kaltenegger}, {Piaulet}, {Roy}, \&
  {Talens}}]{Radica2022}
{Radica}, M., {Albert}, L., {Taylor}, J., {et~al.} 2022, \pasp, 134, 104502,
  \dodoi{10.1088/1538-3873/ac9430}

\bibitem[{{Radica} {et~al.}(2023){Radica}, {Welbanks}, {Espinoza}, {Taylor},
  {Coulombe}, {Feinstein}, {Goyal}, {Scarsdale}, {Albert}, {Baghel}, {Bean},
  {Blecic}, {Lafreni{\`e}re}, {MacDonald}, {Zamyatina}, {Allart1}, {Artigau},
  {Batalha}, {Cook}, {Cowan}, {Dang}, {Doyon}, {Fournier-Tondreau},
  {Johnstone}, {Line}, {Moran}, {Mukherjee}, {Pelletier}, {Roy}, {Talens},
  {Filippazzo}, {Pontoppidan}, \& {Volk}}]{Radica2023}
{Radica}, M., {Welbanks}, L., {Espinoza}, N., {et~al.} 2023, \mnras, 524, 835,
  \dodoi{10.1093/mnras/stad1762}

\bibitem[{{Radica} {et~al.}(2024){Radica}, {Coulombe}, {Taylor}, {Albert},
  {Allart}, {Benneke}, {Cowan}, {Dang}, {Lafreni{\`e}re}, {Thorngren},
  {Artigau}, {Doyon}, {Flagg}, {Johnstone}, {Pelletier}, \&
  {Roy}}]{Radica_2024}
{Radica}, M., {Coulombe}, L.-P., {Taylor}, J., {et~al.} 2024, \apjl, 962, L20,
  \dodoi{10.3847/2041-8213/ad20e4}

\bibitem[{{Rebollido} {et~al.}(2024){Rebollido}, {Stark}, {Kammerer}, {Perrin},
  {Lawson}, {Pueyo}, {Chen}, {Hines}, {Girard}, {Worthen}, {Ingerbretsen},
  {Betti}, {Clampin}, {Golimowski}, {Hoch}, {Lewis}, {Lu}, {van der Marel},
  {Rickman}, {Seager}, {Soummer}, {Valenti}, {Ward-Duong}, \&
  {Mountain}}]{rebollido_2024}
{Rebollido}, I., {Stark}, C.~C., {Kammerer}, J., {et~al.} 2024, \aj, 167, 69,
  \dodoi{10.3847/1538-3881/ad1759}

\bibitem[{Roman \& Rauscher(2019)}]{roman2019modeled}
Roman, M., \& Rauscher, E. 2019, The Astrophysical Journal, 872, 1

\bibitem[{Rothman {et~al.}(1987)Rothman, Gamache, Goldman, Brown, Toth,
  Pickett, Poynter, Flaud, Camy-Peyret, Barbe, Husson, Rinsland, \&
  Smith}]{Rothman_1987}
Rothman, L.~S., Gamache, R.~R., Goldman, A., {et~al.} 1987, Appl. Opt., 26,
  4058, \dodoi{10.1364/AO.26.004058}

\bibitem[{{Rothman} {et~al.}(2010){Rothman}, {Gordon}, {Barber}, {Dothe},
  {Gamache}, {Goldman}, {Perevalov}, {Tashkun}, \& {Tennyson}}]{Rothman_2010}
{Rothman}, L.~S., {Gordon}, I.~E., {Barber}, R.~J., {et~al.} 2010, \jqsrt, 111,
  2139, \dodoi{10.1016/j.jqsrt.2010.05.001}

\bibitem[{{Rothman} {et~al.}(2013){Rothman}, {Gordon}, {Babikov}, {Barbe},
  {Chris Benner}, {Bernath}, {Birk}, {Bizzocchi}, {Boudon}, {Brown},
  {Campargue}, {Chance}, {Cohen}, {Coudert}, {Devi}, {Drouin}, {Fayt}, {Flaud},
  {Gamache}, {Harrison}, {Hartmann}, {Hill}, {Hodges}, {Jacquemart}, {Jolly},
  {Lamouroux}, {Le Roy}, {Li}, {Long}, {Lyulin}, {Mackie}, {Massie},
  {Mikhailenko}, {M{\"u}ller}, {Naumenko}, {Nikitin}, {Orphal}, {Perevalov},
  {Perrin}, {Polovtseva}, {Richard}, {Smith}, {Starikova}, {Sung}, {Tashkun},
  {Tennyson}, {Toon}, {Tyuterev}, \& {Wagner}}]{Rothman_2013}
{Rothman}, L.~S., {Gordon}, I.~E., {Babikov}, Y., {et~al.} 2013, \jqsrt, 130,
  4, \dodoi{10.1016/j.jqsrt.2013.07.002}

\bibitem[{Ruffio {et~al.}(2023)Ruffio, Perrin, Hoch, Kammerer, Konopacky,
  Pueyo, Rickman, Theissen, Agrawal, Greenbaum, Miles, Barman, Balmer,
  Llop-Sayson, Girard, Rebollido, Soummer, Allen, Anderson, Beichman, Bellini,
  Bryden, Espinoza, Glidden, Huang, Lewis, Libralato, Louie, Sohn, Seager,
  van~der Marel, Wakeford, Watkins, Ygouf, \& Mountai}]{ruffio_2023}
Ruffio, J.-B., Perrin, M.~D., Hoch, K. K.~W., {et~al.} 2023, JWST-TST High
  Contrast: Achieving direct spectroscopy of faint substellar companions next
  to bright stars with the NIRSpec IFU.
\newblock \doarXiv{2310.09902}

\bibitem[{{Sarkis} {et~al.}(2021){Sarkis}, {Mordasini}, {Henning}, {Marleau},
  \& {Molli{\`e}re}}]{Sarkis2021}
{Sarkis}, P., {Mordasini}, C., {Henning}, T., {Marleau}, G.~D., \&
  {Molli{\`e}re}, P. 2021, \aap, 645, A79, \dodoi{10.1051/0004-6361/202038361}

\bibitem[{{Schwartz} \& {Cowan}(2015)}]{Schwartz2015}
{Schwartz}, J.~C., \& {Cowan}, N.~B. 2015, \mnras, 449, 4192,
  \dodoi{10.1093/mnras/stv470}

\bibitem[{{Seager}(2010)}]{Seager2010}
{Seager}, S. 2010, {Exoplanet Atmospheres: Physical Processes}

\bibitem[{Sing {et~al.}(2015)Sing, Fortney, Nikolov, Wakeford, Kataria, Evans,
  Aigrain, Ballester, Burrows, Deming, Désert, Gibson, Henry, Huitson,
  Knutson, Etangs, Pont, Showman, Vidal-Madjar, Williamson, \&
  Wilson}]{Sing_2015}
Sing, D.~K., Fortney, J.~J., Nikolov, N., {et~al.} 2015, Nature, 529, 59–62,
  \dodoi{10.1038/nature16068}

\bibitem[{Sing {et~al.}(2019)Sing, Lavvas, Ballester, Lecavelier~des Etangs,
  Marley, Nikolov, Ben-Jaffel, Bourrier, Buchhave, Deming, Ehrenreich,
  Mikal-Evans, Kataria, Lewis, López-Morales, García~Muñoz, Henry,
  Sanz-Forcada, Spake, \& Wakeford}]{Sing_2019}
Sing, D.~K., Lavvas, P., Ballester, G.~E., {et~al.} 2019, The Astronomical
  Journal, 158, 91, \dodoi{10.3847/1538-3881/ab2986}

\bibitem[{Skaf {et~al.}(2020)Skaf, Bieger, Edwards, Changeat, Morvan, Kiefer,
  Blain, Zingales, Poveda, Al-Refaie, Baeyens, Gressier, Guilluy, Jaziri,
  Modirrousta-Galian, Mugnai, Pluriel, Whiteford, Wright, Yip, Charnay,
  Leconte, Drossart, Tsiaras, Venot, Waldmann, \& Beaulieu}]{Skaf_2020}
Skaf, N., Bieger, M.~F., Edwards, B., {et~al.} 2020, The Astronomical Journal,
  160, 109, \dodoi{10.3847/1538-3881/ab94a3}

\bibitem[{Sotzen {et~al.}(2019)Sotzen, Stevenson, Sing, Kilpatrick, Wakeford,
  Filippazzo, Lewis, Hörst, López-Morales, Henry, Buchhave, Ehrenreich,
  Fraine, García~Muñoz, Jayaraman, Lavvas, des Etangs, Marley, Nikolov,
  Rathcke, \& Sanz-Forcada}]{Sotzen_2019}
Sotzen, K.~S., Stevenson, K.~B., Sing, D.~K., {et~al.} 2019, The Astronomical
  Journal, 159, 5, \dodoi{10.3847/1538-3881/ab5442}

\bibitem[{Southworth(2012)}]{southworth2012homogeneous}
Southworth, J. 2012, Monthly Notices of the Royal Astronomical Society, 426,
  1291

\bibitem[{Speagle(2020)}]{Speagle_2020}
Speagle, J.~S. 2020, Monthly Notices of the Royal Astronomical Society, 493,
  3132, \dodoi{10.1093/mnras/staa278}

\bibitem[{Stevenson {et~al.}(2014{\natexlab{a}})Stevenson, Bean, Madhusudhan,
  \& Harrington}]{Stevenson_2014a}
Stevenson, K.~B., Bean, J.~L., Madhusudhan, N., \& Harrington, J.
  2014{\natexlab{a}}, The Astrophysical Journal, 791, 36,
  \dodoi{10.1088/0004-637x/791/1/36}

\bibitem[{Stevenson {et~al.}(2014{\natexlab{b}})Stevenson, Désert, Line, Bean,
  Fortney, Showman, Kataria, Kreidberg, McCullough, Henry, Charbonneau,
  Burrows, Seager, Madhusudhan, Williamson, \& Homeier}]{Stevenson_2014b}
Stevenson, K.~B., Désert, J.-M., Line, M.~R., {et~al.} 2014{\natexlab{b}},
  Science, 346, 838–841, \dodoi{10.1126/science.1256758}

\bibitem[{{Tashkun} \& {Perevalov}(2011)}]{Tashkun_2011}
{Tashkun}, S.~A., \& {Perevalov}, V.~I. 2011, Journal of Quantitative
  Spectroscopy and Radiative Transfer, 112, 1403,
  \dodoi{10.1016/j.jqsrt.2011.03.005}

\bibitem[{Tennyson \& Yurchenko(2018)}]{Tennyson_2018}
Tennyson, J., \& Yurchenko, S. 2018, Atoms, 6, 26, \dodoi{10.3390/atoms6020026}

\bibitem[{Tennyson {et~al.}(2016)Tennyson, Yurchenko, Al-Refaie, Barton, Chubb,
  Coles, Diamantopoulou, Gorman, Hill, Lam, \& et~al.}]{Tennyson_2016}
Tennyson, J., Yurchenko, S.~N., Al-Refaie, A.~F., {et~al.} 2016, Journal of
  Molecular Spectroscopy, 327, 73–94, \dodoi{10.1016/j.jms.2016.05.002}

\bibitem[{{Thorngren} {et~al.}(2019){Thorngren}, {Gao}, \&
  {Fortney}}]{Thorngren2019}
{Thorngren}, D., {Gao}, P., \& {Fortney}, J.~J. 2019, \apjl, 884, L6,
  \dodoi{10.3847/2041-8213/ab43d0}

\bibitem[{{Thorngren} \& {Fortney}(2018)}]{Thorngren2018}
{Thorngren}, D.~P., \& {Fortney}, J.~J. 2018, \aj, 155, 214,
  \dodoi{10.3847/1538-3881/aaba13}

\bibitem[{{Townsend} \& {Lopez}(2023)}]{Townsend_2023}
{Townsend}, R., \& {Lopez}, A. 2023, The Journal of Open Source Software, 8,
  4602, \dodoi{10.21105/joss.04602}

\bibitem[{{Tremblin} {et~al.}(2015){Tremblin}, {Amundsen}, {Mourier},
  {Baraffe}, {Chabrier}, {Drummond}, {Homeier}, \& {Venot}}]{Tremblin2015}
{Tremblin}, P., {Amundsen}, D.~S., {Mourier}, P., {et~al.} 2015, \apjl, 804,
  L17, \dodoi{10.1088/2041-8205/804/1/L17}

\bibitem[{Triaud {et~al.}(2010)Triaud, Collier~Cameron, Queloz, Anderson,
  Gillon, Hebb, Hellier, Loeillet, Maxted, Mayor, Pepe, Pollacco, Ségransan,
  Smalley, Udry, West, \& Wheatley}]{Triaud_2010}
Triaud, A. H. M.~J., Collier~Cameron, A., Queloz, D., {et~al.} 2010, Astronomy
  \& Astrophysics, 524, A25, \dodoi{10.1051/0004-6361/201014525}

\bibitem[{Wende {et~al.}(2010)Wende, Reiners, Seifahrt, \&
  Bernath}]{Wende_2010}
Wende, S., Reiners, A., Seifahrt, A., \& Bernath, P.~F. 2010, Astronomy \&
  Astrophysics, 523, A58, \dodoi{10.1051/0004-6361/201015220}

\bibitem[{{Woitke} {et~al.}(2018){Woitke}, {Helling}, {Hunter}, {Millard},
  {Turner}, {Worters}, {Blecic}, \& {Stock}}]{Woitke2018}
{Woitke}, P., {Helling}, C., {Hunter}, G.~H., {et~al.} 2018, \aap, 614, A1,
  \dodoi{10.1051/0004-6361/201732193}

\bibitem[{Yurchenko {et~al.}(2017)Yurchenko, Amundsen, Tennyson, \&
  Waldmann}]{Yurchenko_2017}
Yurchenko, S.~N., Amundsen, D.~S., Tennyson, J., \& Waldmann, I.~P. 2017,
  Astronomy \& Astrophysics, 605, A95, \dodoi{10.1051/0004-6361/201731026}

\bibitem[{Yurchenko {et~al.}(2011)Yurchenko, Barber, \&
  Tennyson}]{Yurchenko_2011}
Yurchenko, S.~N., Barber, R.~J., \& Tennyson, J. 2011, Monthly Notices of the
  Royal Astronomical Society, 413, 1828–1834,
  \dodoi{10.1111/j.1365-2966.2011.18261.x}

\end{thebibliography}
\bibliographystyle{aasjournal}

\appendix  
\renewcommand\thefigure{\thesection.\arabic{figure}}    

\section{Data reduction intercomparison}

Here, we compare the residuals between different data reduction methods. Figure\,\ref{fig:emission_spec_residuals} displays the residuals between \text{\tt transitspectroscopy}, \text{\tt supreme-SPOON}, and \text{\tt Ahsoka}. The comparison between \text{\tt transitspectroscopy} and \text{\tt supreme-SPOON} is based on fits allowing negative eclipse depths, while the comparison with \text{\tt Ahsoka} uses positive eclipse depths. \text{\tt Ahsoka} shows a notable increase at shorter wavelengths, likely due to poor fits near zero eclipse depth in order 2. The deviations between \text{\tt transitspectroscopy} and the other methods above 2 $\upmu$m may be explained by differences in the pipeline processing steps. 

%\textcolor{red}{Needs at least one sentence in each appendix (AAS Journal policy)}

\setcounter{figure}{0}    
\begin{figure}[htpb]
  \includegraphics[width=\textwidth]{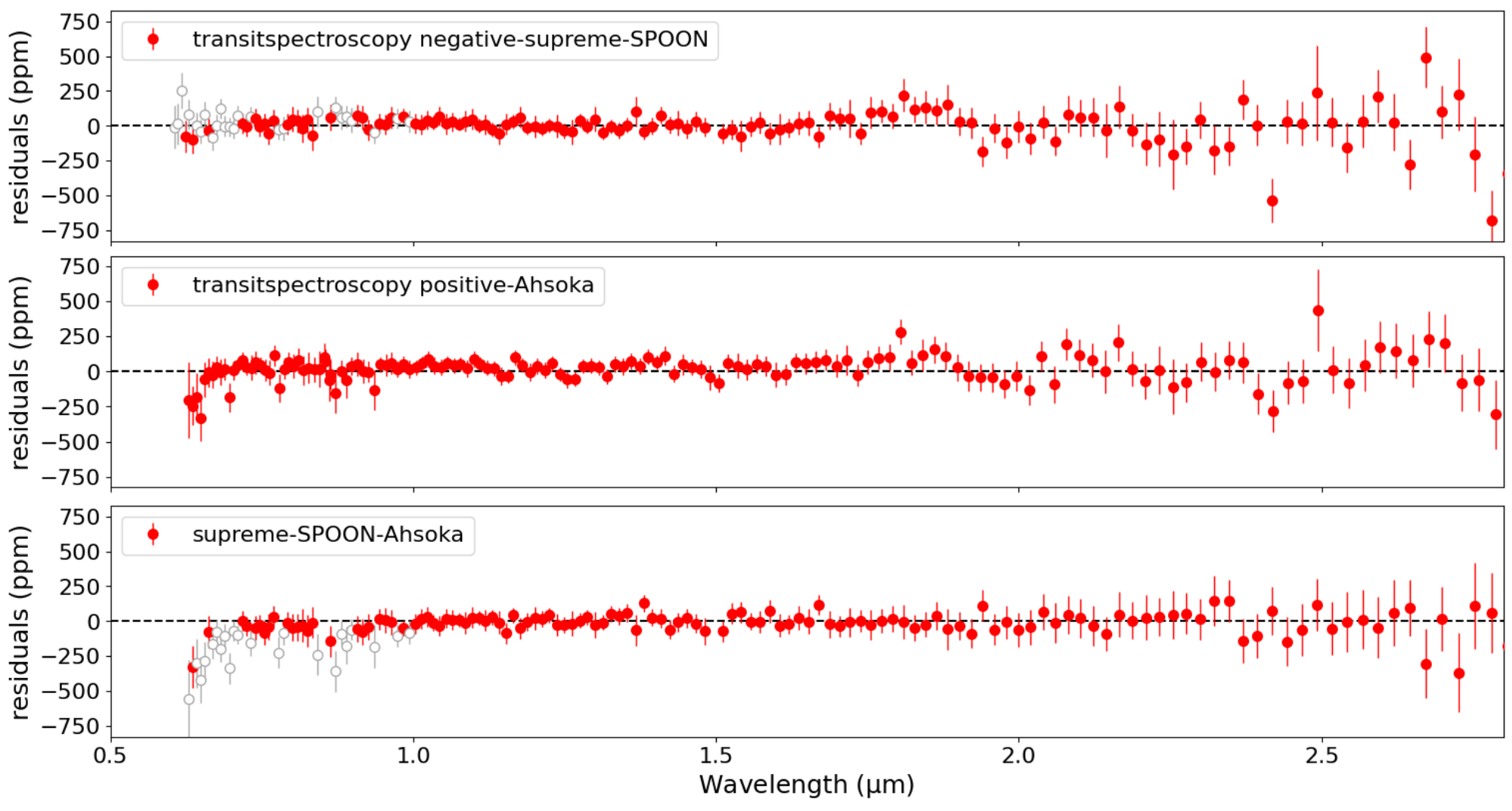}
 \caption{Residuals between reductions. White data points correspond to negative eclipse depth values. }
 \label{fig:emission_spec_residuals}
\end{figure}

%\newpage 
\section{Retrieval priors and results}

This appendix lists the atmospheric retrieval results from our WASP-17b dayside emission spectrum using two Bayesian frameworks: \texttt{TauREx} and \texttt{POSEIDON}. We evaluate the outcomes using two different temperature-pressure (T-P) parameterizations: (i) \citet{Madhusudhan_2009}, and (ii) a custom brown-dwarf-like T-P profile, as described in Sections\,\ref{sec:retrieval:taurex} and\,\ref{sec:retrieval:poseidon}. We ran retrievals for both the positive and negative eclipse depth fitting methods, with both codes, and with the two T-P profile prescriptions, for a total of 8 retrievals (alongside subsidiary nested retrievals for Bayesian model comparison purposes). Table~\ref{table:retrieval_priors} lists the priors for each retrieval, while Table~\ref{table:retrieval_results} presents the results.

\begin{deluxetable*}{lCCCC}
\label{table:retrieval_priors}
\tablecaption{Atmospheric retrieval priors.}
\tablehead{
    Retrieval Code & \multicolumn{2}{|c|}{\texttt{TauREx}} & \multicolumn{2}{c}{\texttt{POSEIDON}} \\
    \hline
    T-P Profile & \multicolumn{1}{|c|}{M\&S (2009) T-P} & \multicolumn{1}{c|}{BD T-P} & \multicolumn{1}{c|}{M\&S (2009) T-P} & \colhead{BD T-P}
}
\startdata
\tablebreak
\hline
Parameter & \multicolumn{4}{|c}{Temperature-Pressure Profile} \\
\hline
$T_{\rm ref}$ (K) & $\mathcal{U}(1000, 3500)$ & \text{---} & $\mathcal{U}(400, 4000)$ & \text{---} \\
$\alpha_{1}$ & $\mathcal{U}(0.02, 1.12)$ & \text{---} & $\mathcal{U}(0.02, 2.0)$ & \text{---} \\
$\alpha_{2}$ & $\mathcal{U}(0.02, 1.12)$ & \text{---} & $\mathcal{U}(0.02, 2.0)$ & \text{---} \\
log(P$_{1}$ / bar) & $\mathcal{U}(-7, 1)$ & \text{---} & $\mathcal{U}(-6, 2)$ & \text{---} \\
log(P$_{2}$ / bar) & $\mathcal{U}(-7, 1)$ & \text{---} & $\mathcal{U}(-6, 2)$ & \text{---} \\
log(P$_{3}$ / bar) & $\mathcal{U}(-7, 1)$ & \text{---} & $\mathcal{U}(-2, 2)$ & \text{---} \\
\tablebreak
$T_{\rm top}$ (K) & \text{---} & $\mathcal{U}(1000, 3500)$ & \text{---} & \text{---} \\
$T_{\rm phot}$ (K) & \text{---} & \text{---} & \text{---} & $\mathcal{U}(400, 3000)$ \\
$\Delta T_{1}$ (K) & \text{---} & $\mathcal{U}(-500, 1000)$ & \text{---} & $\mathcal{U}(0, 1000)$ \\
$\Delta T_{2}$ (K) & \text{---} & $\mathcal{U}(0, 1000)$ & \text{---} & $\mathcal{U}(0, 1000)$ \\
$\Delta T_{3}$ (K) & \text{---} & $\mathcal{U}(0, 1000)$ & \text{---} & $\mathcal{U}(0, 1000)$ \\
$\Delta T_{4}$ (K) & \text{---} & $\mathcal{U}(0, 1000)$ & \text{---} & $\mathcal{U}(0, 1000)$ \\
$\Delta T_{5}$ (K) & \text{---} & $\mathcal{U}(0, 1000)$ & \text{---} & $\mathcal{U}(0, 1000)$ \\
$\Delta T_{6}$ (K) & \text{---} & \text{---} & \text{---} & $\mathcal{U}(0, 1000)$ \\
$\Delta T_{7}$ (K) & \text{---} & \text{---} & \text{---} & $\mathcal{U}(0, 2000)$ \\
$\Delta T_{8}$ (K) & \text{---} & \text{---} & \text{---} & $\mathcal{U}(0, 2000)$ \\
$\Delta T_{9}$ (K) & \text{---} & \text{---} & \text{---} & $\mathcal{U}(0, 2000)$ \\
\tablebreak
\hline
Parameter & \multicolumn{4}{|c}{Chemical Composition} \\
\hline
log($X_i$) & $\mathcal{U}(-12, -1)$ & $\mathcal{U}(-12, -1)$ & $\mathcal{U}(-14, -0.3)$ & $\mathcal{U}(-14, -0.3)$ \\
\tablebreak
\hline
Parameter & \multicolumn{4}{|c}{Other Parameters} \\
\hline
R$_{\rm P, ref}$ (R$_{\rm Jup}$) & $\mathcal{U}(1.12, 2.15)$ & $\mathcal{U}(1.12, 2.15)$ & $\mathcal{U}(1.12, 2.15)$ & $\mathcal{U}(1.12, 2.15)$ \\
\tablebreak
\hline
\enddata
\tablecomments{For the M\&S (2009) T-P profile, $T_{\rm ref}$ is defined at the top of the atmosphere ($10^{-7}$\,bar) for \texttt{TauREx} and 10\,mbar for \texttt{POSEIDON}. For the BD T-P profile, \texttt{TauREx} uses a reference temperature at the top of the modeled atmosphere ($T_{\rm top}$ at $10^{-7}$\,bar) at 5 $\Delta T$ parameters (spaced between consecutive pressure pairs in $\log (P / \mathrm{bar}) = [-7, -4, -2, -1, 0, 1]$), while \texttt{POSEIDON} uses a reference photosphere pressure ($T_{\rm phot}$ at $10^{-1.5}$\,bar) and 9 $\Delta T$ parameters (using consecutive pressure pairs in $\log (P / \mathrm{bar}) = [-6, -5, -4, -3, -2, -1.5, -1, 0, 1, 2]$). The mixing ratio parameters, log($X_i$), include H$_2$O, CO, CO$_2$, CH$_4$, HCN, H$^{-}$, NH$_3$, FeH, VO, and TiO. R$_{\rm P, ref}$ is defined at the 10\,mbar pressure level for \texttt{POSEIDON} and at the bottom of the atmosphere (10\,bar) for \texttt{TauREx}.}
\end{deluxetable*}

\begin{deluxetable*}{lCCCCCCCCC}
\label{table:retrieval_results}
\tablecaption{Atmospheric retrieval results for WASP-17b's dayside emission spectrum using two different retrieval codes, different T-P profile parameterizations, and for both the positive and negative eclipse depth fitting methods.}
\tablehead{
    Retrieval Code & \multicolumn{4}{|c|}{\texttt{TauREx}} & \multicolumn{4}{c}{\texttt{POSEIDON}} \\
    \hline
    T-P Profile & \multicolumn{2}{|c|}{M\&S (2009) T-P} & \multicolumn{2}{c|}{BD T-P} & \multicolumn{2}{c|}{M\&S (2009) T-P} & \multicolumn{2}{c}{BD T-P} \\
    \hline
    Eclipse Depths & \multicolumn{1}{|c}{Positive} & \multicolumn{1}{c|}{Negative} & \colhead{Positive} & \multicolumn{1}{c|}{Negative} & \colhead{Positive} & \multicolumn{1}{c|}{Negative} & \colhead{Positive} & \colhead{Negative}
}
\startdata
\tablebreak
\hline
Parameter & \multicolumn{8}{|c}{Temperature-Pressure Profile} \\
\hline
T$_{\rm ref}$ (K) & $1354^{+73}_{-76}$ & $1103^{+73}_{-62}$ & \text{---} & \text{---} & $2468^{+101}_{-129}$ & $1920^{+96}_{-63}$ & \text{---} & \text{---} \\
$\alpha_{1}$ & $0.74^{+0.16}_{-0.16}$ & $0.60^{+0.05}_{-0.04}$ & \text{---} & \text{---} & $0.14^{+0.01}_{-0.01}$ & $0.12^{+0.03}_{-0.03}$ & \text{---} & \text{---} \\
$\alpha_{2}$ & $0.35^{+0.04}_{-0.03}$ & $0.41^{+0.12}_{-0.13}$  & \text{---} & \text{---} & $1.09^{+0.50}_{-0.52}$ & $1.20^{+0.46}_{-0.50}$ & \text{---} & \text{---} \\
log(P$_{1}$ / bar) & $-6.66^{+0.39}_{-0.23}$ & $-2.14^{+0.72}_{-0.60}$ & \text{---} & \text{---} & $-1.32^{+0.59}_{-0.33}$ & $-2.98^{+0.55}_{-0.52}$ & \text{---} & \text{---} \\
log(P$_{2}$ / bar) & $-6.26^{+0.39}_{-0.32}$ & $-1.21^{+0.95}_{-0.95}$ & \text{---} & \text{---} & $-1.32^{+0.59}_{-0.33}$ & $-2.96^{+1.75}_{-1.68}$ & \text{---} & \text{---} \\
log(P$_{3}$ / bar) & $-1.23^{+0.88}_{-0.37}$ & $0.04^{+0.65}_{-0.98}$  & \text{---} & \text{---} & $0.64^{+0.78}_{-0.96}$ & $0.33^{+0.98}_{-1.19}$ & \text{---} & \text{---} \\
\tablebreak
T$_{\rm top}$ (K) & \text{---} & \text{---} & $1123^{+101}_{-79}$  & $1110^{+109}_{-72}$  & \text{---} & \text{---} & \text{---} & \text{---} \\
T$_{\rm phot}$ (K) & \text{---} & \text{---} & \text{---} & \text{---} & \text{---} & \text{---} & $2503^{+159}_{-126}$ & $2331^{+318}_{-251}$ \\
$\Delta T_{1}$ (K) & \text{---} & \text{---} & $564^{+96}_{-119}$ & $588^{+104}_{-161}$ & \text{---} & \text{---} & $519^{+314}_{-323}$ & $464^{+318}_{-251}$ \\
$\Delta T_{2}$ (K) & \text{---} & \text{---} & $315^{+52}_{-51}$ & $338^{+103}_{-88}$ & \text{---} & \text{---} & $516^{+243}_{-199}$ & $490^{+249}_{-195}$ \\
$\Delta T_{3}$ (K) & \text{---} & \text{---} & $527^{+63}_{-68}$ & $439^{+334}_{-289}$ & \text{---} & \text{---} & $67^{+109}_{-49}$ & $109^{+120}_{-75}$ \\
$\Delta T_{4}$ (K) & \text{---} & \text{---} & $134^{+131}_{-87}$ & $498^{+301}_{-301}$ & \text{---} & \text{---} & $457^{+177}_{-363}$ & $190^{+234}_{-129}$ \\
$\Delta T_{5}$ (K) & \text{---} & \text{---} & $474^{+305}_{-291}$ &$513^{+296}_{-312}$ & \text{---} & \text{---} & $216^{+285}_{-168}$ & $219^{+270}_{-156}$ \\
$\Delta T_{6}$ (K) & \text{---} & \text{---} & \text{---} & \text{---} & \text{---} & \text{---} & $108^{+183}_{-80}$ & $242^{+281}_{-173}$ \\
$\Delta T_{7}$ (K) & \text{---} & \text{---} & \text{---} & \text{---} & \text{---} & \text{---} & $213^{+256}_{-151}$ & $267^{+326}_{-189}$ \\
$\Delta T_{8}$ (K) & \text{---} & \text{---} & \text{---} & \text{---} & \text{---} & \text{---} & $310^{+349}_{-218}$ & $273^{+321}_{-194}$ \\
$\Delta T_{9}$ (K) & \text{---} & \text{---} & \text{---} & \text{---} & \text{---} & \text{---} & $354^{+407}_{-253}$ & $343^{+404}_{-247}$ \\
\tablebreak
\hline
Parameter & \multicolumn{8}{|c}{Chemical Composition} \\
\hline
log(H$_{2}$O) & $-1.46^{+0.31}_{-0.45}$ & $-1.50^{+0.34}_{-0.54}$ & $-2.01^{+0.24}_{-0.23}$  & $-1.37^{+0.25}_{-0.43}$ & $-0.89^{+0.35}_{-0.46}$ & $-1.61^{+0.70}_{-0.69}$ & $-1.40^{+0.94}_{-0.85}$ & $-1.15^{+0.50}_{-0.47}$ \\
log(CO) & $<-1.0$ & $<-1.0$ & $<-1.0$ & $<-1.0$ & $< -2.04$ & $< -1.36$ & $< -1.85$ & $< -1.62$ \\
log(CO$_2$) & $-1.96^{+0.47}_{-0.69}$ & $-2.10^{+0.52}_{-0.76}$ & $-2.63^{+0.42}_{-0.67}$  & $-2.26^{+0.51}_{-1.53}$ & $< -1.95$ & $< -1.12$ & $< -2.20$ & $< -1.41$ \\
log(CH$_4$) & $<-2.05$ & $<-1.86$ & $<-2.97$ & $<-1.89$ & $<-2.90$ & $< -2.76$ & $< -3.40$ & $< -3.25$ \\
log(HCN) & $<-1.13$ & $<-0.87$  & $<-1.60$ & $<1.0$ & $< -3.21$ & $< -3.22$ & $< -2.67$ & $< -2.61$ \\
log(H$^-$) & $<-1.0$ & $<-1.0$ & $<-1.0$ & $<1.0$ & $< -9.07$ & $< -3.18$ & $< -8.21$ & $< -7.07$ \\
log(NH$_3$) & $-2.04^{+0.34}_{-0.43}$ & $-2.43^{+0.42}_{-0.59}$  & $-2.44^{+0.26}_{-0.28}$ & $-2.21^{+0.32}_{-0.57}$  & $< -2.42$ & $< -1.88$ & $-2.15^{+0.50}_{-0.42}$ & $-2.13^{+0.49}_{-0.44}$ \\
log(FeH) & $-4.75^{+0.77}_{-1.33}$ & $< -2.81$ & $-7.06^{+0.95}_{-2.05}$& $-6.40^{+1.74}_{-2.89}$ & $-3.72^{+1.06}_{-1.66}$ & $< -3.58$ & $-5.25^{+1.20}_{-2.99}$ & $< -2.41$ \\
log(VO) & $<-7.96$ & $<-5.58$  & $<-8.40$  & $< -5.47$ & $< -8.33$ & $< -6.66$ & $< -7.82$ & $< -5.77$ \\
log(TiO) & $<-9.31$ &$<-5.25$ &$<-9.98$ & $< -5.69$ & $< -9.03$ & $< -5.70$ & $< -9.33$ & $-6.83^{+0.96}_{-2.80}$ \\
\tablebreak
\hline
Parameter & \multicolumn{8}{|c}{Other Parameters} \\
\hline
R$_{\rm P, ref}$ (R$_{\rm Jup}$) & $2.03^{+0.08}_{-0.12}$ & $2.04^{+0.07}_{-0.12}$ & $2.02^{+0.08}_{-0.11}$ & $2.04^{+0.07}_{-0.12}$ & $1.49^{+0.06}_{-0.04}$ & $1.8^{+0.2}_{-0.2}$ & $2.06^{+0.06}_{-0.09}$ & $2.1^{+0.1}_{-0.1}$ \\
\tablebreak
\hline
\tablebreak
\hline
\multicolumn{9}{c}{Statistics} \\
\hline
reduced $\chi^2$ & $1.56$ & $1.86$ & $1.52$ & $1.86$ & $1.61$ & $1.86$ & $1.48$ & $1.87$ \\
ln$\mathcal{Z}$ & $1256.5$ & $1210.7$ & $1260.6$ & $1213.0$ & $1249.6$ & $1205.0$ & $1269.2$ & $1214.0$ \\
BIC & $301$ & $343$ & $296$ &  $342$  & \text{---} & \text{---} & \text{---} & \text{---} \\
%BIC & $301$ & ADD & ADD & ADD & $314$ & ADD & $284$ & ADD \\
DoF & $138$ & $138$  & $138$ & $138$ & $138$ & $138$ & $134$ & $134$ \\
\tablebreak
\hline
\enddata
\tablecomments{For the M\&S (2009) T-P profile, $T_{\rm ref}$ is defined at the top of the atmosphere ($10^{-7}$\,bar) for \texttt{TauREx} and 10\,mbar for \texttt{POSEIDON}. For the BD T-P profile, \texttt{TauREx} uses a reference temperature at the top of the modeled atmosphere ($T_{\rm top}$ at $10^{-7}$\,bar) at 5 $\Delta T$ parameters (spaced between consecutive pressure pairs in $\log (P / \mathrm{bar}) = [-7, -4, -2, -1, 0, 1]$), while \texttt{POSEIDON} uses a reference photosphere pressure ($T_{\rm phot}$ at $10^{-1.5}$\,bar) and 9 $\Delta T$ parameters (using consecutive pressure pairs in $\log (P / \mathrm{bar}) = [-6, -5, -4, -3, -2, -1.5, -1, 0, 1, 2]$). Constraints for non-detected chemical species (i.e. not favored by Bayesian model comparisons) are quoted as 2\,$\sigma$ abundance upper limits. R$_{\rm P, ref}$ is defined at the 10\,mbar pressure level for \texttt{POSEIDON} and at 10\,bar for \texttt{TauREx}.}
\end{deluxetable*}

\end{document}